\documentclass[10pt]{article}

\usepackage{graphicx,capt-of}
\usepackage{color}
\usepackage{bm}
\usepackage[font=Large]{subfig}
\usepackage{tabularx,ragged2e,booktabs,caption}
\usepackage{grffile}
\usepackage{amsmath}
\usepackage{cite}
\usepackage[toc]{appendix}
\usepackage{float}

\def\beq{\begin{equation}}
\def\eeq{\end{equation}}
\def\ba{\begin{eqnarray}}
\def\ea{\end{eqnarray}}

\usepackage[english]{babel}

\begin{document}

\begin{center}
{\Large{\bf Enhancement of Thermal Conductivity in Polymer Composites
 by Maximizing Surface-Contact Area of Polymer-Filler Interface}}\\
\ \\
by \\

Vijendra Kumar$^1$, Abhishek Barnwal$^2$, R. K. Shukla$^1$ and Jyoti Shakya$^3$ \\
$^1$Department of Physics, University of Lucknow, Lucknow -- 226007, India. \\
$^2$Department of Chemical Engineering, Indian Institute of Technology Delhi, New Delhi -- 110016, India.\\
$^3$Department of Physical Sciences, Indian Institute of Sciences, Bangalore -- 560012, India.
\end{center}

\begin{abstract}

In this article we discuss in detail the effective approaches to enhance the thermal conductivity in polymer composites. It is shown from numerical simulations that maximizing interfacial area between filler and polymer enhances very significantly the effective thermal conductivity in composites. Our study outlines two main facts.
(a) Although the nature of the filler's geometry plays an important role in the effective thermal conductivity, we show that among the different geometries thermal conductivity is high for those geometries for which the ratio of surface-area to volume is high. Thus non-spherical shaped fillers show high thermal conductivity compared to the spherical fillers.
(b) For fillers of a particular geometry, by maximizing its surface area without changing the volume fraction of the metallic filler, the effective thermal conductivity increases. Thus, the interfacial area between filler and polymer plays an important role in the enhancement of thermal conductivity. Maximizing interfaces facilitates more routes for heat conduction through the metallic filler. Thus filler material can be transformed to result into more surface such that more interfaces between the filer polymer can be obtained. It is also observed that as this interfacial area increases, increase in effective thermal conductivity follows from linear to the logarithmic growth. It should be noted that to inherit the polymer properties there is a restriction on the upper bound of volume fraction of the fillers. The current study bring out an important step in this direction. Our results are technologically very important in designing composite polymers for better heat conduction, and are very cost-effective. This study also provides a connection between the bulk and the surface area in effectively determination of the thermal conductivity.
\end{abstract}
\newpage
\section{Introduction}
Polymers are the integral part of day to day life in the modern era\cite{APS1}. They are extraordinarily functional materials which span a wide range and a wide class of materials based on their physical and chemical properties \cite{APS2}. The properties of a given type of polymer may strikingly different for different polymers\cite{APS3,APS4}. The molecular structure of a polymer is one of the main factor deciding its chemical and the physical properties. Due to their easy
synthesis processes, low ­cost fabrication, light ­weight, robustness, and excellent chemical stability, polymers are now being used heavily in many engineering and mechanical applications\cite{APS4,APS5}. In this regard the recent studies show that polymers have emerged as very good heat exchangers by addition of metallic fillers in the base polymer, and are being used in many electronic and mechanical devices\cite{APS2,APS3,APS4,APS5,APS6}. Thermal conductivity is one of the most important property to be calibrated in the heat transfer process\cite{APS7,APS8,APS9,APS10}.

Temperature dependence of polymers originates from their molecular structures. Polymers in their pure state have low thermal conductivity. This is due to slow propagation of phonons in the polymers and the movement is limited upto short distances. On the other hand, heat conduction in solids such as metals is carried out mainly by phonons and these phonons can move from one end to other end instantly. Due to the poor crystalline structure and the complex morphology of polymer chains phonons can not diffuse easily in the polymer \cite{APS11,APS12,APS13}. The thermal conductivity in bulk polymers is usually very low on the order of 0.1 ­ 0.5 W/m/K\cite{APS2,APS3,APS4,APS5,APS6}.

Due to the immense utility of polymers in heat conduction numerous research have focussed on enhancing the thermal conductivity of polymers by synthesizing their composites. This is achieved by introducing high conducting metallic fillers in matrix polymer. Although the polymers with metallic fillers have high thermal conductivity, there are limitations on the use of amount of filler (i.e., volume fraction of the fillers). This also limits the maximum increase in the thermal conductivity\cite{APS14,APS15,APS16,APS17,APS18,APS19,APS20,APS21,APS22}.
To address this issue, there have been many developments to increase the conductivity of polymers based on the geometry and shape of fillers. It has been observed that shapes of fillers considerably change the polymer conductivity. In this regard there has been search of the best shapes for fillers. In general, cubes and spherical fillers have been used to enhance the conductivity. Apart from these general shapes there are other shapes in the form of alphabets such as Y, I, T. The shape dependence of thermal conductivity has much quest for the search of best shape fillers\cite{APS5}.

However, the basic question remains the same, i.e., what is the best shape filler?. The important question that should be asked is why fillers with different shapes give different thermal conductivity and what is specific about this shape dependence thermal conductivity. What is the physical structure of filler that changes the thermal conductivity.
There is also an interesting fact that the symmetry in the shape of fillers plays an important role in heat conduction. It is observed that more asymmetricity in the filler shape brings more increase in the thermal conductivity. These geometries and shapes of fillers suggest that heat conduction in the composite system is path dependent \cite{APS6}. In this article we address these issues. We investigate the possible routes to enhance the thermal conductivity in polymer composites keeping the same amount (i.e., the constant volume fraction) of the filler material. 

The basic idea is to show that the thermal conductivity in the composites can be maximized by maximizing the interfacial surface area between the metallic filler and polyer material \cite{APS23,APS24,APS25,APS26,APS27,APS28,APS29,APS30,APS31}. We show that this can be achieved by using hollow fillers instead of solid fillers. In this case polymer is filled within as well outside the hollow filler. By adopting this procedure we create more interfaces between the polymer and filler. We generalize this mechanism to the arbitrary shape fillers. It is easy to observe that the hollow shape fillers facillitate more contact with the base polymer in the composite materials \cite{APS6,APS7}. For example, when we have solid filler in the shape of a cylinder only the outer surface of the filler is in contact with polymer matrix. But when we have a hollow cylinderical filler the interfacial area is doubled because now polymer is filled within as well as outside the hollow cylinder. Thus the interface area between the polymer and filler is increased thereby facilitating more routes for heat conduction through the metallic path leading to enhancement in the conductivity. Thus if we take a solid cylinderical filler, transform it into a thin sheet and again replace it inside the base polymer then it is observed taht there is a drastic incerase in the thermal conductivity of the composite. We can further reduce the thickness of this sheet by expanding it to get more surface area. However, there is a limitation on the minimum thickness. The thickness of the filler material sould be great enough such that it retains its basic property.

The fundamental cause behind this increase is accounted to the thermal excitations of molecules in the polymer chains. As the polymer molecules come into contact with a heat bath they gain local thermal excitations. 
If the number of molecules coming into contact with heat bath is large the larger excitations is possible. This is possible when the contact area/inetrface between the polymer and metallic filer is large. Note that the metallic filler works here as a heat bath for the polymer molecules. If the interfacial area is doubled the number of polymer molecules interacting with the metallic filler (i.e., the heat bath) is doubled. 
When we use thick solid filler, such as a solid cylinder or a solid cube, only the outer surface of the fillers are in contact with the base polymer and therefore there is less inetraction between filler-polymer. Thus interfacial area at the boundaries is an essential property for facillitating larger 
heat conduction. 
Indeed the material property and the amount of the filler (i.e., the bulk or volume fraction of the filler) are important fcators for thermal conduction,\cite{APS32,APS33,APS34,APS35,APS36,APS37,APS38,APS39} however it is the interfacial area between the polymer and the filler material in the composite which plays an extra role in maximization of the thermal conductivity. To establish these facts we replace the solid cubical fillers in the form of multiple sheets having gap in-between them and then these sheets are embedded in the matrix polymer. By doing so we have more parallel routes for heat conduction.

Polymer composites are multi­phase systems consisting of matrix polymer and reinforcing material. In polymer composites matrix polymer is an homogeneous phase which includes fillers such as metals, inorganic non­metallic materials etc. Note that one of the main reasons of using composite polymers in different chemical and mechanical systems is to avoid corrosion\cite{APS40,APS41,APS42,APS43,APS44,APS45,APS46}. In some cases ceramics has been used as filler material to avoid carbon or metallic fillers to get an electrical isolator. In this article we discuss the polymer composites having only metallic fillers.  
We would like to stress the fact that composite material is a new phase which consists of materials of different components and phases with distinct physical and chemical properties. The important thing to note is that, composites not only constitute the basic properties of the original component, but they also show new behaviours which may not be constituted by any of the original components. A composite material should have the following important properties (a) they should be microscopically nonhomogeneous and should constitute distinct interfaces (b) they should have improved performance which might be clearly distinct from the original base material and (c) the volume fraction of each component are such that they should not change the properties of base polymer. The volume fraction of filler in the current study does not exceeded $10\%$ of the whole composite.
\section{Heat Transfer Equations}

\begin{figure}[H]
\begin{center}
\includegraphics*[width=0.75\textwidth]{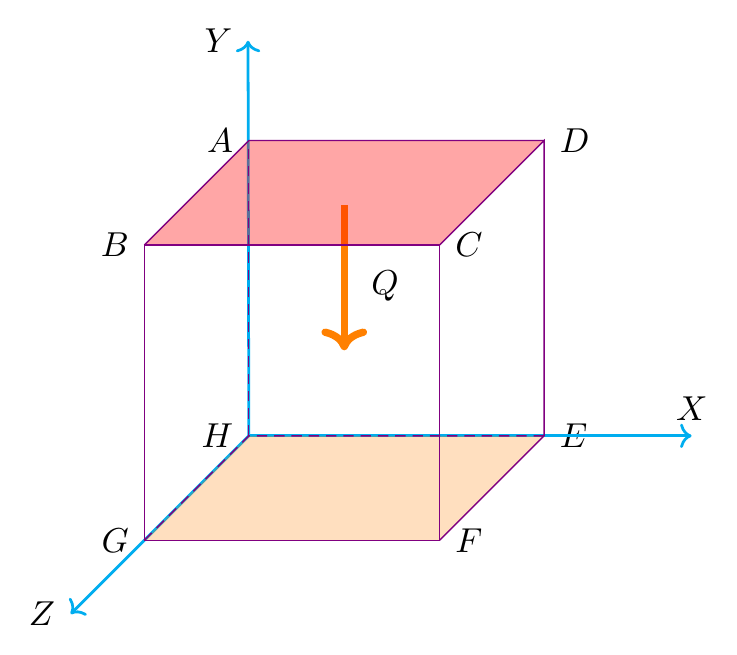}
\end{center}
\caption{ A schematic diagram showing the polymer-composite in the shape of cubical box. The box is made of polymer and metallic fillers are used within (as discussed and shwon in later discussions). The top and bottom surfaces are shown with the two different colors, representing the two different temperatures which provide the gradient for heat flow. The direction of heat-flux is from top to the bottom surface in the negative z-direction. Top surface is kept at the constant temperature at $T=400$ K.}
\label{f1}
\end{figure}

In this section we describe the equations governing the heat transfer in 3-D geometries. Figure.~\ref{f1} shwows a schematic diagram which has been used as our model system to describe the heat conduction from the top to the bottom surface in the composite-polymers considered in the present manuscript. The heat-flux is constrained in only one direction (i.e., from top to bottom surface or from left to the right surface) and the correponding heat equations are modified with the suitable boundary conditions. All the other surfaces are considered to be insulated and do not conduct heat flux, hence the corresponding boundary conditions for heat flux from these surfaces are set to be zero.

As shown in fig.~\ref{f1}, in the generation of the three-dimensional cubic cell, the heat transfer in the base or at the bottom surface of the polymer matrix is satisfied by the following equation:
\begin{equation}\label{eq1}
\frac{K_{p}} {(\rho c_{p})_{p}} \Bigg[\frac{ \partial^{2} T_{p}}{\partial x^{2}} + \frac{\partial^{2} T_{p}}{\partial y^{2}} + \frac{\partial^{2} T_{p}}{\partial z^{2}} \Bigg]=0
\end{equation}
where $T$ is the temprature and $x,y,z$ are the spatial coordinates. The subscript $p$ here refers the base polymer-matrix. The parameter such as $K$ refers the thermal conductivity, $\rho$ is the density and $c_{p}$ is the specific heat. These parameters better reflect the local properties of the $3D$ grid cell. In our numerical simulations the thermal conductivity of the matrix-polymer $K_{p}$ is set to the $0.29$ W/m-K which corresponds to the thermal conductivity of the \textit{polyethylene}.

The heat transfer equation corresponding to the high conducting filler material is similar to the Eq.~(\ref{eq1}) and is given by:
\begin{equation}\label{eq2}
\frac{K_{f}} {(\rho c_{p})_{f}} \Bigg[\frac{ \partial^{2} T_{f}}{\partial x^{2}} + \frac{\partial^{2} T_{f}}{\partial y^{2}} + \frac{\partial^{2} T_{f}}{\partial z^{2}} \Bigg]=0.
\end{equation}
Here the subscipt $f$ refers the filler. The filler material used in the current study is the \textit{Aluminium}, and the corresponding thermal conductivity $K_{f}$ is given as $205.0$ W/m-K. Note that Aluminium is very soft and non-magnetic metal. It can be given any shape due to its soft and ductile nature. Also, Aluminium is used as a good heat-exchanger in many devices due to it's high conductivity and low-cost alternative. Although copper is the excellent thermal conductor ($385.0$ W/m-K) but it is less cost-effective and hence not widely used on the large scale. Other metals which can be used as filler materials are brass ($109.0$ W/m-K), iron ($79.50$ W/m-K), and steel ($50.20$ W/m-K) but their thermal conductivities are very low. 
The boundary condition for the top surface ABCD corresponds to the constant temperature which is set at $T=400$ K. Another boundary condition for the bottom surface EFGH is through convective heat transfer with a constant heat-transfer coefficient. 

\begin{equation}\label{eq4}
K_{p} \frac{\partial T}{\partial n} \bigg|_{\Gamma} = h(T-T_{f})|_{\Gamma}. 
\end{equation}
Here the convective heat transfer coefficient $h$, and the ambient extensive temperature $T_{f}$ are set to be constants. In our numerical simulations, we have set $h$ to be $20$ W/\(m^{2}\)k, and $T_{f}$ is set to be $300$K. Apart from these two specific bounday conditions (i.e., for top and bottom surfaces), we have four more boundary conditions for the remaining planes corresponding to each of the following surfaces EDAH, CDEF, BCFG, and ABGH, which can be written as,
\begin{equation}\label{eq5}
\frac{\partial T}{\partial n}\bigg|_{\Gamma} = 0.
\end{equation}
Here $\Gamma$ represents the four surfaces mentioned above. These boundary conditions clearly mention that the surfaces are adibatic and there is no heat flux through these surfaces. 
\subsection{Heat Transfer Coupling}
Note that at the interface of the polymer and filler material, i.e., at the contact surface area, the heat-flux is same. Heat-flow in the base polymer and the filler are thus coupled. This coupling is very important to determine the effective thermal conductivity of the composite polymer-filler material. This can be written explicity as follows
\begin{equation}\label{eq6}
{-K_{p}} \frac{\partial T_{p}}{\partial n}\bigg|_{\Gamma} = {-K_{f}} \frac{\partial T_{f}}{\partial n}\bigg|_{\Gamma}
\end{equation}
where subscipt $\Gamma$ represents the contact surface amid the base matrix and the fillers. When the temperature fields in the cell are calculated, the effective thermal conductivity in $z$ direction for the cell is estimated by the following equation 
\begin{equation}\label{eq7}
{K_{c}}= \frac{L_{y}}{\bigtriangleup T_{cell}}~Q
\end{equation}
where $\Delta T$ is the mean temperature difference between the top and bottom surface of the cell. The surface represents the area of the plane such as \(L_{x}\)-\(L_{y}\). Dimensions along the x axis is \(L_{x}\), dimensions along the y axis is \(L_{y}\) and dimensions along z axis is \(L_{z}\) which are the
cell length, height and width respectively. We also write the thermal conductivity explicitly as
 \begin{equation}\label{eq8}
 Q= -KA \frac{\Delta T}{\Delta X}
 \end{equation}
\begin{equation}\label{eq9}
K= -\frac{Q}{A} \frac{\Delta X}{\Delta T}.
\end{equation} 
where $K$ is thermal conductivity (W/m-k), and $Q$ is the total heat transfer. $\Delta T$ is the difference in temperature $T_{2}-T_{1}$. $T_{1}$ is fixed at 400K, and $T_{2}$ is variable.
 A = area = 0.01 x 0.01 \(m^{2}\). $\Delta X$ is thickness through which heat conduction takes place(m)=(0.01 m). here K=\(K_{c}\) (thermal conductivity of composites material, and $K_{c}/K_{p}$ is the relative thermal conductivity.
\subsection{Geneartion of Grid-Cell }
We consider the polymer-composite in the form of a cubical box of side $a=0.01$m. 
This is used as the model system, and is taken as the 3-D simulation box.
In our numerical simulation the volume $V$ of the box is $0.01\times 0.01\times 0.01$ $m^{3}$. For our purpose the voulme of the box is sufficient for the 3-D numerical simulations in the thermal conductivity calculations. Therefore the volume $V$ defines the total volume of the composite polymer. In absence of any filler material this volume corresponds to the volume of the matrix-polymer. 
Therefore volume fraction $(V_{f})$ of filler material is zero in this case, and the thermal conductivity corresponds to the bare matrix polymer. Note that the $V_{f}$ denotes the volume of metalic filler inserted in the matrix polymer. Therefore the remaining polymer matrial in the composite thus formed is $(V-V_{f})$. 
The fillers used in this manuscript vary in shape and size. The general shape of the fillers are spherical, cylinderical, and planar sheets. In this category we use both hollow as well as solid shapes. Also, the number of these fillers vary, i.e., a single cylinder as well as multiple cylinders, single sheet as well as multiple sheets. We also use multiple boxes which are isolated. The cylinders may be concentric as well as isolated. Apart from these general shapes we use fillers in alphabatical forms such as I, T, X, and Y shapes. These filler may take random positions or they can be put in more symmetric fashion\cite{APS5,APS6,APS7}.
The matrix-polymer used in our numerical simulations is polyethylene whose thermal conductivity is
$0.29$ W/m-K. The density of the polyethylene is $938$ Kg/m$^{3}$, and the specific heat is $1900$ J/Kg/K. The filler material as described before is taken as aluminium whose thermal conductivity is $205$ W/m-K with the density $2700$ Kg/m$^{3}$, and specific heat $990$ J/Kg/K.  
Once the base polymer and the filler materials are slected we use numerical simulation technique described in next subsection to study the thermal conductivity of the composite material.
We calculate the relative thermal conductivity $K_{c}/K_{p}$ for the different cases considered in this manuscript. Note that $K_{c}$ is the thermal conductivity of the composite material after using the filler, and $K_{p}$ denotes the thermal conductivity of the base polymer without any filler.
We mention that the volume percent of the filler material varies from $1\%$ to $9\%$ which are below $10\%$. The reason behind using less volume percent is clear in the sense that is does not destroy the inheretic properties of the base polymer in the composite polymer mixture.  
The variation in shape and size of the fillers constitute the building blocks in the study of thermal conductivity. Also, in the generation of fillers we have used various geometries with different width and height. The height (i.e., the vertical length) of the cylinderical shape fillers varies from $0.5$ cm to $0.8$ cm. Similarly the width (i.e., the horizontal size or the diameter) of the cylinderical shape filler varies again from $0.5$ to $0.8$cm. We have taken hollow multiple boxes of volume $0.1\times0.1\times0.1$ cm$^{3}$, with thickness of $0.05$cm, and solid multiple box of volume fraction $0.1\times0.1.\times0.1$ cm$^{3}$.
The fillers with the shapes in alphabetical forms such as I, X, Y, T are also generated with vertical length $0.5$ cm and horizontal length $0.65$ cm with varying thicknesses according to the filler content.

Cylinderical shape fillers are generated with volume fractions $(1\%$-$9\%)$ and with varying radius $($0.135$, $0.169$, $0.195$, $0.212$, $0.2285$, $0.2429$, $0.278$)$ in cm.
Fillers such as planar sheet with varying numbers $1, 2, 4, 8, 12, 16 , 20, 30, 40$, and with varying thicknesses $0.05$, $0.025$, $0.0125$, $0.00625$, $0.00425$, $0.003125$, $0.0025$, 
$0.001666667$, $0.00125$ have been used also. The vertical length of the these sheets is $0.5$ cm.
These calculations are repeated with different voulme fractions.
As an example hollow cylinderical fillers with a voulme fraction of $4\%$ and vertical length 0.5cm, varying in numbers  $1$, $2$, $4$, $6$, $8$, $10$, $12$, $16$, $20$, $24$, $30$ with thicknesses in order given as $0.11723$, $0.732$, $0.0416$, $0.03212$, $0.025382$ $0.021034$, $0.1797$, $0.13965$, $0.011427$, $0.009675$, $0.007869$ (cm) have been studied. Again these calculation are repeated with different volume fractions. All the calculations correspond to the 3-D filler model.

\subsection{Numerical Simulation}
We numerically study the  effect of filler on thermal conductivity of polymer in a composite material by Ansys 18.0 using three dimensional models. Almunium (Al) is used as a filler material to study the effect of thermal conductivity. The aim of this study is to observe the effect of shape, size, the composition, and the position of the filler in the thermal conductivity of the composites.
By varying these parameters one can control the thermal conductivity. We use fillers of various shapes and sizes in the current study. The fillers are, single and multiple solid box (cube), single and multiple hollow box, single solid cylinder and single hollow cylinder. We also use fillers with different symmetric properties such as I, T, X, Y, and spherical shape.
The polymer matrix is filled in a cubical box of of size 0.01$\times$0.01$\times$0.01 m$^3$. 
Inside this polymer matrix the fillers are inserted to study the conducting behaviour of composite system. 
We consider our system (which is made of polymer and filler ) as a large mesh composed of small cubical cells also called the grids. The temperature of each cell is kept at a constant value, e.g., $T=300K$ initially. Therefore initially the whole mesh and all the grid points are at the same temperature. 
There is no gradient for flux at this stage. 
Now we define the boundary conditions for the top and bottom surfaces which create a gradient for the flow of heat. Top surface is fixed at a constant high tempearture $(400K)$. The bottom surface has a convective heat flow beacuse it is surrounded by the air which is at the $300K$ $(T_\infty)$. The heat transfer coefficient at the bottom surface is $20 W/m^2/K$. All the other surfaces are insulated and do not transfer heat. 
The finite element method is used to solve the heat Eqns.~(\ref{eq1})-(\ref{eq5}) in different region starting from polymer to filler\cite{APS1,APS22,APS27}.
The heat equation is solved in each grid, i.e., in each of the cell. Initially the top surface which is at $400K$, also corresponds to the initial temperature of the each cell $(T_{i+1})$ which are on the top surface. All the regions bellow the top surface or the cells are initially at 300K.  Now the intermediate temperature at the center of each cell which is $T_{i }$ (which is also a variable quantity at each time) is calculated by the iteratative heat equation using finite element method.  
The heat equation is solved simulatneously in each cell for the steady state. The simulation stops when the residual (i.e., the error in temperature calibration calculated from the heat equation) attains the minimum values (which is set to a fixed value in the begining of the experiment). 
From the average temperature of the surfaces we get the thermal conductivity of composite polymer by using the Fourier law of heat transfer equation.  
The three dimensional  conduction heat transfer Eqns.~(\ref{eq1})-(\ref{eq2}) are solved for the base polymer where $K_{f} = K_{p}$. The boundary condition to solve above equation is that between the planes $\Gamma$= ABCD, EFGH heat transfer is by convection.
\begin{equation}
K_{p}\frac{\partial T}{\partial n}|_{\Gamma}= h(T-T_{\infty}).
\end{equation}
The top surface is maintained at a temperature $400$K. The other four surface of the composite cube are assumed adiabatic. i.e  for all four planes $\Gamma^{\prime}$= EHDA, HGCD, FGCB, and EFBA, we have
\begin{equation}
\frac{\partial T}{\partial n}|_{\Gamma^{\prime}}=0.
\end{equation}                       
Heat flux at the interface, i.e., at the contact between filler and polymer surfaces are same. This can be expressed as $-K_{p}(\frac{\partial T}{\partial n})|_{\gamma} =-K_{f}(\frac{\partial T}{\partial n})|_{\gamma}$. Now by using heat transfer equations and the above mentioned boundary conditions the average temprature in the composite is calculted using the flux between the top and bottom of the polymer composites. The effective thermal conductivity is calculated in the negative $Z$-direction.
\section{Results and Discussion}
\begin{figure}[H]
\begin{center}
\includegraphics*[width=.8\textwidth]{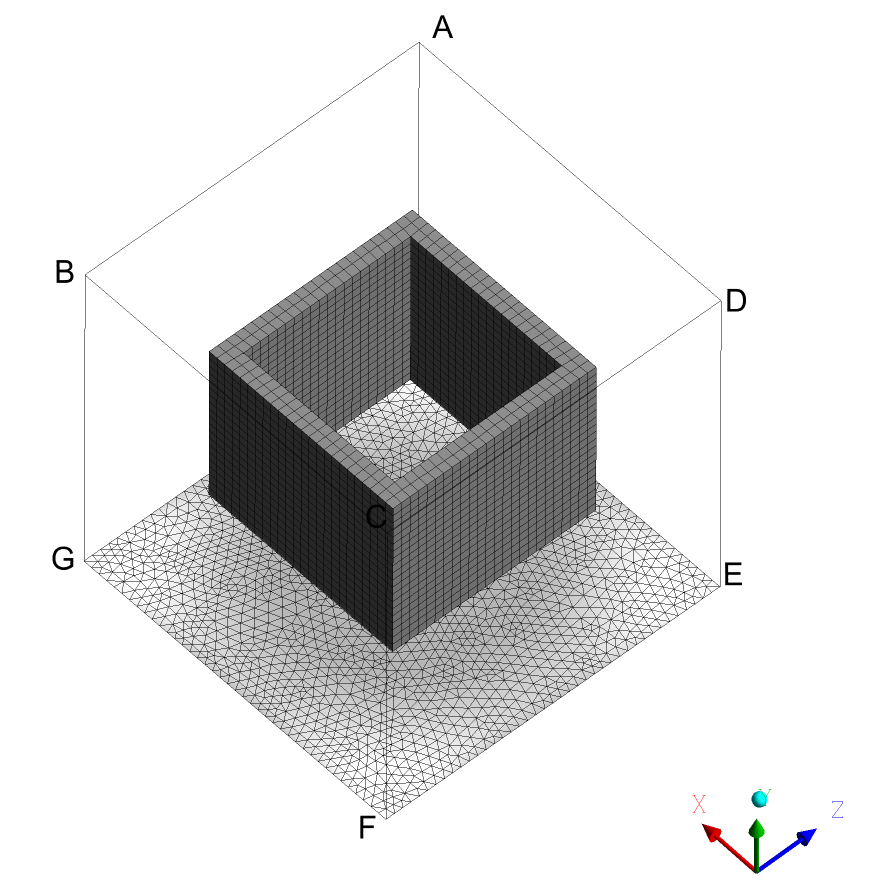}
\end{center}
\caption{ The mesh structure of a composite polymer which has a metalic filler in the shape of a hollow cubical box. Note that the empty volume is filled matrix of polymer. the base polymer is filled within and outside the filler. The contact surface area between polymer and filler has been maximized in this construction since the contact surface has been increased two fold by making the box hollow.The volume percent of the filler is $6\%$. Note that top and bottom contact surface area is reduced but the sideways has two-fold. This has remarkabe effect in the enhancement of thermal conductivity of the composite polymer.}
\label{f2}
\end{figure}

\begin{figure}[H]
\begin{center}
\includegraphics*[width=1.1\textwidth]{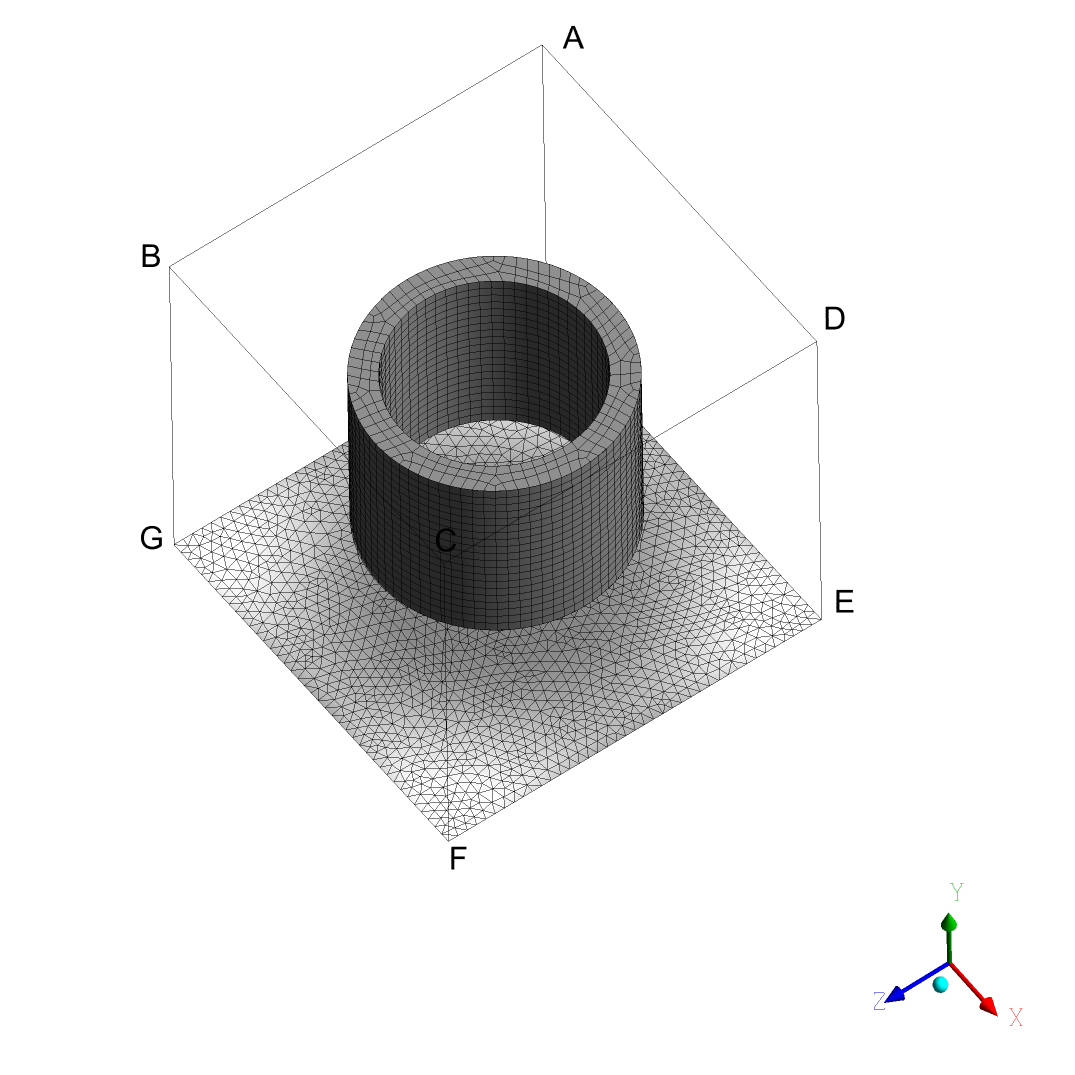}
\caption{ The mesh structure of a composite polymer which has a metalic filler in the shape of hollow cylinder. In the whole empty voulme the base polymer is filled again. Therefore polymer is filled within and outside the cylinderical area. Note that contact area between polymer and filler has been two-fold increased in this case also. The bulk inside is replaced by the base polymer. This also suggests the procedure to maximize the surface area by changing the bulk (which is not in contact)  into surface by increasing the diameter of the cylinder or by increasing it's vertical height by keeping the same volume percent. The volume percent of filler in this case is only $6\%$. }
\label{f3}
\end{center}
\end{figure}

\begin{figure}[H]
\centering
   \includegraphics[width=0.66\textwidth]{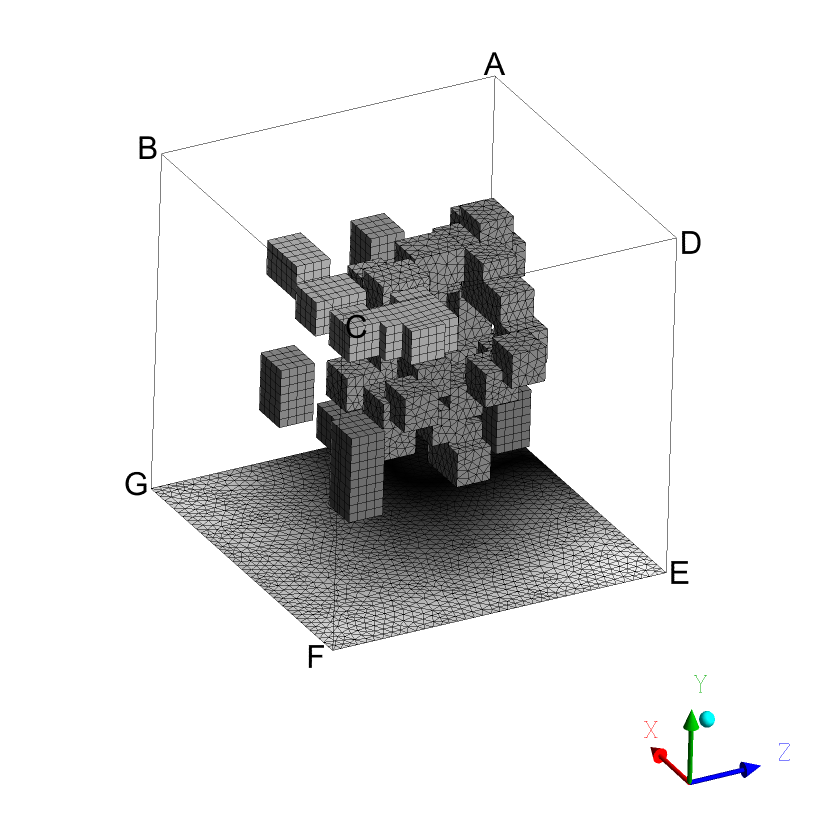}
   \caption {mesh structure of polymer composite having fillers in the form of the generalized box. Only bottom surface is shown. In the present geometry the conatct surface area is low. The surface contact area between polymer and filler is low since the boxes are connected, and polyer-filler intrface is reduced. Note that the thermal conductivity of generalized box and multiple box clearly indicate that it is not the bulk but the surface area in the contact which maximizes the thermal conductivity. 
The mesh structure of generalized box figure showimg only bottom surface and filler mesh in the cell with 6\% volume fraction with box size 0.1x0.1x0.1 \(cm^{3}\) and filled filler in cell of 0.5cm.}
   \label{fig4} 
   \includegraphics[width=0.66\textwidth]{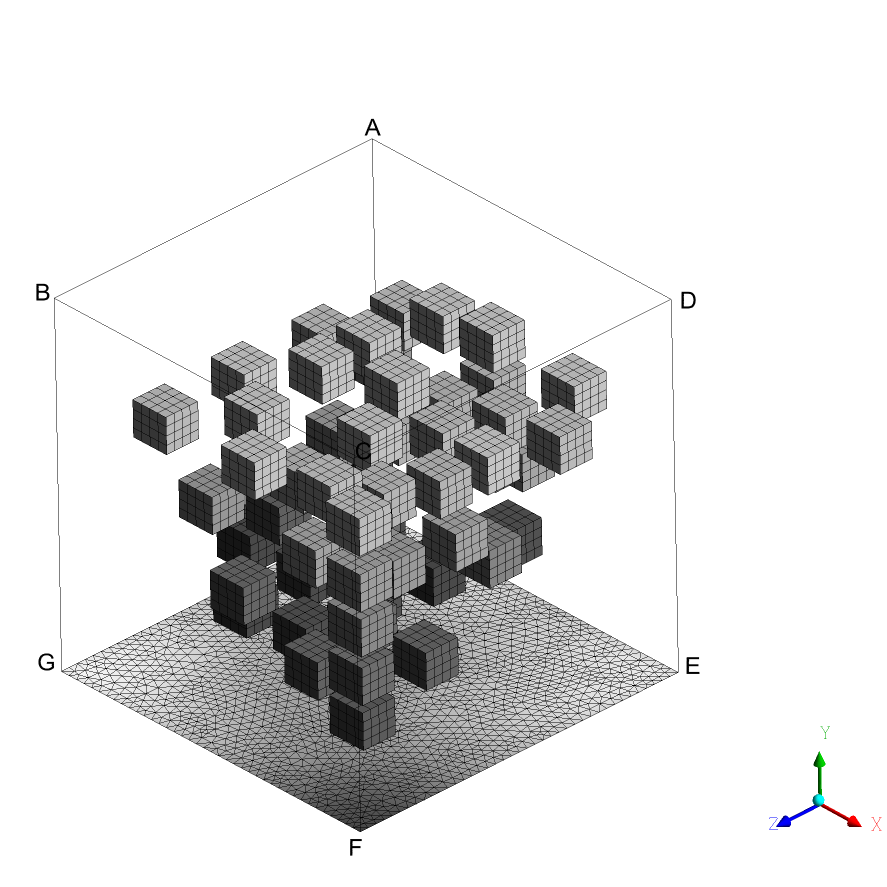}
 \caption{ The figure showing the composite polymer (filler in the shape of multiple box and base polymer). The box fillers are distributed randomly with the gap in between. 
We note that the The mesh structure of solid multiple box figure showimg only bottom surface and filler mesh in the cell with 6\% volume fraction with box size 0.1x0.1x0.1 \(cm^{3}\) and filled filler in cell 0f 0.75cm.}
   \label{fig5}
\end{figure}

\begin{figure}[H]
\centering
   \includegraphics[width=0.58\textwidth]{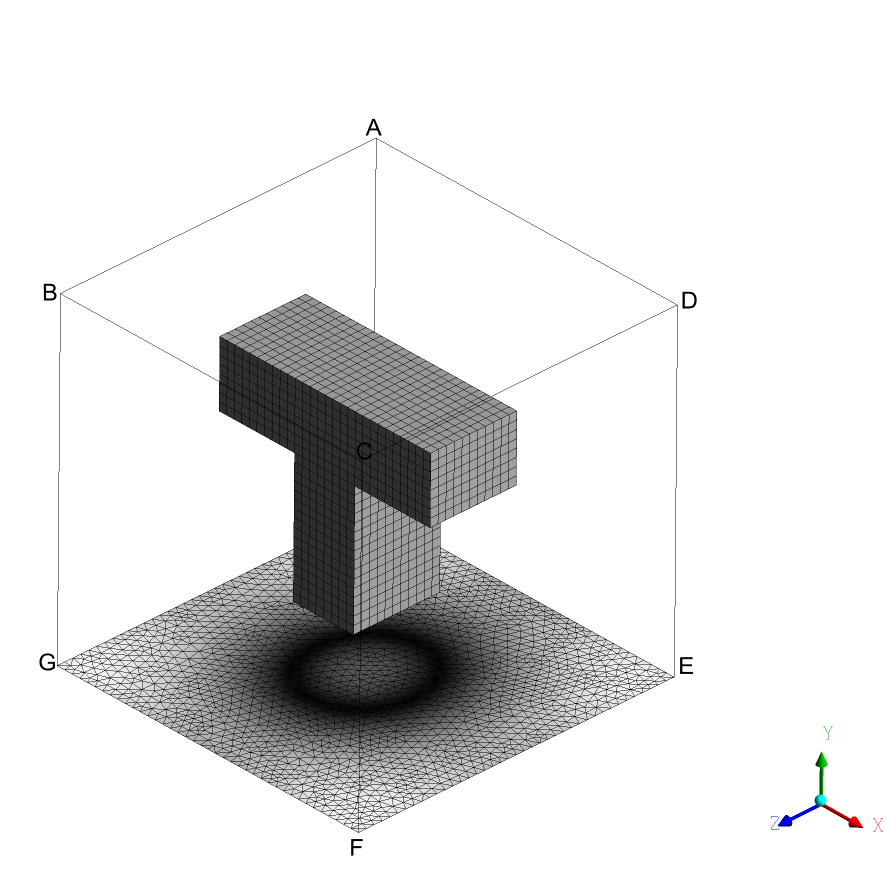}
   \caption  {The mesh structure of T-shape figure showimg only bottom surface and filler mesh in the cell with 6\% volume fraction and filler length 0.5cm with 0.25cm thickness.}
   \label{fig6} 
   \includegraphics[width=0.6\textwidth]{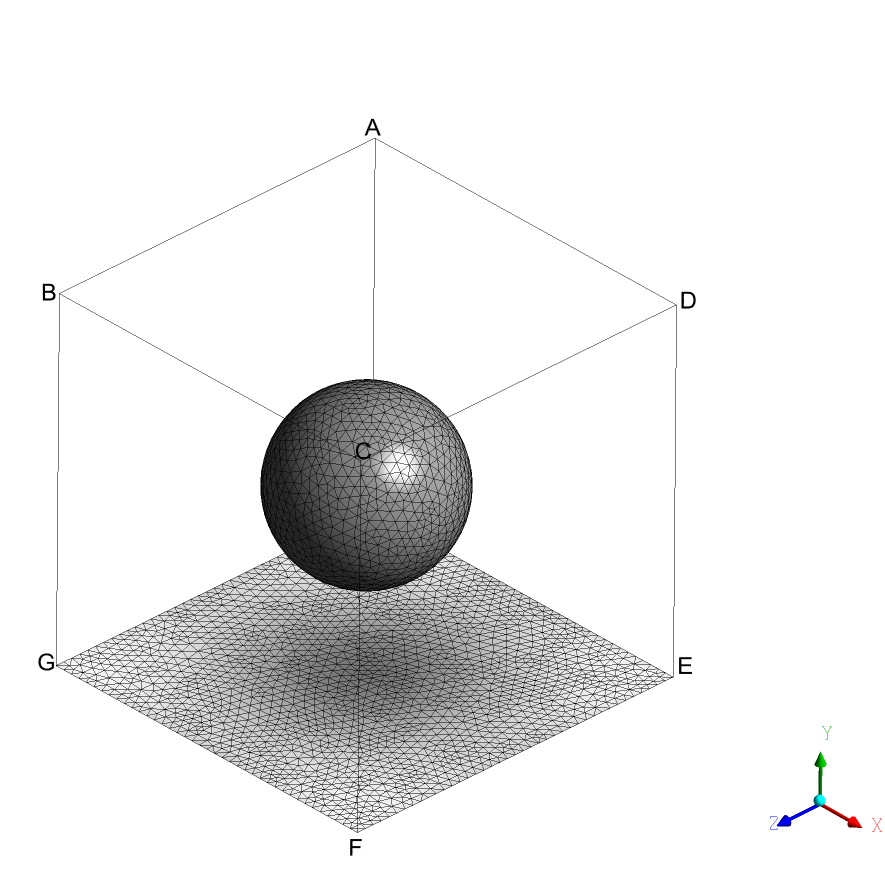}
 \caption{The mesh structure of composite polymer with filler having the shape of solid sphere. Note that spherical filler is uniformy extended in the space of matrix polymer. The contact surface area is very low compared to the vertical shape fillers (such as cubical box and cylinderical fillers). This characterisc is clearly visible in the effective thermal conductivity of the composite polymer for spherical shape filer. It can be observed that the effective thermal conductivity is lowest compared to all the fillers. Sphere shape figure showimg only bottom surface and filler mesh in the cell with 6\% volume fraction with 0.25cm radius. It can be shown that if we use hollow spherical fillers the thermal conductivity is increased very significantly.}
   \label{fig7}
\end{figure}

In this section we discuss the detailed numerical results.  We start with the filler geometries of different shape and sizes and study their effect in the thermal conductivity of the resultant composite polymers.
In figs.~(\ref{f2}-\ref{fig7}) we have shown the schematic diagrams of different fillers with their characteristic structures. The different geometries chosen are designed to pass more heat flux. The fundamental assumption behind these geometries is to get more interface, i.e., the more contact surface area between the polymer and the filler material. The conducting behaviour of these polymers is studied by their corresponding temperature-contour diagrams (both in 2-D and 3-D representing the gradient of heat flux), and by calculating their effective thermal conductivities.  

Note the schematic geometries of the filler in the figures \ref{f1} and \ref{f2} are hollow. In figure \ref{f1} the cubical box has only four surface without any bulk inside, and the bare polymer is filled within. Similarly in figure 2 we have a hollow cylinder instead of the solid cylinder. In this case also the bare polymer is filled within the cylinder. The thermal conductivity measurement will demonstrate that these hollow geometries give better result in the enhancement of the effective thermal conductivity compared to their solid counterpart. 
Apart from these hollow fillers we use other shapes such as $T$, $I$, and $Y$ to observe and compare their effect on the thermal conductivity. The volume percent of these different fillers is kept constant to compare the effective thermal conductivity. 
Apart from these single geometries we use multiple fillers toghether to observe the heat-flux and thermal conductivity behaviour. In figures \ref{fig4} and \ref{fig5} we show the fillers in the form of generalized and non-genralized boxes. These fillers are solid geometries and does not have empty space within them. 
The fillers used in the composite polymers are upto $10\%$ by volume in all the cases whether a single filler or multiple fillers, whether it is a hollow or the solid filler.  

\begin{figure}[H]
\centering
   \includegraphics[width=0.8\textwidth]{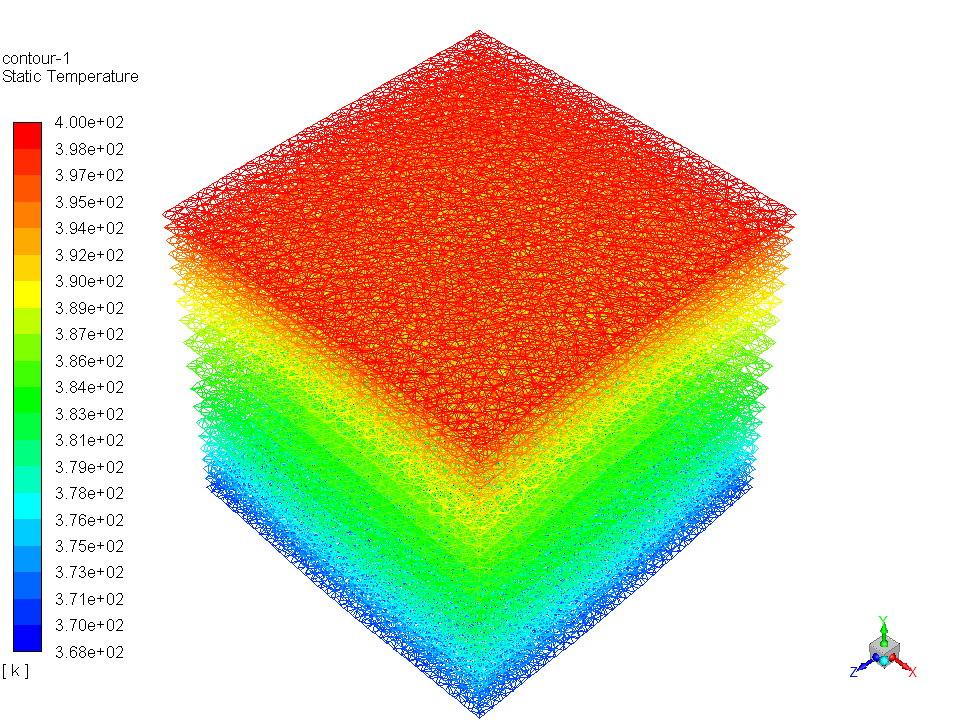}
   \caption{ The three dimensional contour representing the temperature gradient from top to the bottom surface in the static condition for a filler in the shape of \textit{hollow single box}. The top surface represents the maximum temperature kept constant at $400K$ which is uniform across the surface. Filler of size 0.5x0.5x0.5 \(cm^{3}\).}
   \label{f8} 
   \includegraphics[width=0.8\textwidth]{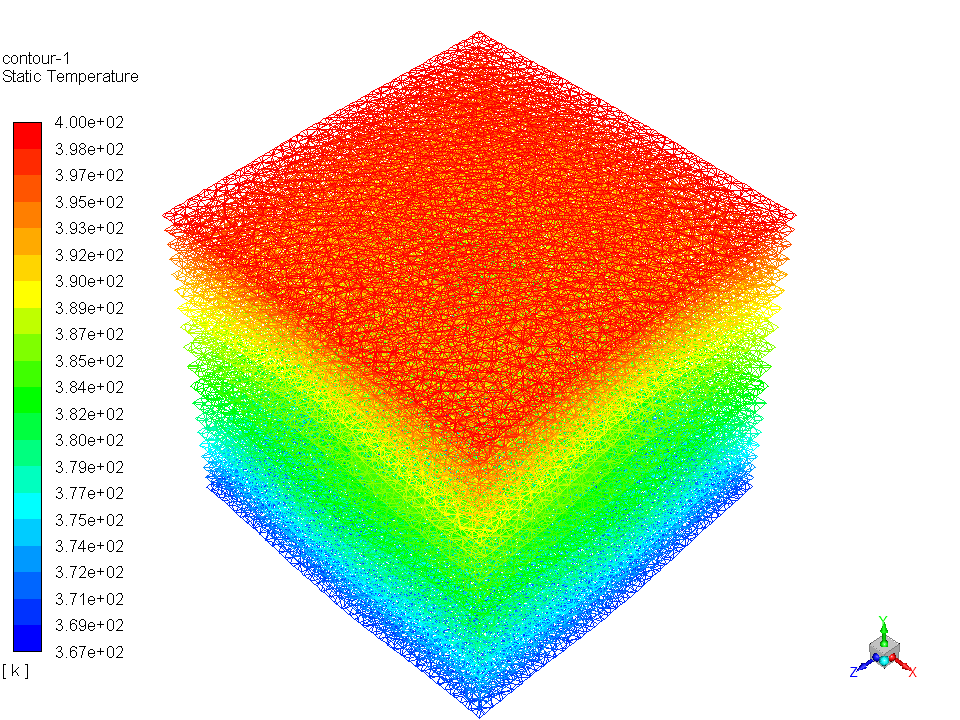}
   \caption {three dimensional contour representing the temperature gradient from top to the bottom surface in the static condition for a filler in the shape of \textit{hollow single cylinder}. The top surface represents the maximum temperature kept constant at $400K$ which is unfiorm across the surface. Filler of size 0.5x0.5x0.5 cm$^{3}$.}
 \label{f9}
\end{figure}

\begin{figure}[H]
\centering
   \includegraphics[width=0.78\textwidth]{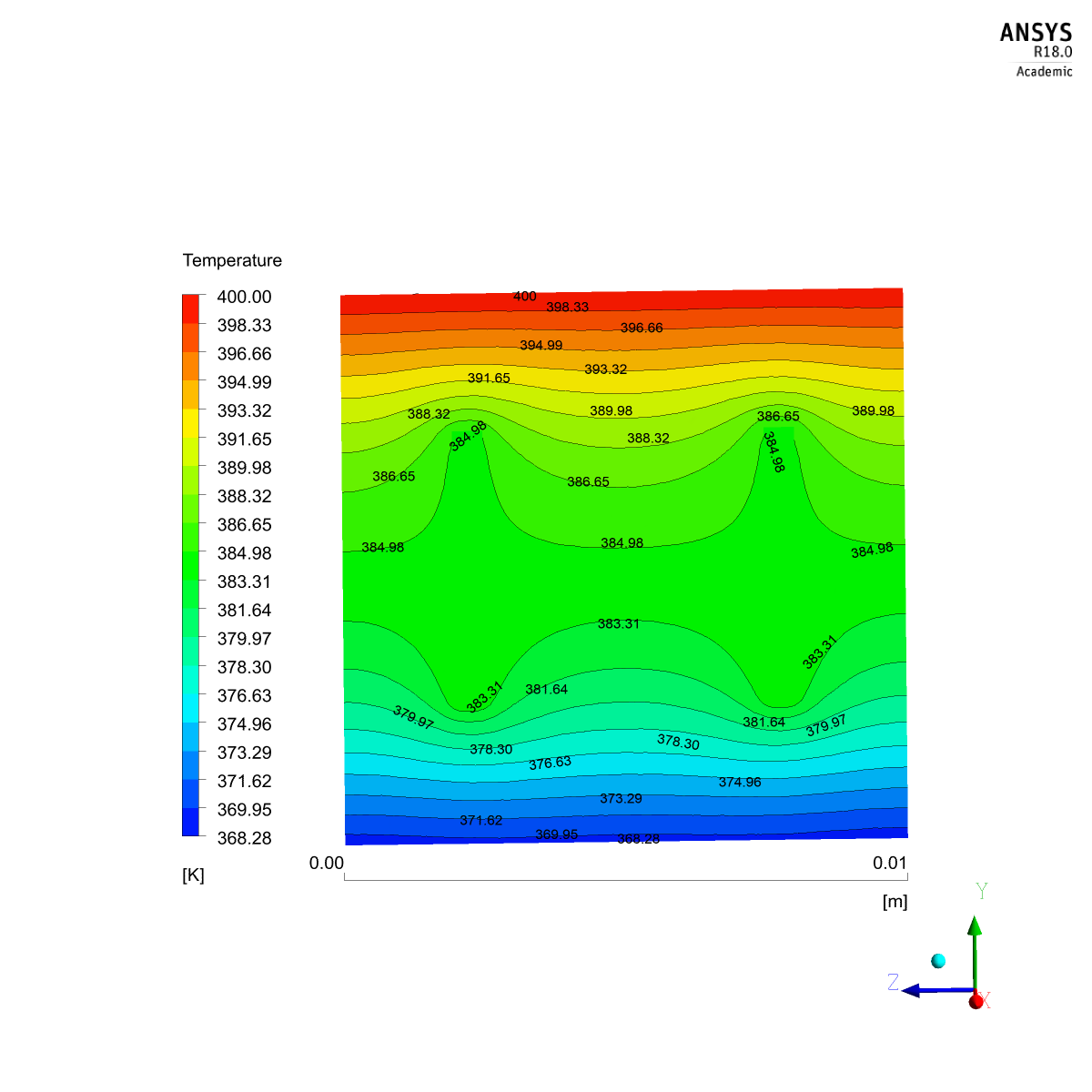}
   \caption{ The two dimensional Contours of temperature(k) of the cross-section z-y plane ($x = 0.5$ $l_{x}$) for the composite polymer representing the heat flux from the top to the bottom surface. The horizontal lines show the variations of temperature across the surface at different vertical length. The figure corresponds to the \textit{hollow box} filler with filler content of $6\%$, and of size $0.5\times0.5\times0.5$ cm$^{3}$.}
   \label{f10} 
   \includegraphics[width=0.72\textwidth]{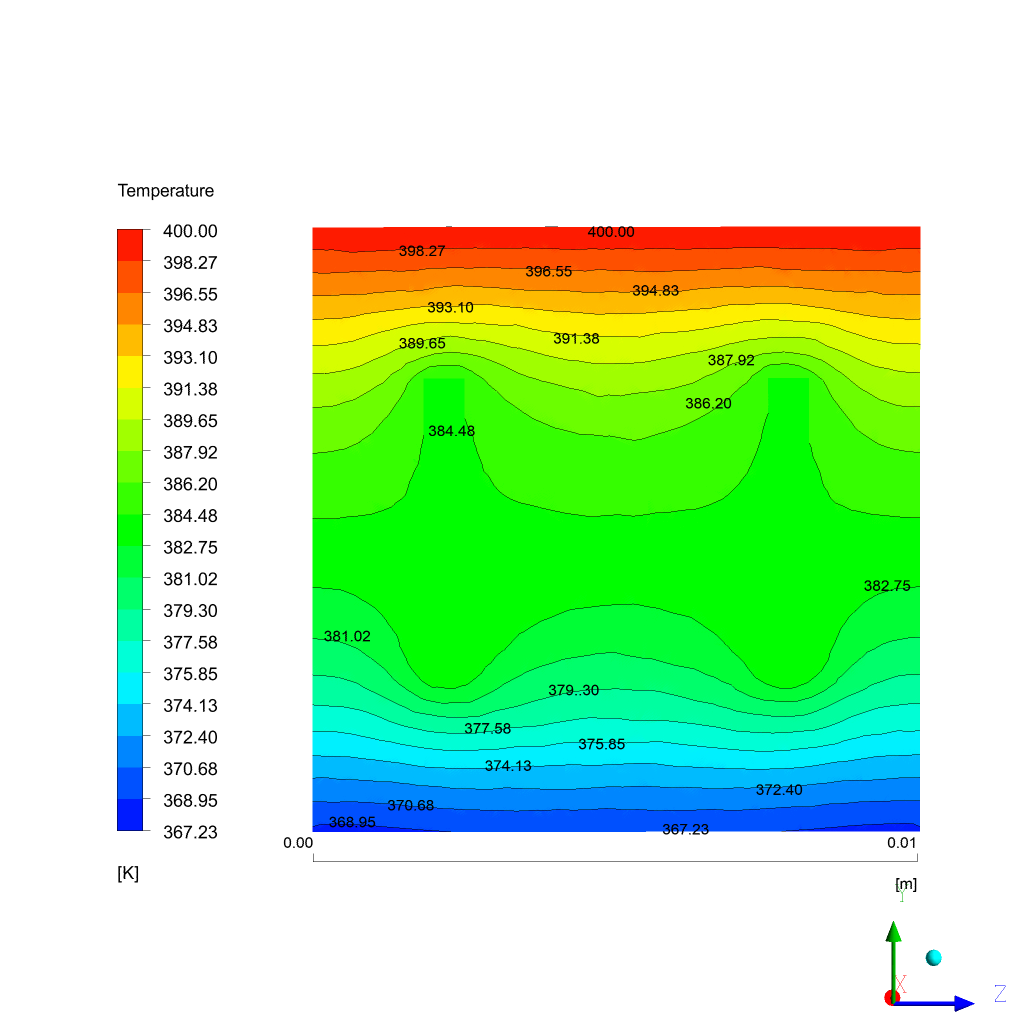}
   \caption { The two dimensional Contours of temperature(k) of the cross-section z-y plane ($x = 0.5$ $l_{x}$) for the composite polymer representing the heat flux from the top to the bottom surface. The horizontal lines show the variations of temperature across the surface at different vertical length. The figure corresponds to the \textit{hollow cylinder} filler with filler content of $6\%$, and of filler height 0.5cm.}
 \label{f11}
\end{figure}

\begin{figure}[H]
\centering
   \includegraphics[width=0.7\textwidth]{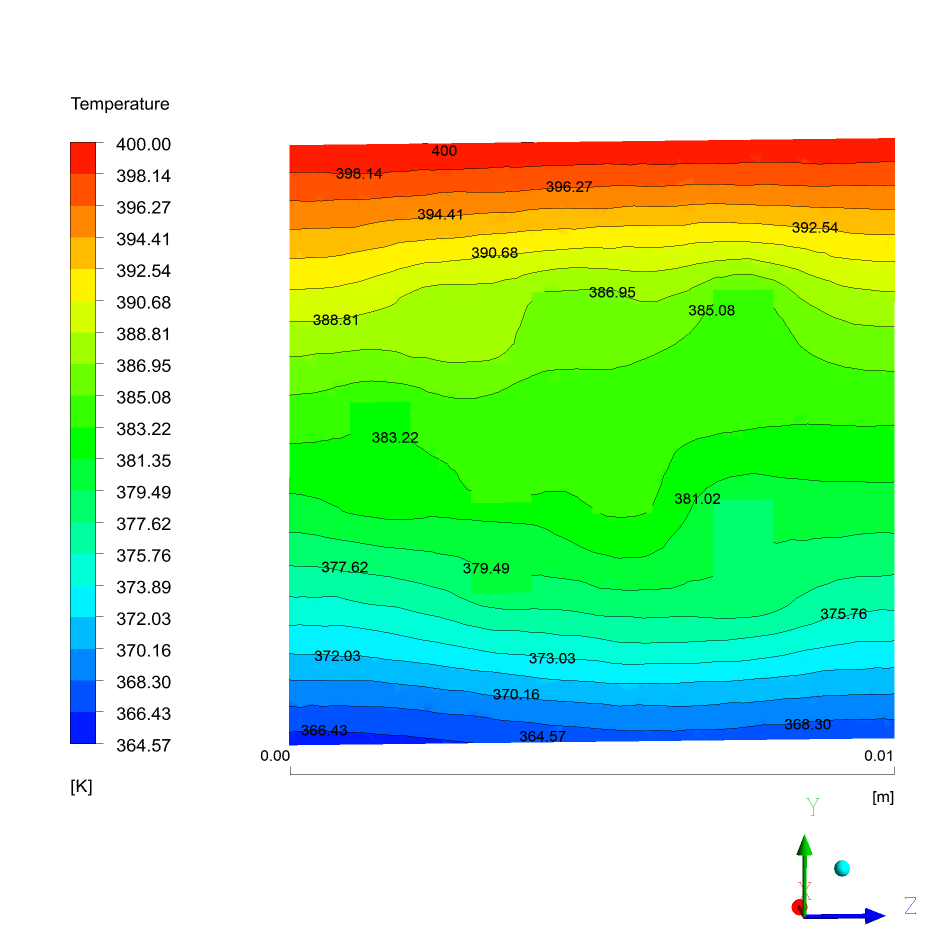}
   \caption{  The two dimensional Contours of temperature(k) of the cross-section z-y plane ($x = 0.5$ $l_{x}$) for the composite polymer representing the heat flux from the top to the bottom surface. The horizontal lines show the variations of temperature across the surface at different vertical length. The figure corresponds to the \textit{genrelized box} filler with filler content of $6\%$, and of filled filler height 0.5cm.}
   \label{f12} 
   \includegraphics[width=0.75\textwidth]{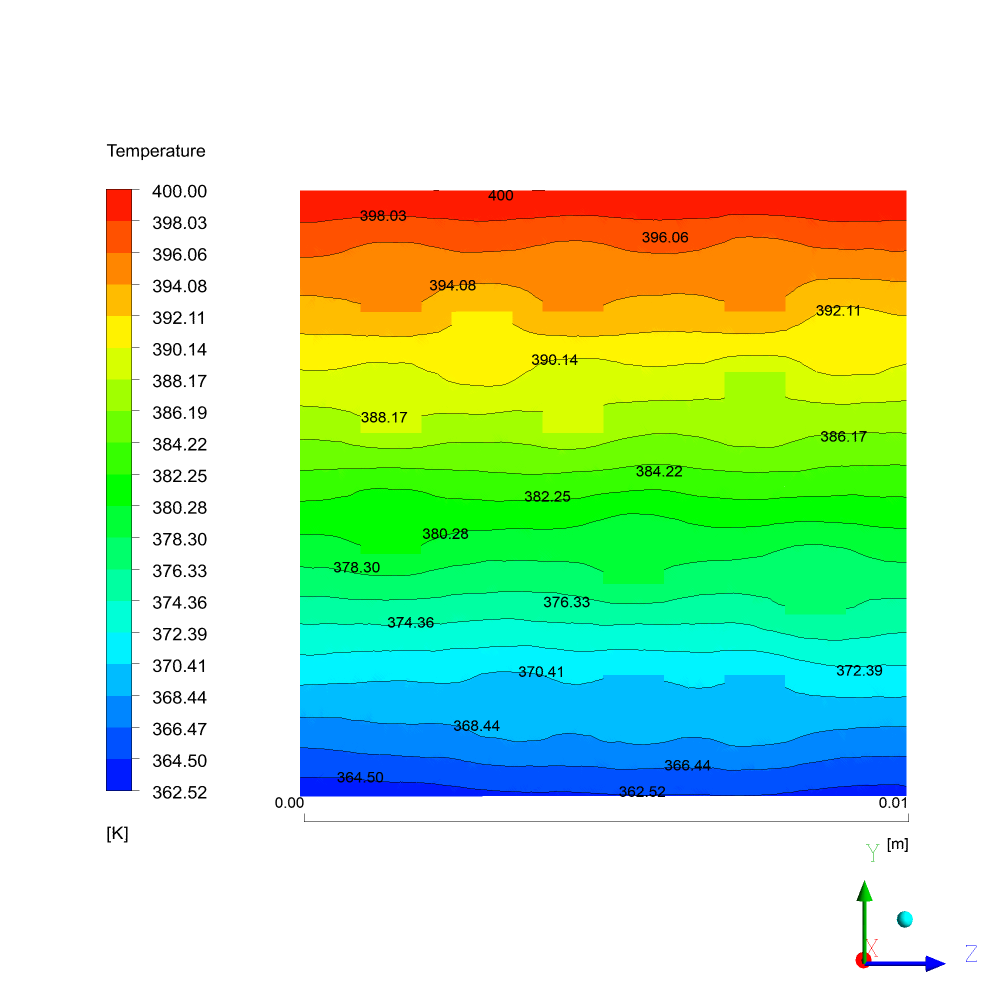}
   \caption { The two dimensional Contours of temperature (k) of the cross-section z-y plane ($x = 0.5$ $l_{x}$) for the composite polymer representing the heat flux from the top to the bottom surface. The horizontal lines show the variations of temperature across the surface at different vertical length. The figure corresponds to the \textit{solid multiple box} filler with filler content of 6\%, and of filled filler height 0.5cm.}
 \label{f13}
\end{figure}

\begin{figure}[H]
\centering
   \includegraphics[width=0.75\textwidth]{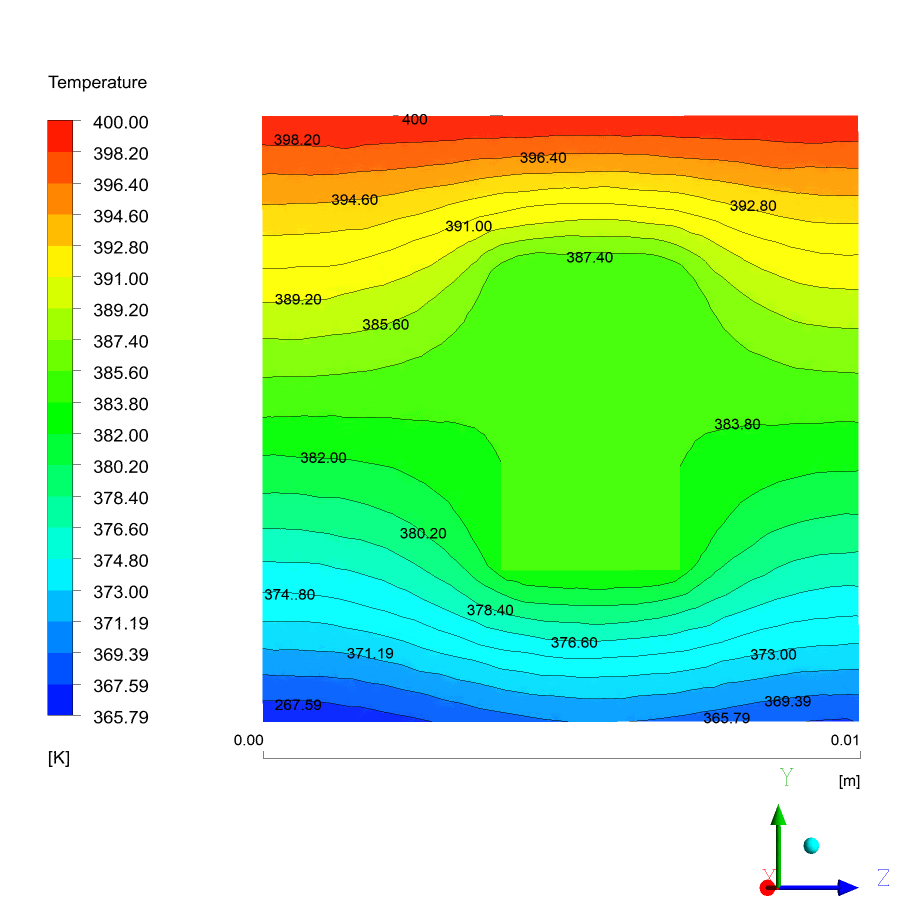}
   \caption{The two dimensional contour of temperature (k) of the cross-section z-y plane ($x = 0.5$ $l_{x}$) for the composite polymer representing the heat flux from the top to the bottom surface. The horizontal lines show the variations of temperature across the surface at different vertical length. The
figure corresponds to the \textit{I-shape} filler with filler content of 6\%, and of filler height
0.5cm.}
   \label{f14} 
   \includegraphics[width=0.75\textwidth]{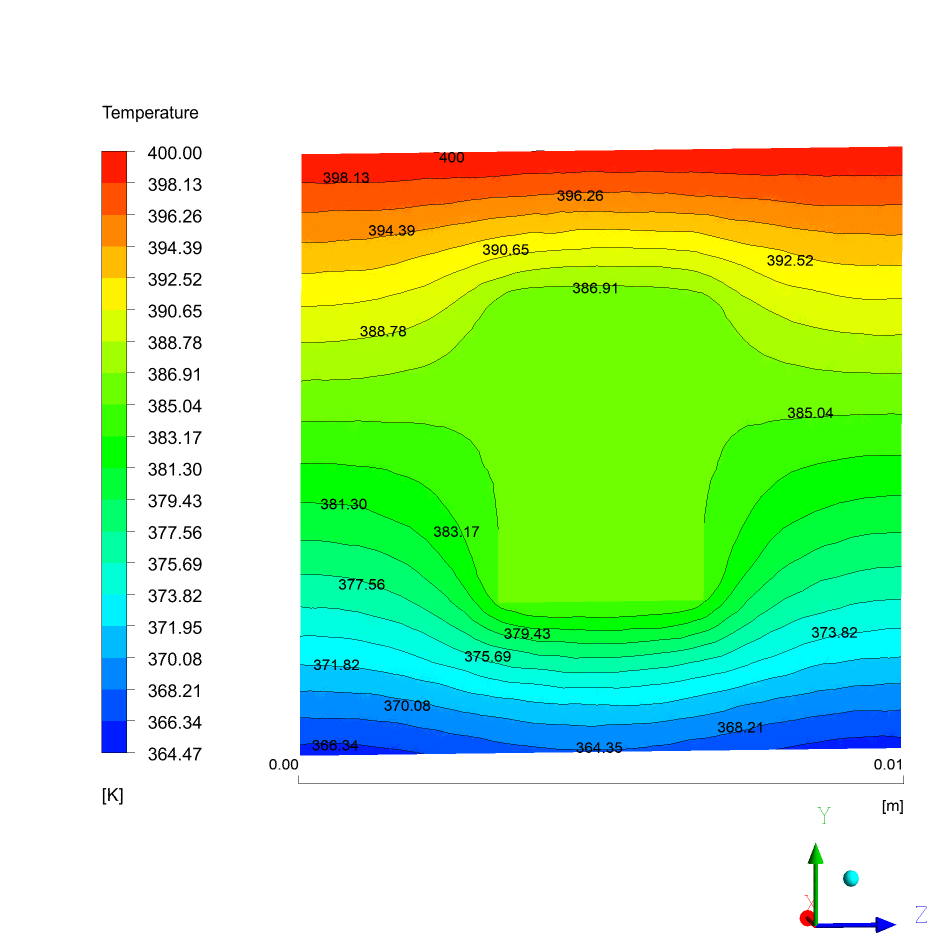}
   \caption { The two dimensional contour of temperature (k) of the cross-section z-y plane ($x = 0.5$ $l_{x}$) the composite polymer representing the heat flux from the top to the bottom surface. The horizontal lines show the variations of temperature across the surface at different vertical length. The
figure corresponds to the \textit{T-shape} filler with filler content of 6\%, and of filler height
0.5cm.}
 \label{f15}
\end{figure}

\begin{figure}[H]
\centering
   \includegraphics[width=0.75\textwidth]{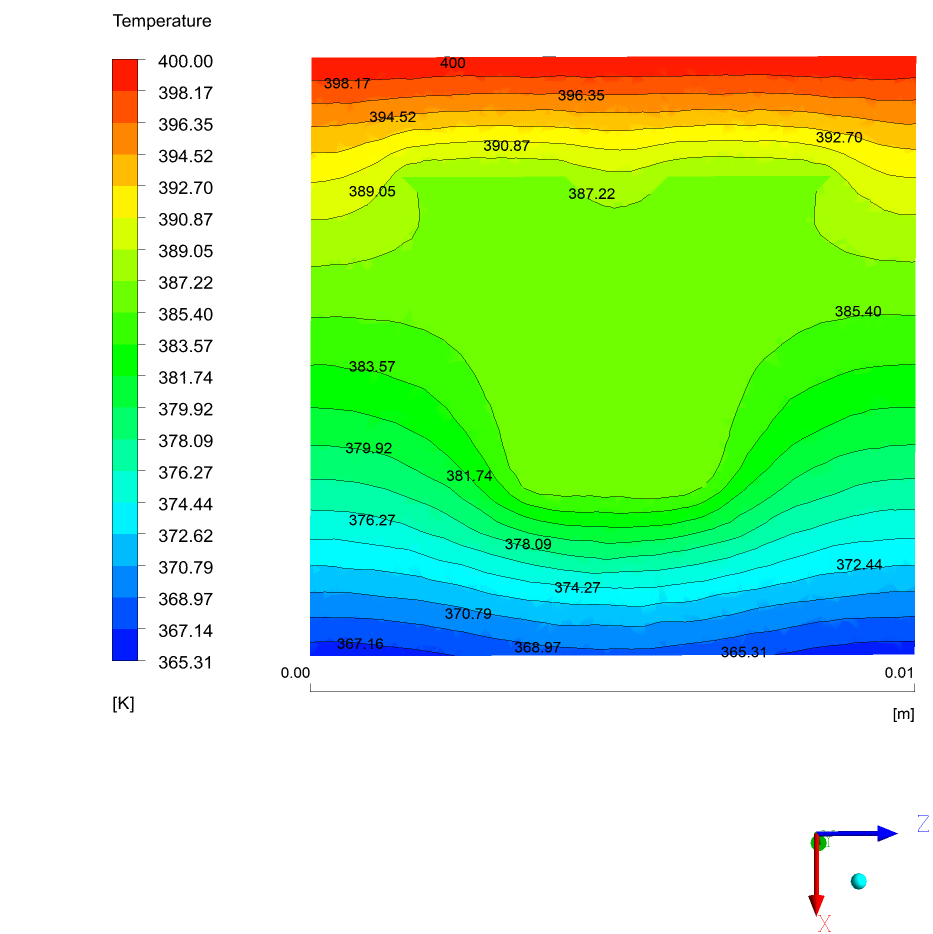}
   \caption{The two dimensional contour of temperature(k) of the cross-section z-x plane ($y = 0.5$ $l_{y}$) for the composite polymer representing the heat flux from the top to the bottom surface. The horizontal lines show the variations of temperature across the surface at different vertical length. The
figure corresponds to the \textit{Y-shape} filler with filler content of 6\%, and of filler height 0.5cm.}
   \label{f16} 
   \includegraphics[width=0.75\textwidth]{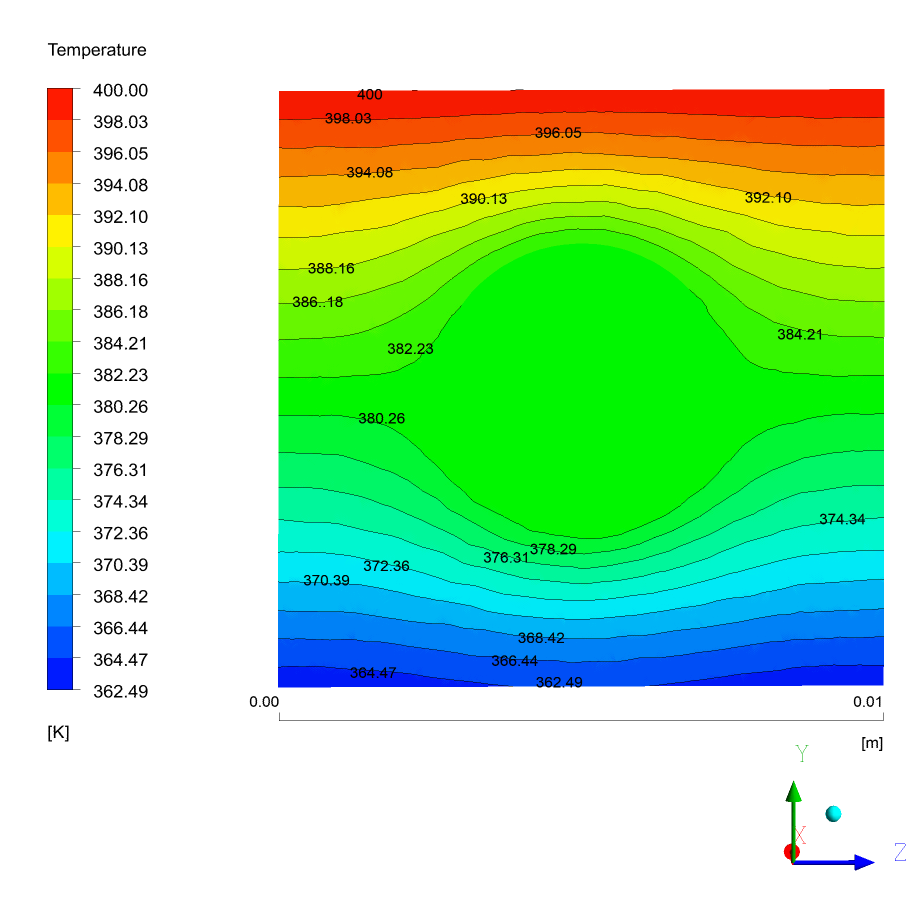}
   \caption {The two dimensional contour for of temperature(k) of the cross-section z-y plane ($x = 0.5$ $l_{x}$) the composite polymer representing the heat flux from the top to the bottom surface. The horizontal lines show the variations of temperature across the surface at different vertical length. The
figure corresponds to the \textit{sphere-shape} filler with filler content of 6\%, and of radius 0.278cm.}
 \label{f17}
\end{figure}

\begin{figure}[H]
\centering
   \includegraphics[width=0.75\textwidth]{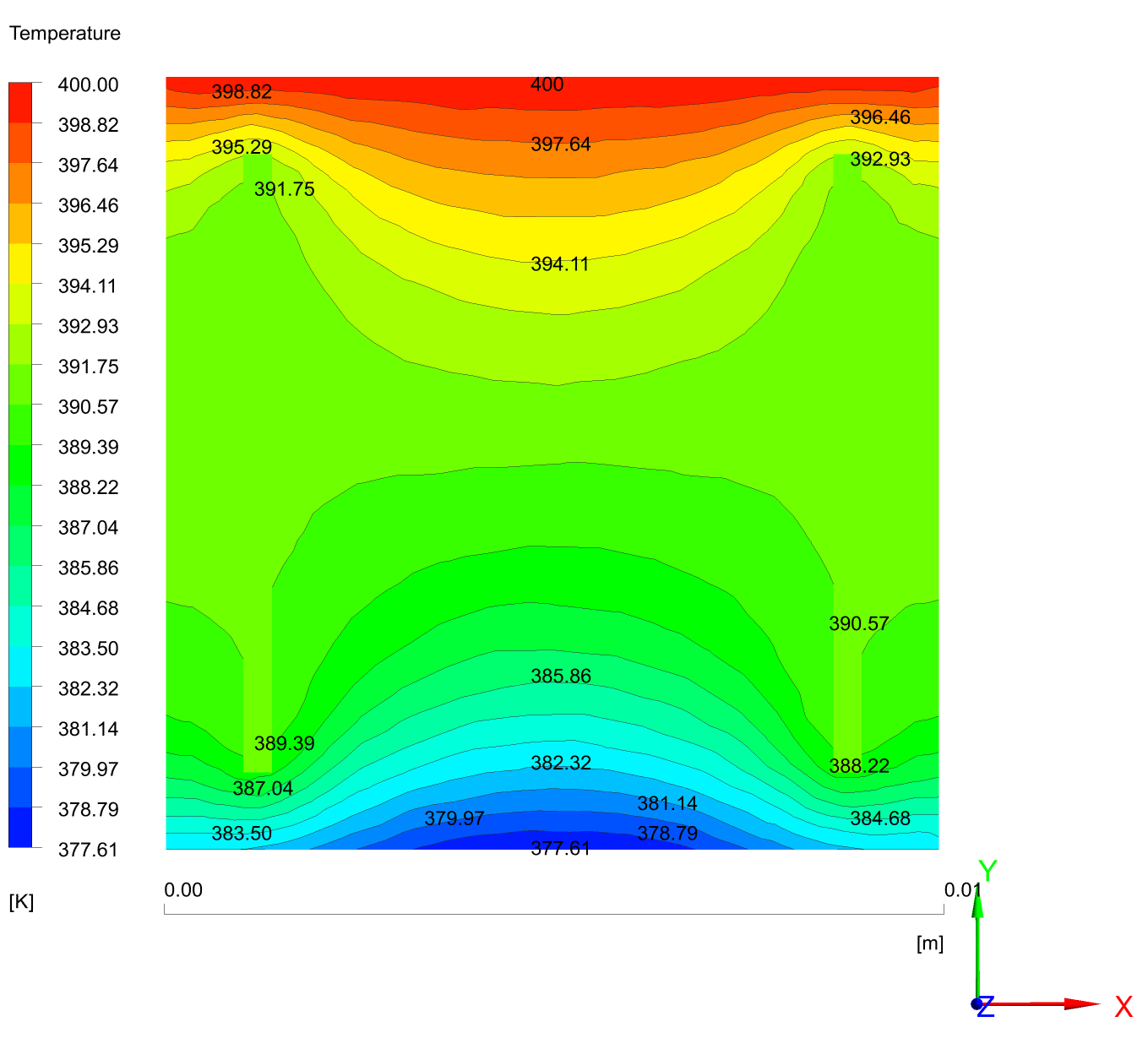}
   \caption{ The two dimensional Contours of temperature(k) of the cross-section x-y plane ($z = 0.5$ $l_{z}$) for the composite polymer representing the heat flux from the top to the bottom surface. The horizontal lines show the variations of temperature across the surface at different vertical length. The figure corresponds to the \textit{hollow cylinder} filler with filler content of 9\%, and of filler height 0.8cm.}
   \label{f18} 
   \includegraphics[width=0.65\textwidth]{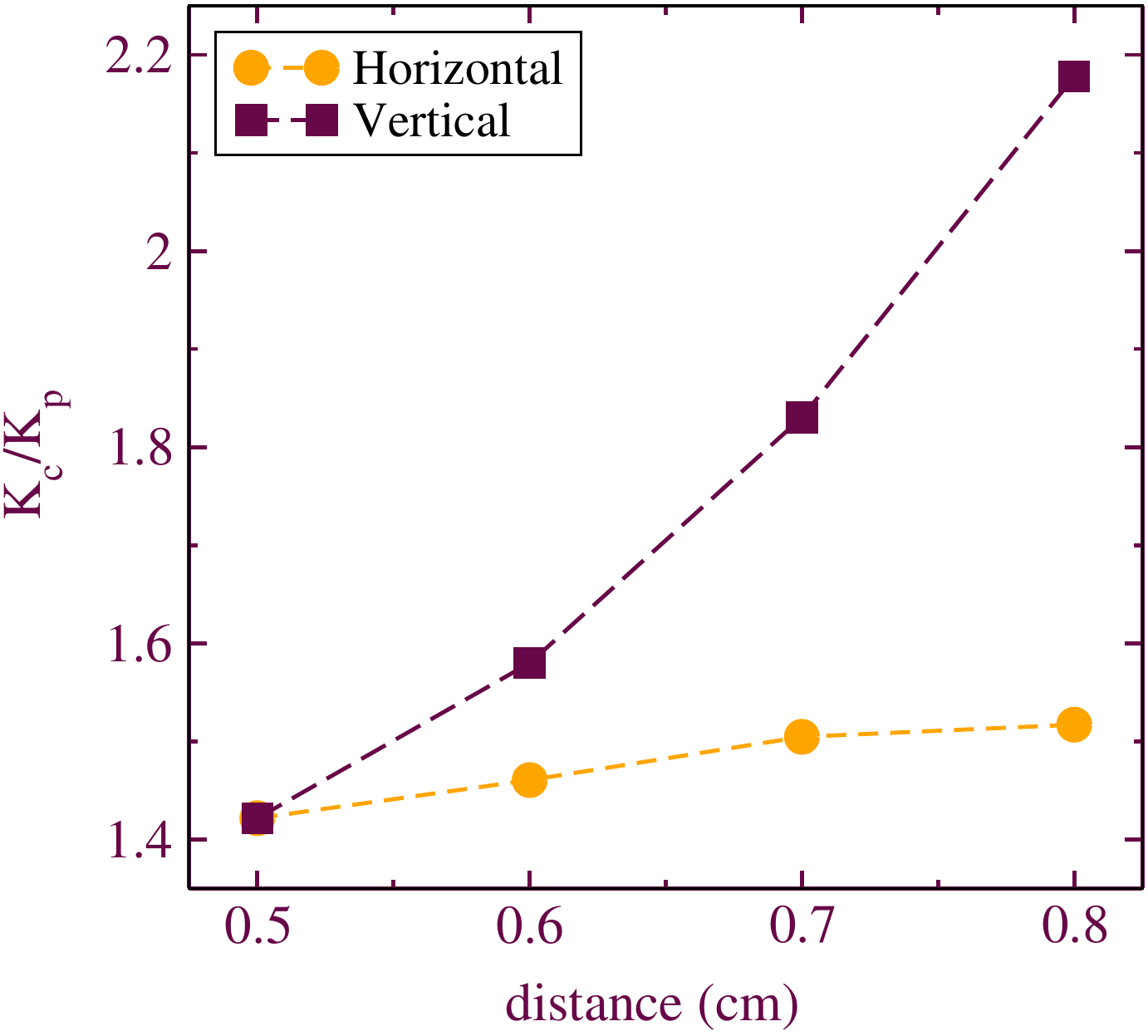}
   \caption {Effect on relative thermal conductivity by increasing horizontal and vertical length of 2\% volume fraction of hollow single box.}
 \label{f19}
\end{figure}

In figures(\ref{f8}-\ref{f9}) we have shown the 3-D temperature contours for the composite polymer comprising a single hollow box and a single hollow cylinder respectively. Figures~(\ref{f10}-\ref{f18}) correspond to the 2-D temperature contours for the fillers discussed in figsures~(\ref{f2}-\ref{fig7}). Note that figure~\ref{f18} corresponds to the single hollow cylinder of volume percent $9\%$. Figure~\ref{f19} shows the effect of the symmetry or alignment of the filler in the composite. It is shown that when the filler has vertical aligemnt, i.e., in the direction of the heat flow the thermal conductivity is very high. 
This behaviour can be understood easily if we observe the effective path avialable for heat-flux. When the filler is kept horizontal the heat-flux is through only a smaller width of the metallic filler, and hence heat can not flow for a longer distance through the filler. This is due to the fact that the filler is horizontal and flux is from the top to bottom. Therefore symmetry is an very imoprtant factor to determine the effective thermal conductivity. 
On the other hand when the filler is parallel to the heat-flux, the path for the heat-flow through the filler is longer and heat can pass to longer distances using the metallic path. \textit{The other aspect of this fact is that the surface area for heat flux should be longer to get the maximum effective thermal conductivity $K_{c}/K_{p}$.}

\begin{figure}[H]
\begin{center}
\includegraphics*[width=0.65\textwidth]{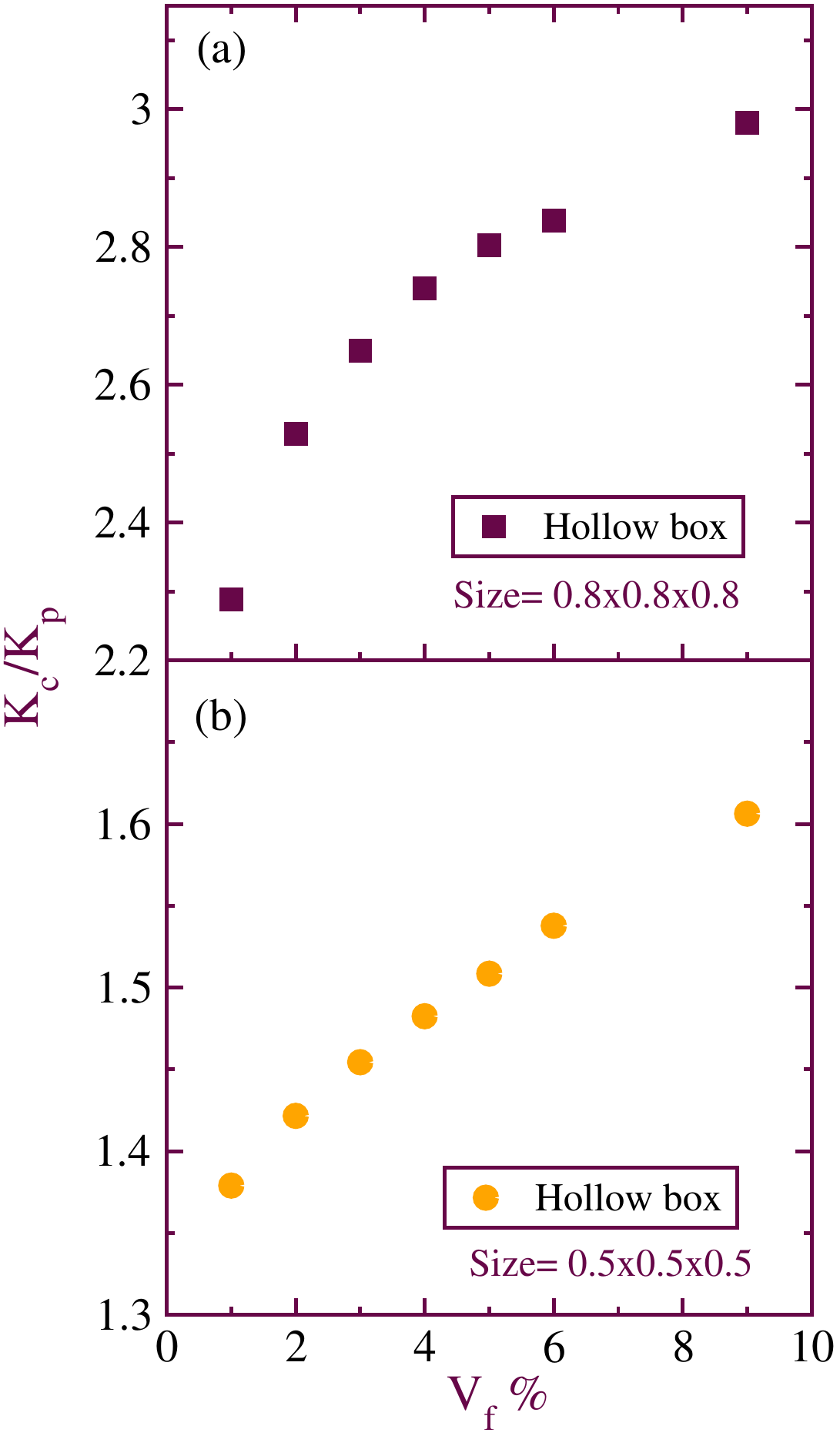}
\end{center}
\caption{ Effect on relative thermal conductivity by increasing the size of hollow single box of constant volume with fixed 6\% volume fraction.} 
\label{f20}
\end{figure}

\begin{figure}[H]
\begin{center}
\includegraphics*[width=0.67\textwidth]{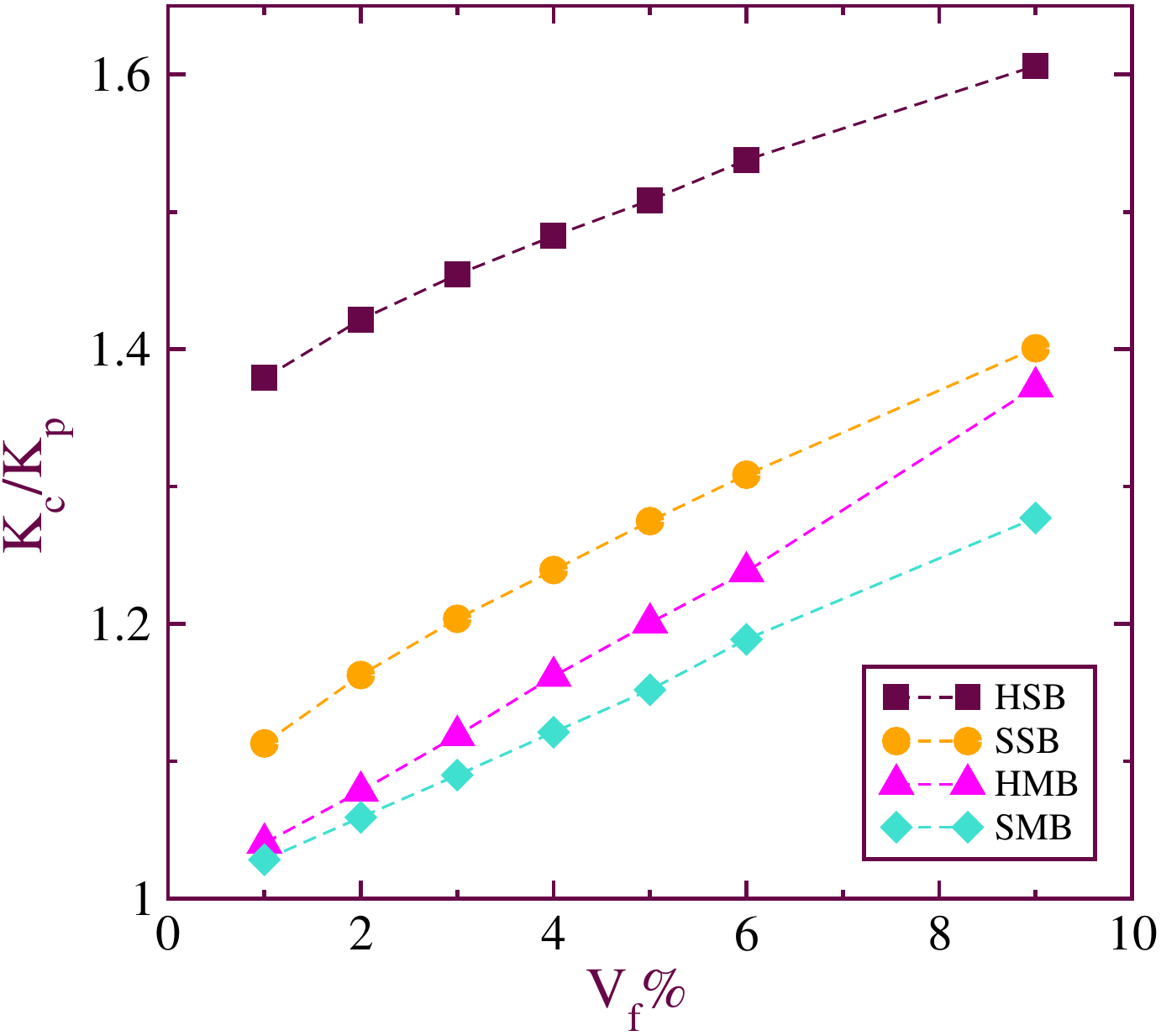}
\end{center}
\caption{ Effect on relative thermal conductivity with different shape of filler hollow single box,solid single box, hollow multiple box and solid multiple box with constant (vertical) length 0f 0.5cm by increasing volume fraction of filler.}
\label{f21}
\end{figure}

\begin{figure}[H]
\begin{center}
\includegraphics*[width=0.67\textwidth]{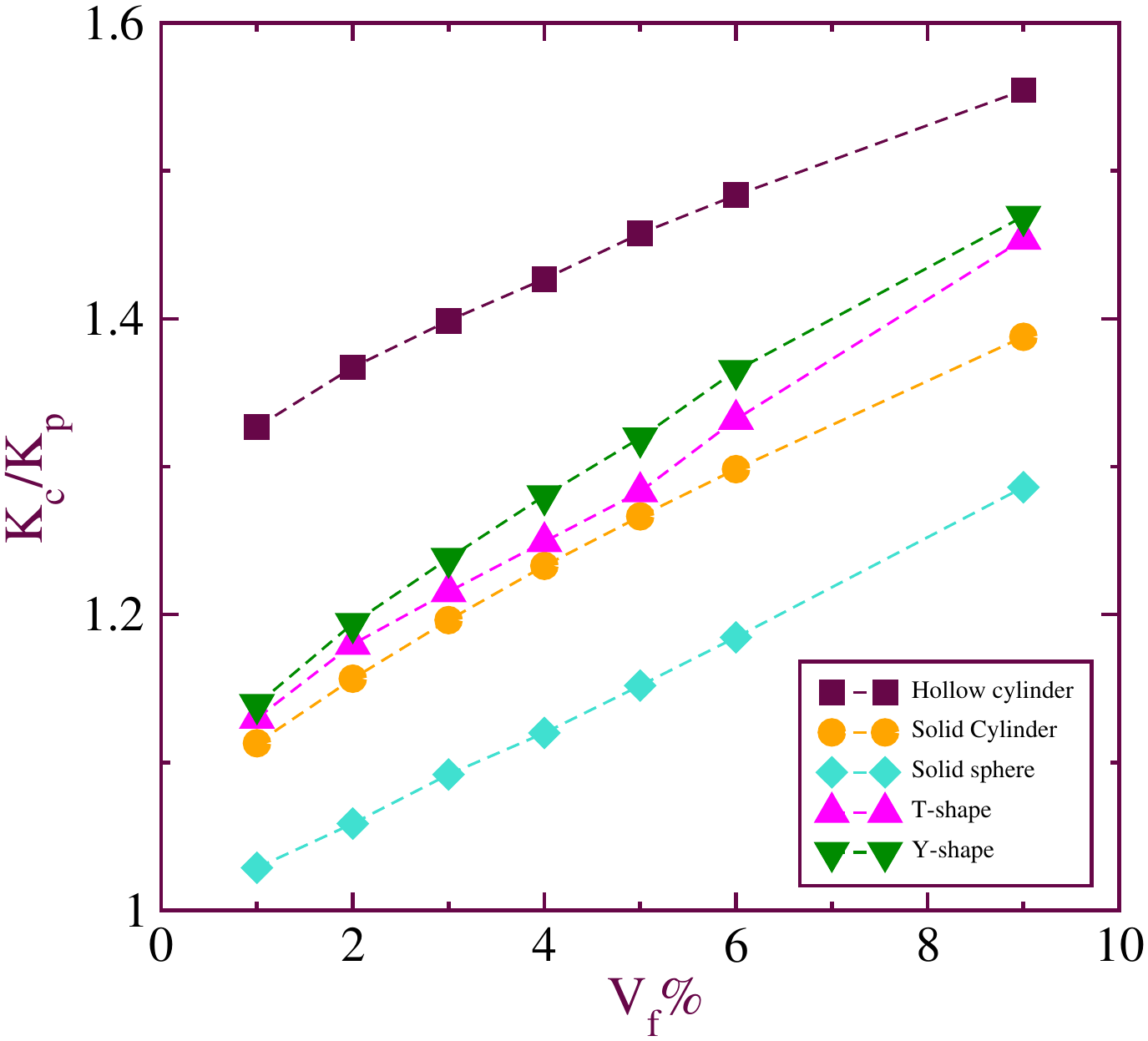}
\end{center}
\caption{ Effect on relative thermal conductivity of different shape of filler hollow cylinder,solid cylinder, solid sphere, T shape and Y shape with constant (vertical) length 0f 0.5cm by increasing volume fraction of filler.}
\label{f22}
\end{figure}

\begin{figure}[H]
\begin{center}
\includegraphics*[width=0.67\textwidth]{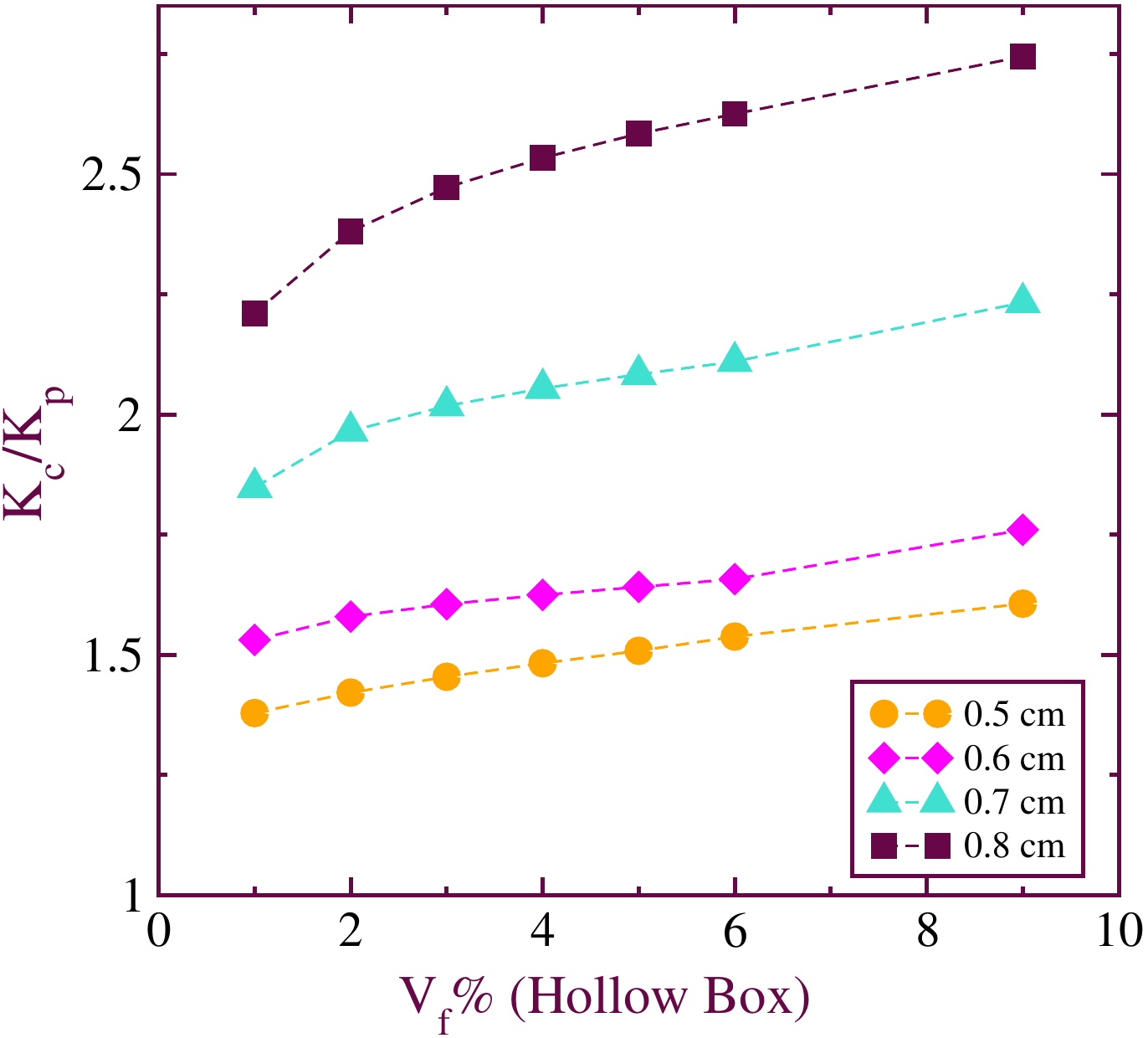}
\end{center}
\caption{( Effect on relative thermal conductivity by increasing the volume fraction of filler with different diameter of hollow single box.}
\label{f23}
\end{figure}

\begin{figure}[H]
\begin{center}
\includegraphics*[width=0.67\textwidth]{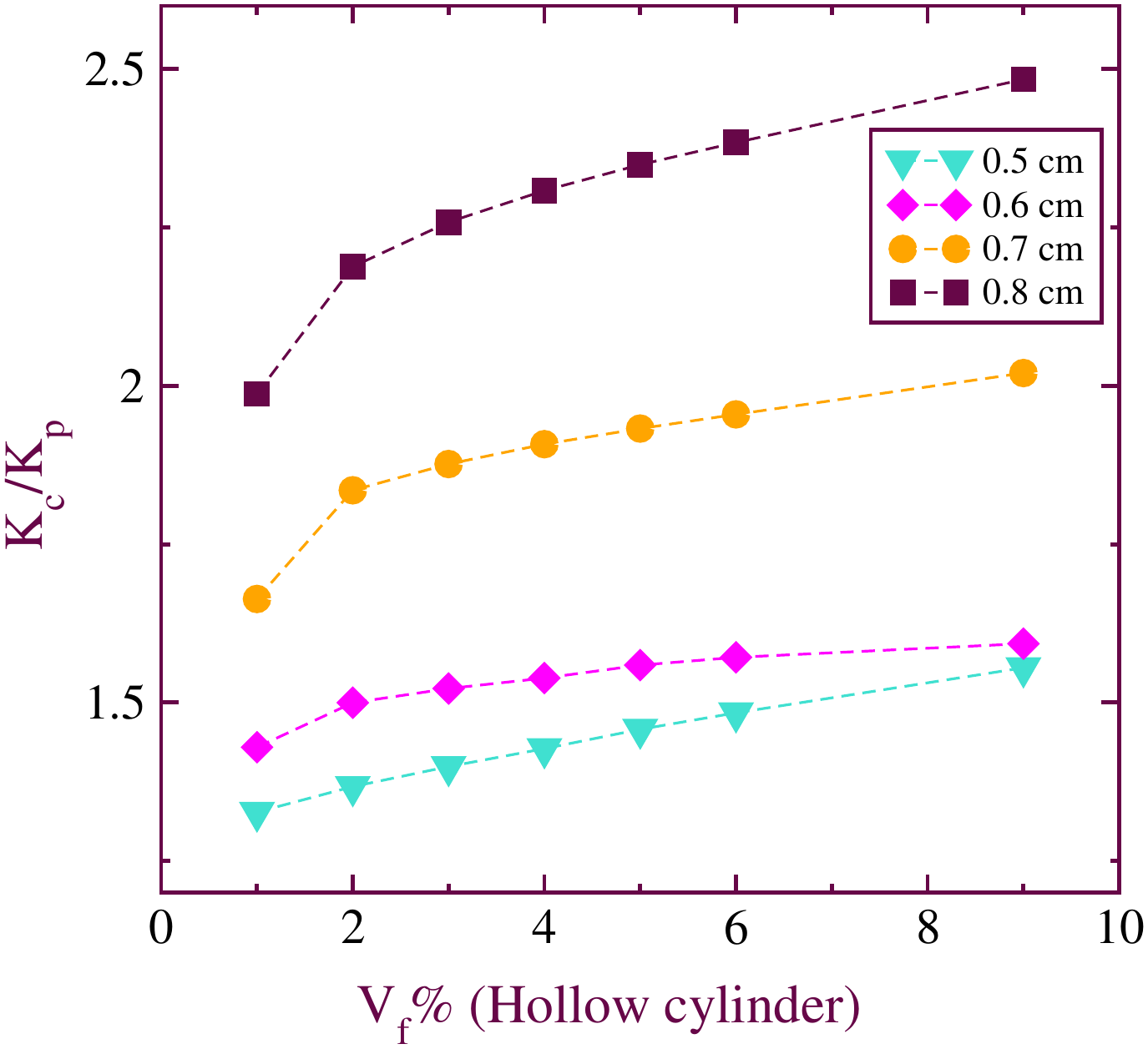}
\end{center}
\caption{ Effect on relative thermal conductivity by increasing the volume fraction of filler with different diameter of hollow single cylinder.}
\label{f24}
\end{figure}

\begin{figure}[H]
\begin{center}
\includegraphics*[width=0.67\textwidth]{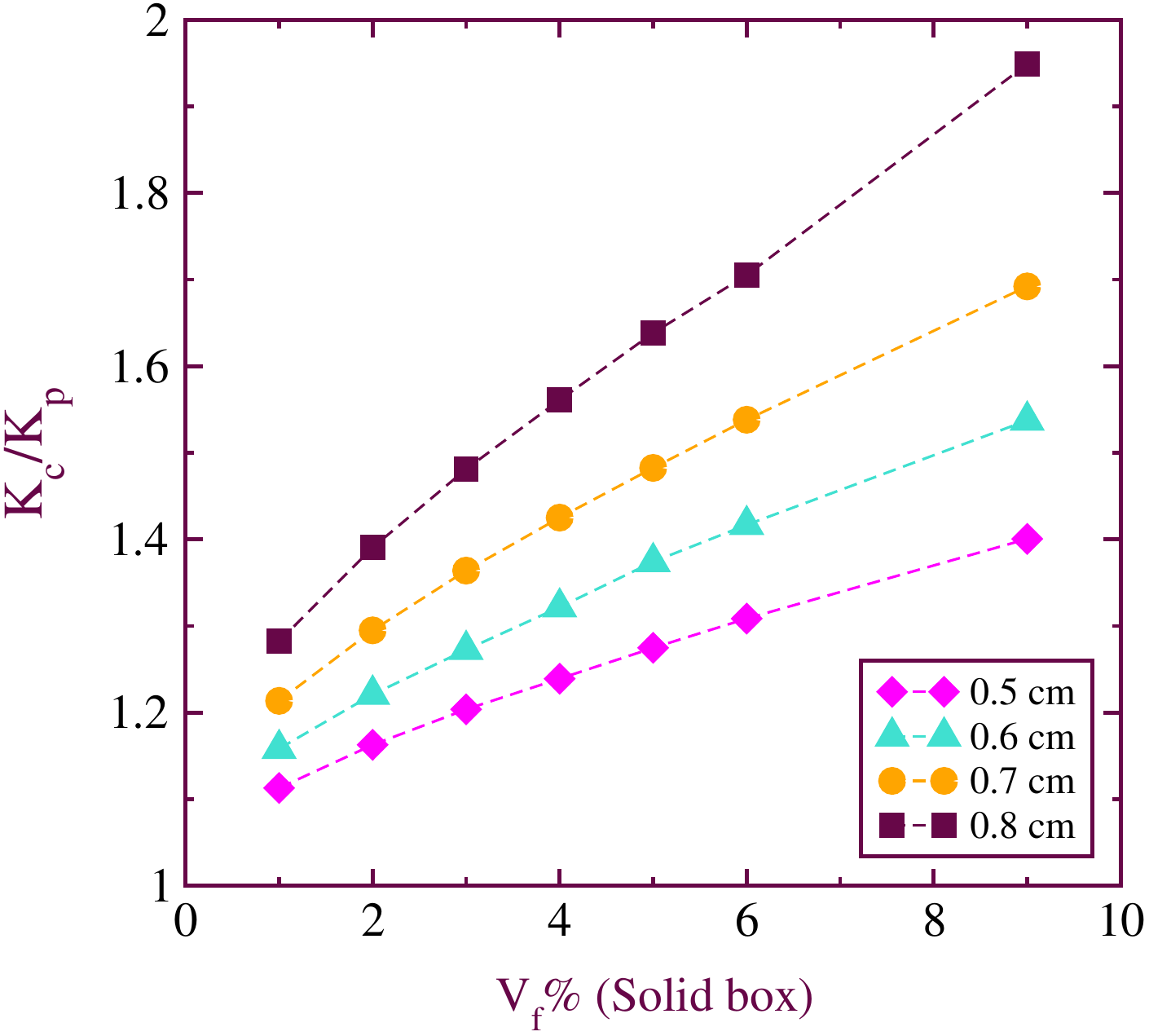}
\end{center}
\caption{ Effect on relative thermal conductivity by increasing the volume fraction of filler with differnt (verticle) length of solid single box.}
\label{f25}
\end{figure}

\begin{figure}[H]
\begin{center}
\includegraphics*[width=0.67\textwidth]{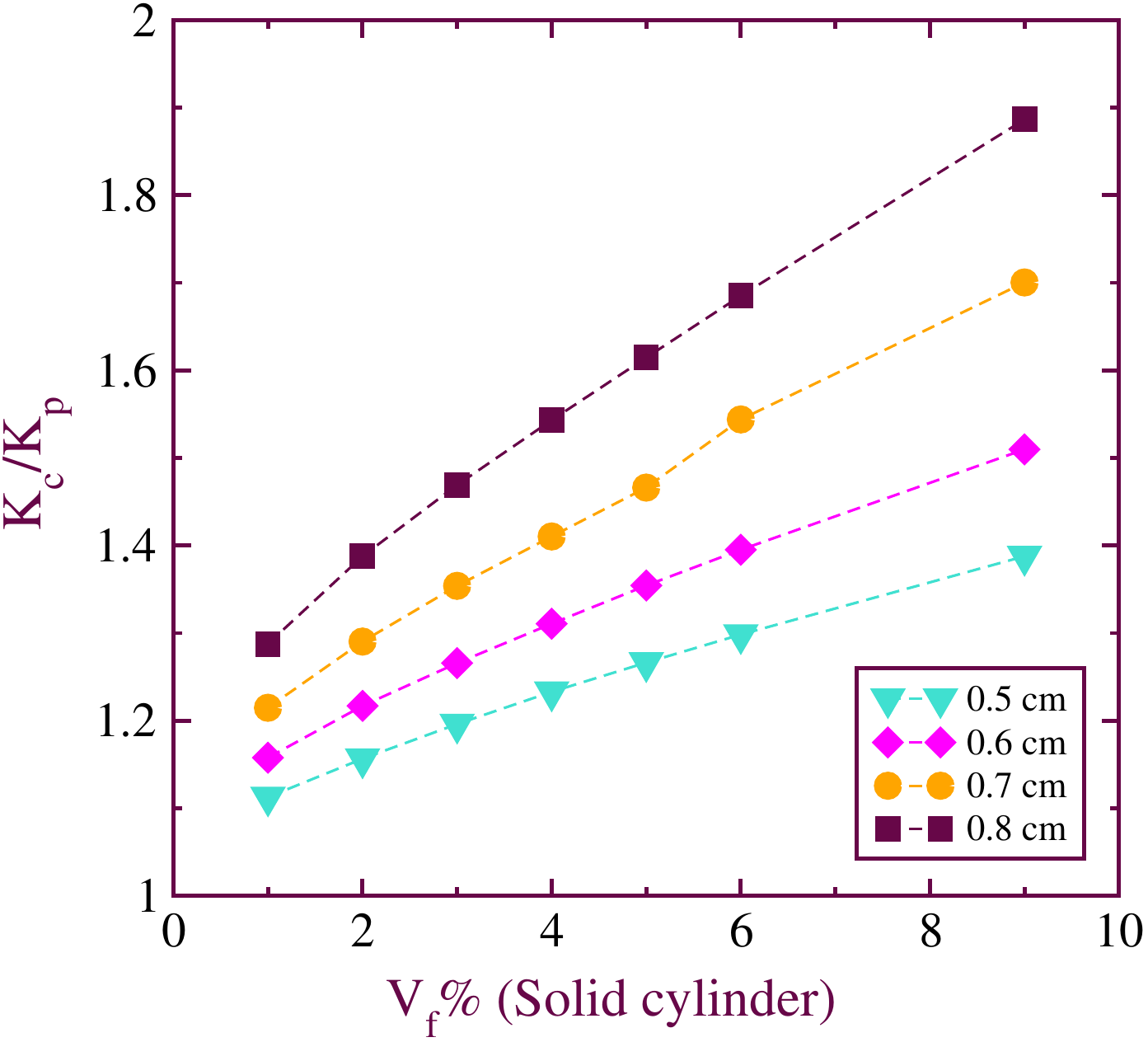}
\end{center}
\caption{Effect on relative thermal conductivity by increasing the volume fraction of filler with differnt length(verticle) of solid single cylinder.}
\label{f26}
\end{figure}

\begin{figure}[H]
\begin{center}
\begin{tabular}{cc}
\includegraphics*[width=0.7\textwidth]{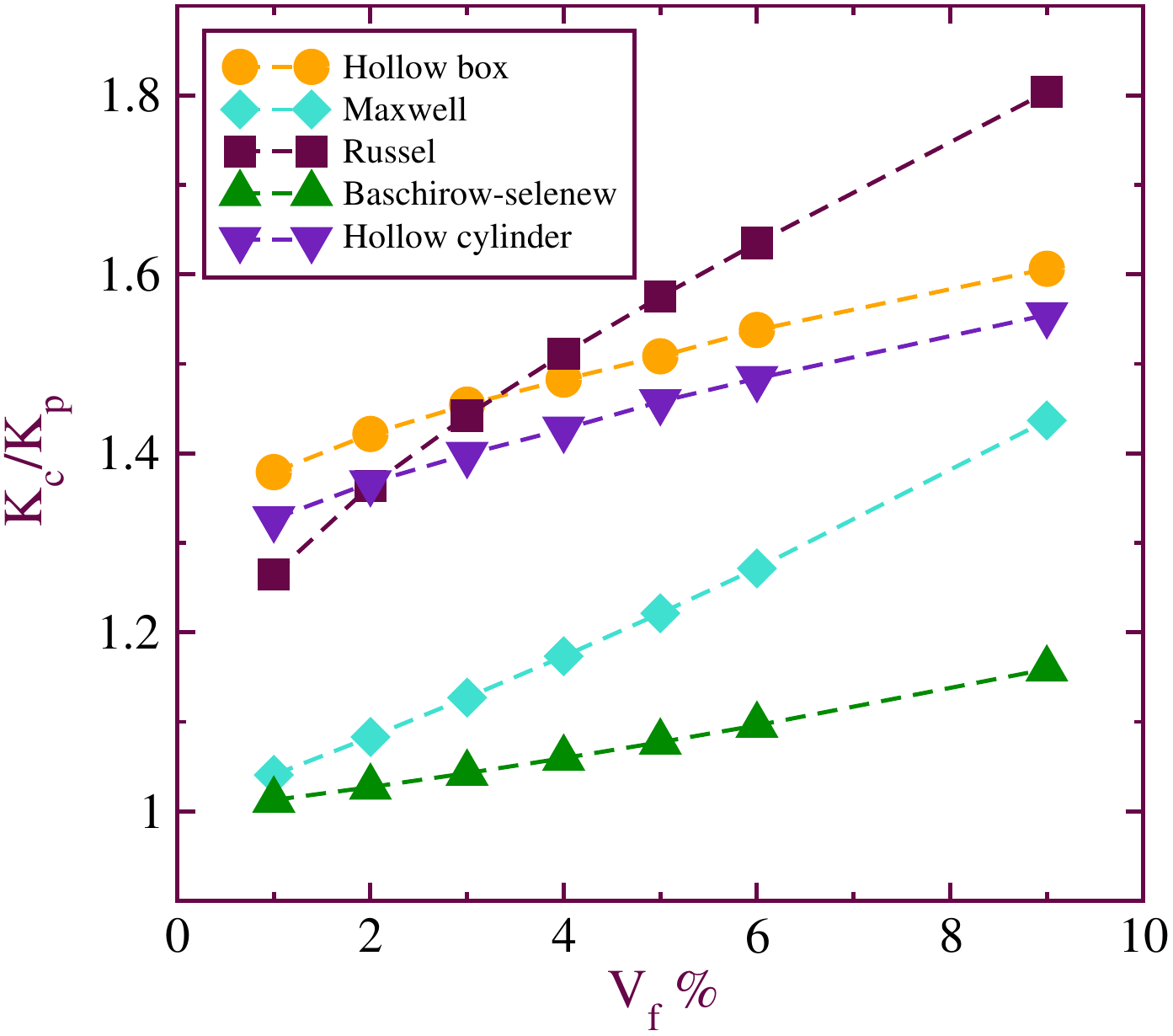} & \ 
 \end{tabular}
\end{center}
\caption {Comparision of the relative thermal conductivity estimated by correlations and modeled }
\label{f27}
\end{figure}

Figure~\ref{f20} shows the effective thermal conductivity for a hollow box with the two different dimensions (a) $0.8\times 0.8\times 0.8$ cm$^{3}$, and (b) $0.5\times0.5\times0.5$ cm$^{3}$ with same amount of filler material. It is clear that when the constant volume percent is used to create more surface area the thermal conductivity increases.
Figures~(\ref{f21}, and \ref{f22}) demonstrate the comparison of the effective thermal conductivities for the fillers of different geometries. The voulme percent of fillers vary from $1\%$ -$10\%$. Comparing the hollow and solid fillers, we have high thermal coductivity for the hollow geometries due to the more surface area, i.e., more interface between the filler and the base polymer.
In fig.~\ref{f23} we show the effective thermal conductivity for a single \textit{hollow cubical box} with diferent sizes. It can be cleary seen from the figure that for the same volume percent of the filler, the box with larger size and hence the larger surface area has the larger effective thermal conductivity. 
In fig.~\ref{f24} we show the effective thermal conductivity for a single \textit{hollow cylinder} with varying diameters. It can be cleary seen from the figure that for the same volume percent of the filler, the \textit{cylinder} with larger size and hence the larger surface area has the highest effective thermal conductivity. 
In fig.~\ref{f25} we show the effective thermal conductivity for a single \textit{solid cubical box} with diferent sizes similar to the figure~\ref{f23}. The behaviour of the solid box is also similar to the corresponding hollow box. However, the one to one comparison between figs.~\ref{f23} and \ref{f25} cleary shows that hollow box has higher effective thermal conductivity in comparison to the solid one for the same volume percent. 
Fig.~\ref{f26} shows the effective thermal conductivity for a single \textit{solid cylinder} with diferent sizes similar to the figure~\ref{f24}. It is observed that variation in effective thermal conductivity in solid cylinder is also similar to the corresponding hollow cylinder. The one to one comparison between figs.~\ref{f24} and \ref{f26} cleary shows that composite polymer with hollow cylinderical filler has higher effective thermal conductivity than to the composite polymer having solid cylinderical filler for the same volume percent.
Figure~\ref{f27} shows the comparison of effective thermal conductivity estimated by the different correlation models. These models give an approximation and the applicability in various conditions using different estimates. The description of these models have been given in the table \ref{table:t8} \cite{APS47,APS48,APS49,APS50,APS51,APS52,APS53,APS54,APS55,APS56,APS57}.

\begin{figure}[H]
\centering
   \includegraphics[width=0.69\textwidth]{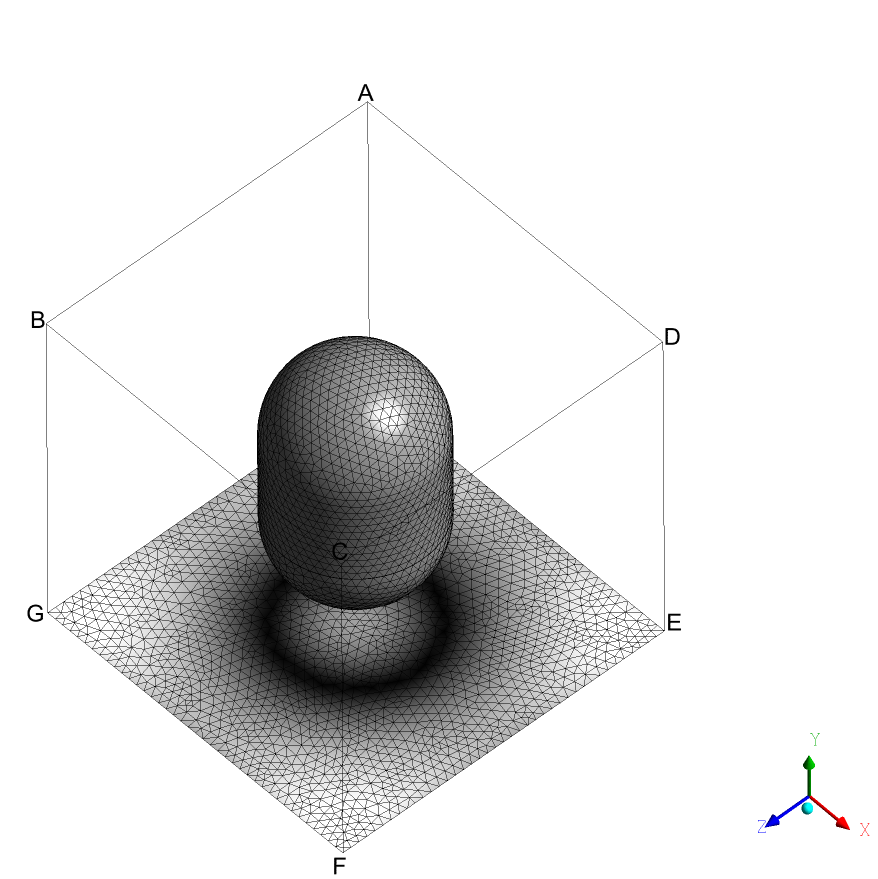}
   \caption{The mesh structure of Capsule-shape figure showimg only bottom surface and filler mesh in the cell with 9\% volume fraction and surface area is 0.75438 $cm^{2}$, and base length 0.21242cm.}
   \label{f28} 
   \includegraphics[width=0.69\textwidth]{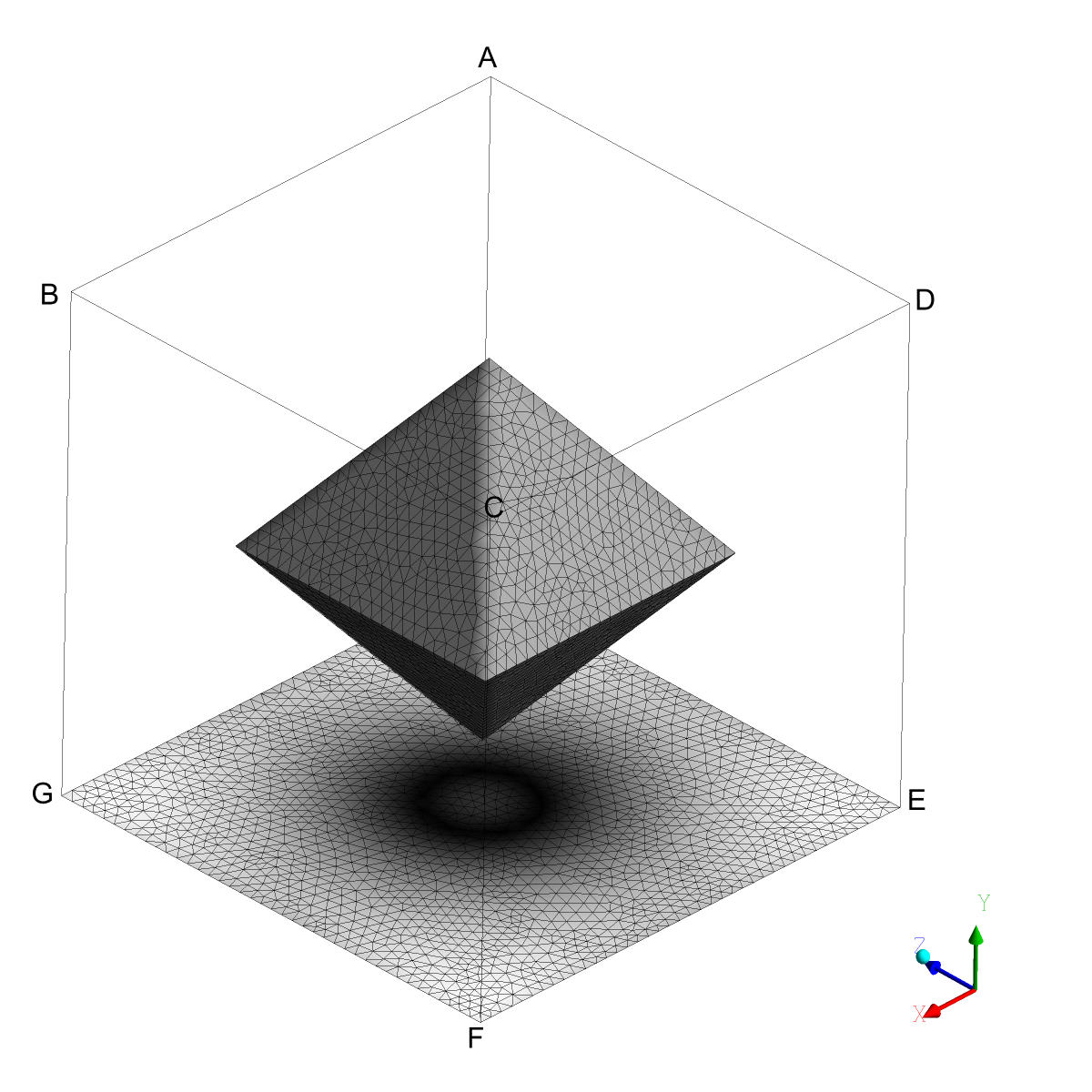}
   \caption {The mesh structure of Octahedron-shape figure showimg only bottom surface and filler mesh in the cell with 9\% volume fraction and surface area is 1.1513 $cm^{2}$, and base length 0.596cm.}
 \label{f29}
\end{figure}

\begin{figure}[H]
\centering
   \includegraphics[width=0.7\textwidth]{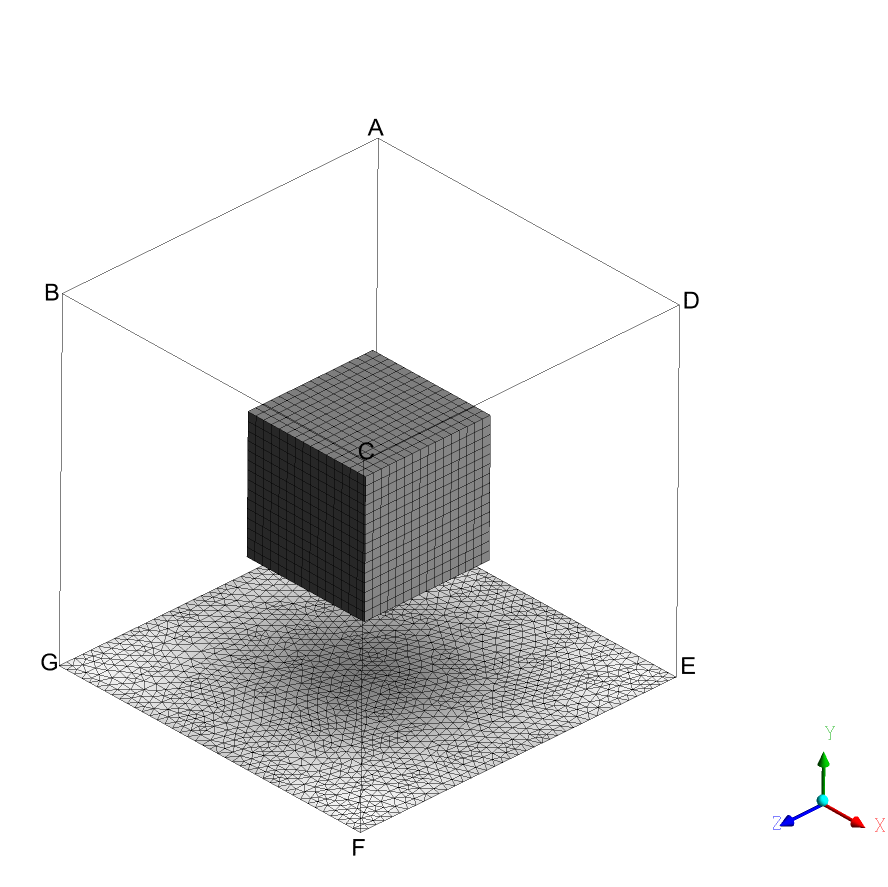}
   \caption{The mesh structure of Cube-shape figure showimg only bottom surface and filler mesh in the cell with 9\% volume fraction and surface area is 1.205 $cm^{2}$, and base length 0.44814cm.}
   \label{f30} 
   \includegraphics[width=0.7\textwidth]{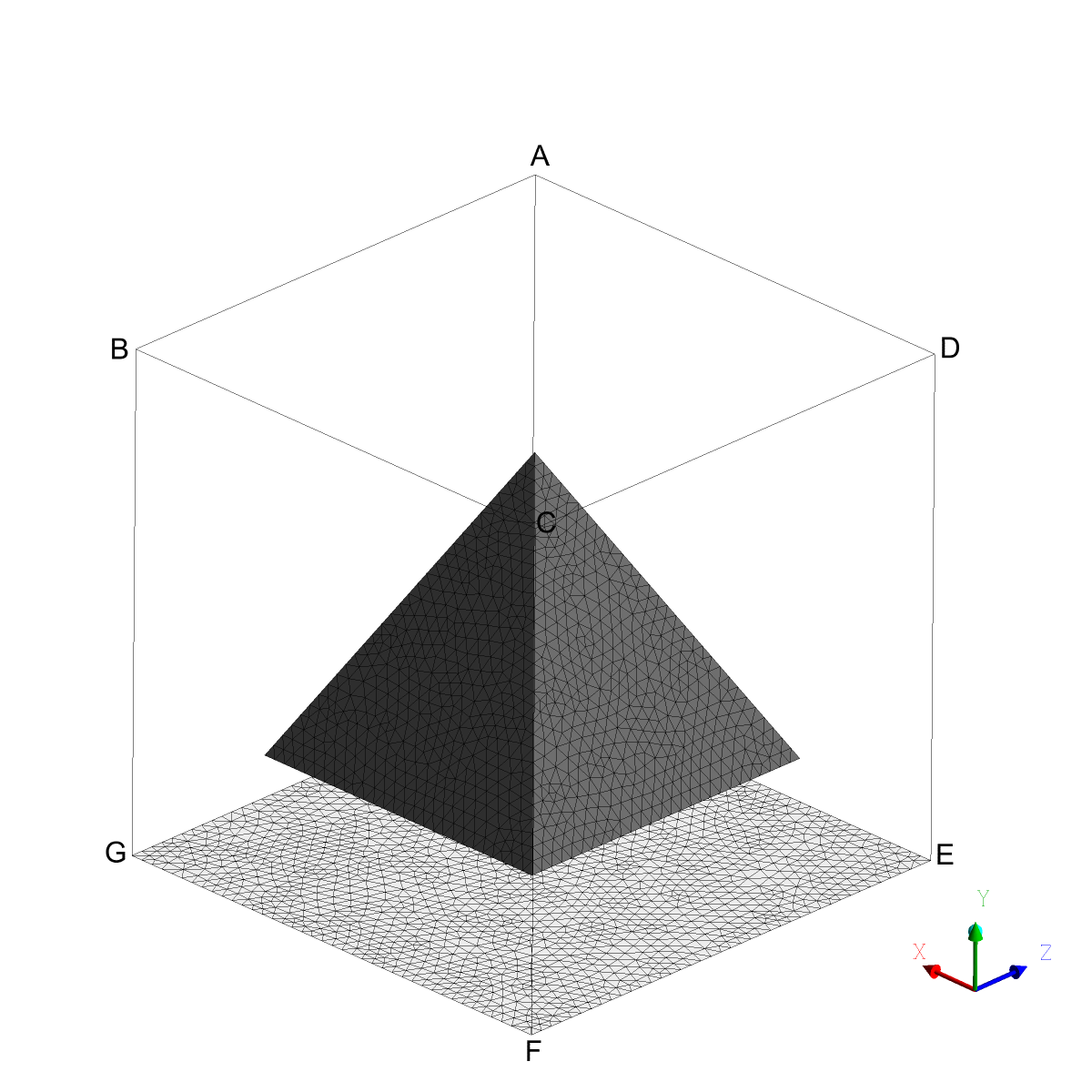}
   \caption {The mesh structure of Tetrahedron-shape figure showimg only bottom surface and filler mesh in the cell with 9\% volume fraction and surface area is 1.3523 $cm^{2}$, and base length 0.647cm.}
 \label{f31}
\end{figure}

\begin{figure}[H]
\centering
   \includegraphics[width=0.83\textwidth]{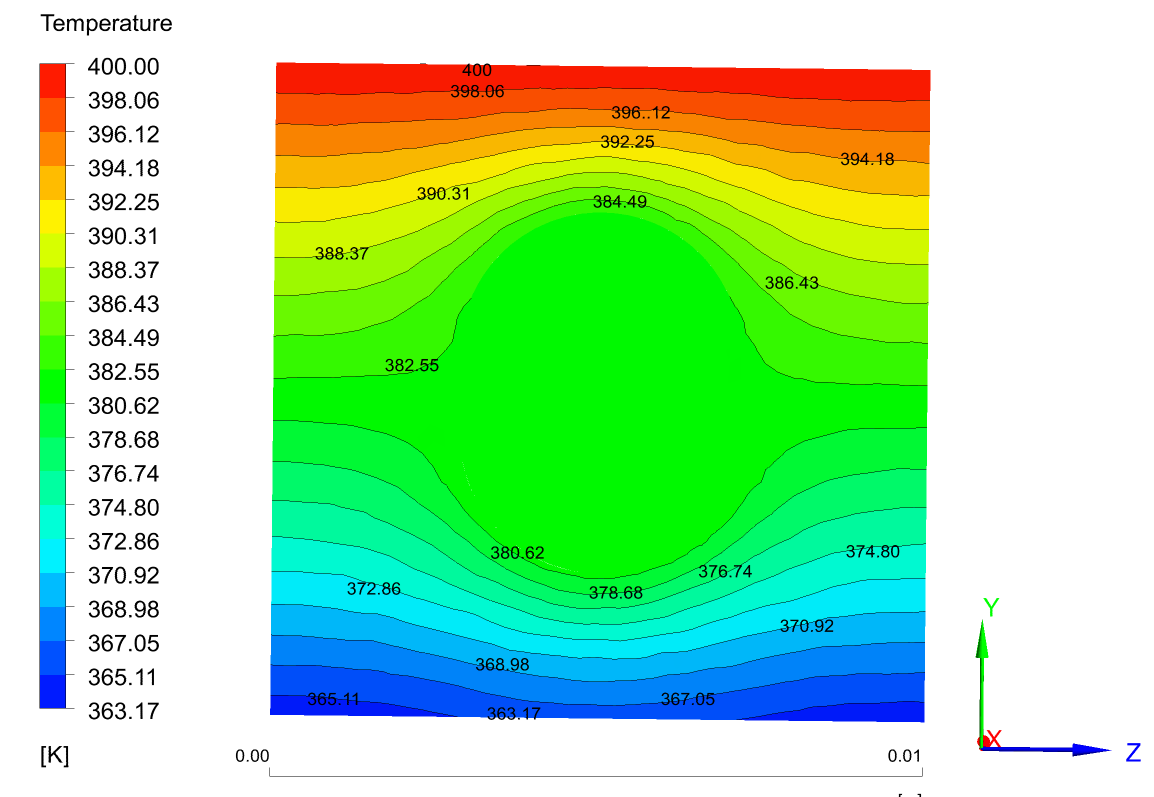}
   \caption{The two dimensional contour of temperature(k) of the cross-section z-y plane ($x = 0.5$ $l_{x}$) for the composite polymer representing the heat flux from the top to the bottom surface. The horizontal lines show the variations of temperature across the surface at different vertical length. The
figure corresponds to the \textit{Capsule-shape} filler with filler content of 9\%, and of base length 0.22242cm.}
   \label{f32} 
   \includegraphics[width=0.85\textwidth]{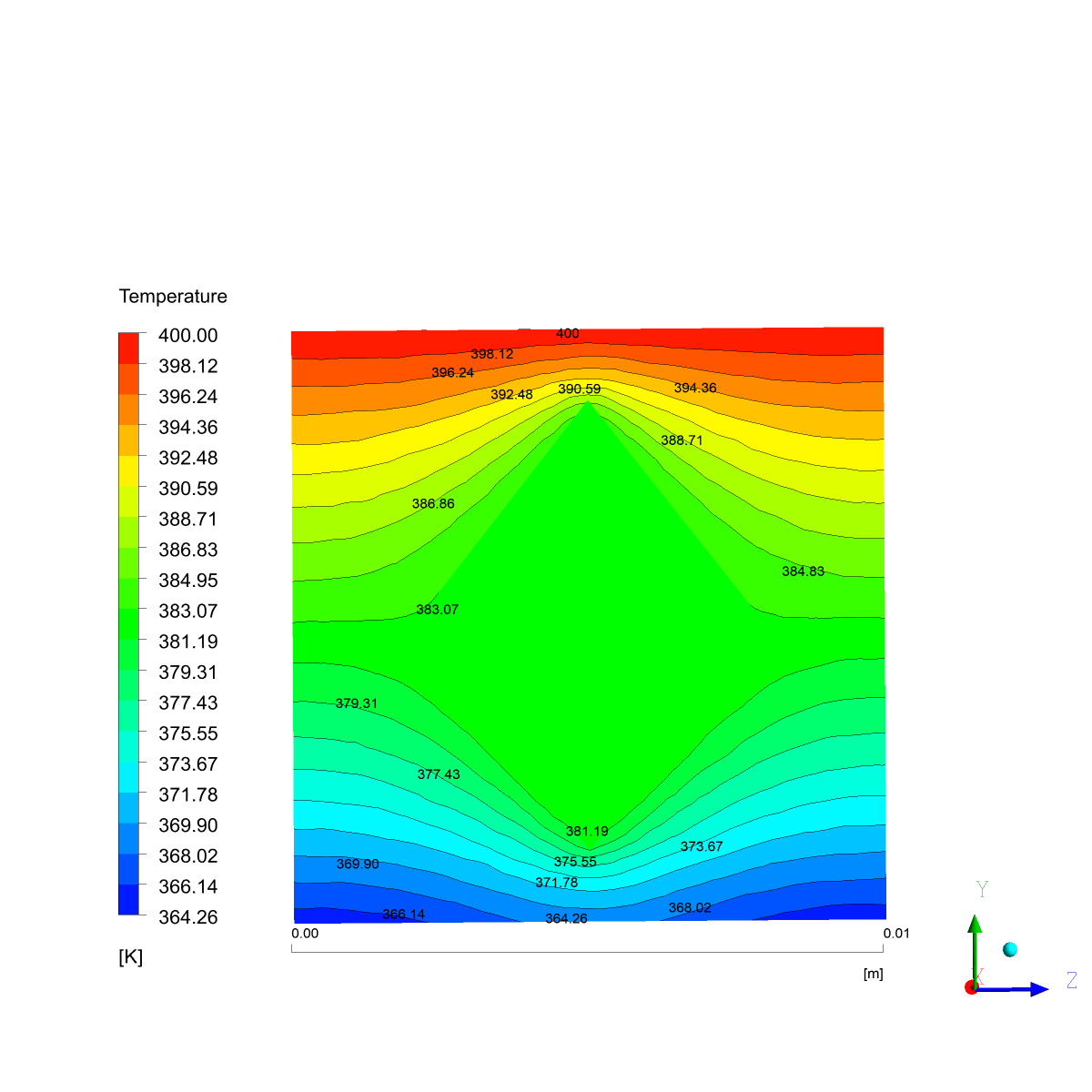}
   \caption {The two dimensional contour of temperature (k) of the cross-section z-y plane ($x = 0.5$ $l_{x}$) for the composite polymer representing the heat flux from the top to the bottom surface. The horizontal lines show the variations of temperature across the surface at different vertical length. The
figure corresponds to the \textit{Octahedron-shape} filler with filler content of 9\%, and of base length 0.647cm.}
 \label{f33}
\end{figure}

\begin{figure}[H]
\centering
   \includegraphics[width=0.78\textwidth]{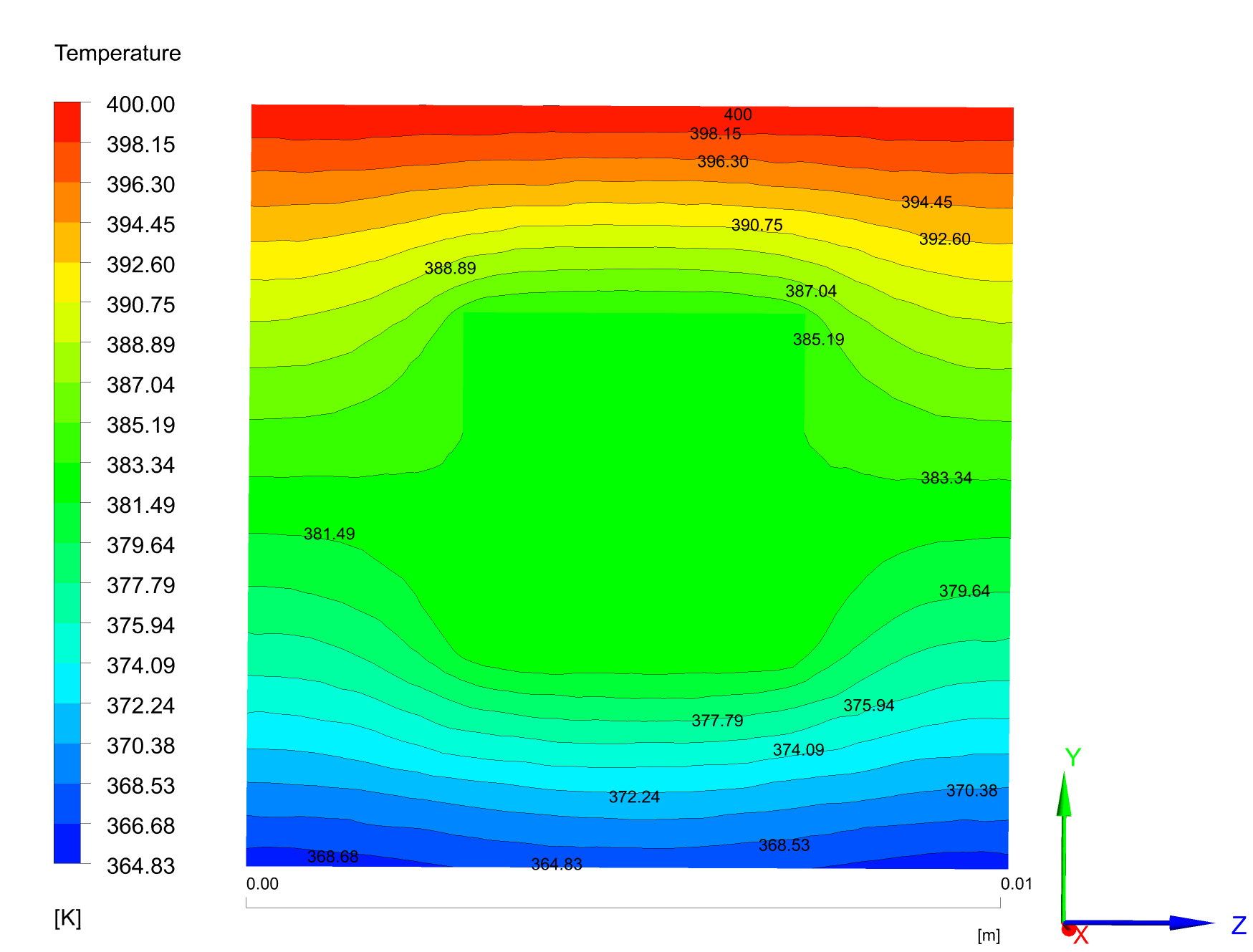}
   \caption{The two dimensional contour of temperature(k) of the cross-section z-y plane ($x = 0.5$ $l_{x}$) for the composite polymer representing the heat flux from the top to the bottom surface. The horizontal lines show the variations of temperature across the surface at different vertical length. The
figure corresponds to the \textit{Cube-shape} filler with filler content of 9\%, and of filler base length 0.596cm.}
   \label{f34} 
   \includegraphics[width=0.78\textwidth]{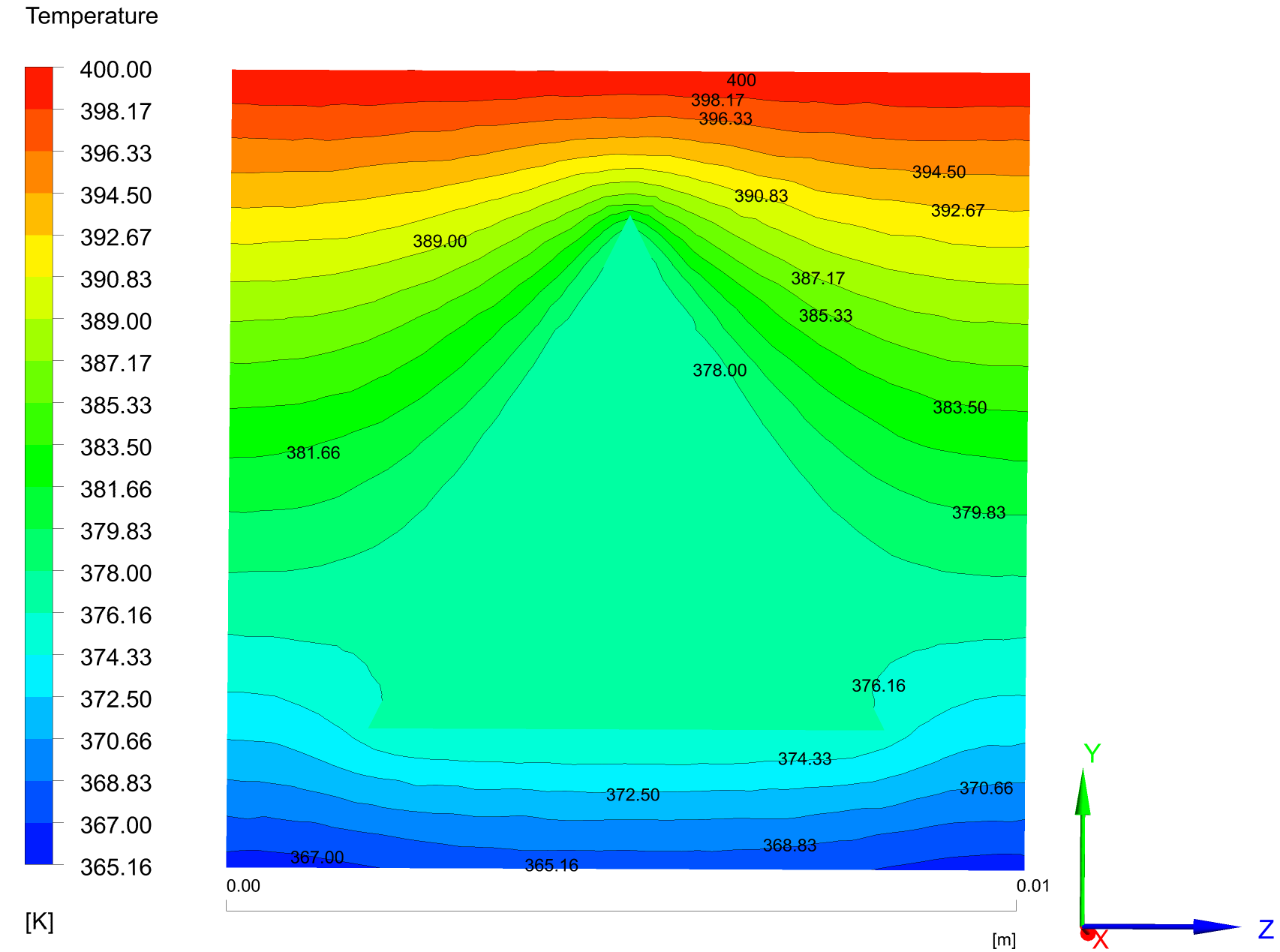}
   \caption {The two dimensional contour of temperature (k) of the cross-section z-y plane ($x = 0.5$ $l_{x}$) for the composite polymer representing the heat flux from the top to the bottom surface. The horizontal lines show the variations of temperature across the surface at different vertical length. The
figure corresponds to the \textit{Tetrahedron-shape} filler with filler content of 9\%, and of base length  0.44814cm.}
 \label{f35}
\end{figure}

\begin{figure}[H]
\begin{center}
\includegraphics*[width=0.6\textwidth]{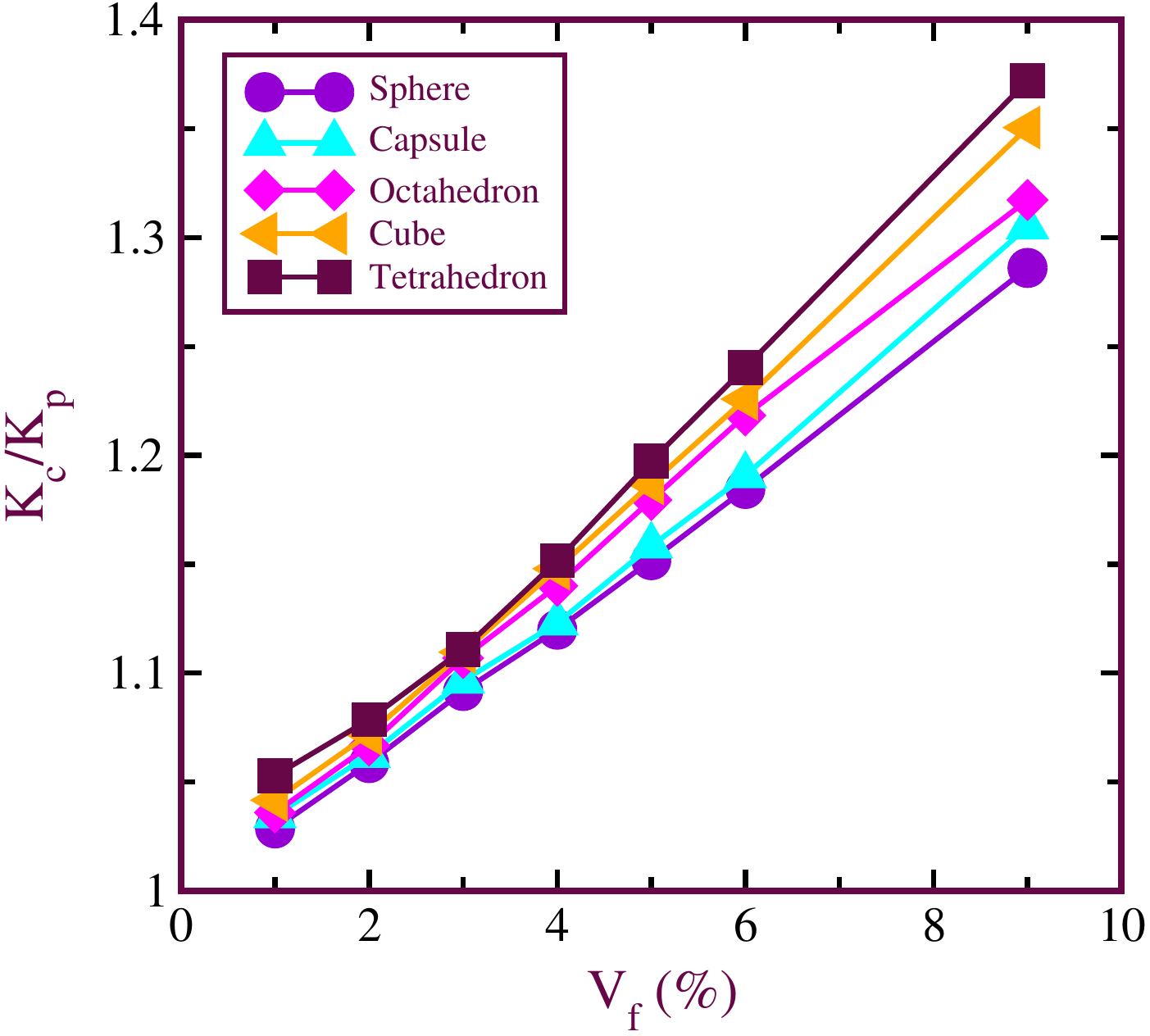}
\end{center}
\caption{ Effect on relative thermal conductivity by increasing the volume fraction of filler with different shape.}
\label{f36}
\end{figure}

We know that among the different geometries the surface area decreases gradually for the rounder shapes, and hence the ratio of surface area ($S$) to volume ($V$) decreases as we keep increasing volume in such cases. It is well known that the surface area of a spherical object is minimum among other geometries for a given volume. Therefore if we take the same volume percent for different geometries in bulk the $S/V$ ratio is lesser for the spherical object if the geometries are designed such that total volume is same in all the cases. Now if we use the spherical objects as fillers in the matrix polymer, we should get the minimum thermal conductivity for the spherical object due to the lower $S/V$ ratio, and hence the minimum interface contact area between the polymer and filler.
Therefore the basic question or the quest, that which geometry should give the maximum thermal conductivity, can be generalized to find out the geometries for which the $S/V$ ratio is maximum with same amount of the bulk of the filler. We have outlined some of the geometries and their $S/V$ ratio.
Figures (\ref{f28}- \ref{f31}) show the various geometries (e.g., capsule, octahedron, cube, tetrahedron) having different surface to the volume $S/V$ ratio.  The corresponding 2-D contour plots for the effective thermal conductivities have been shown in figs.~(\ref{f32}-\ref{f35}).
Figure~\ref{f36} shows the comparison between effective thermal conductivities for the different geometries corresponding to figs.~(\ref{f28}- \ref{f31}). 
We can see that the best geometries are those with multiple faces and sides. However these geometries are not very simple to design due to multiples faces and due to the complex orientations of these faces. 

We also mention that although the bulk (i.e. solid metallic filler) plays an important role in maximizing the thermal conductivity in polymer composites, however as the bulk of the filler increases it replaces the bare polymer and hence the composite behaves more like the metallic phase.
Note that by using more bulk/material of the filler we are allowing more path or extra routes for the passage of the heat flux through the metal in the composite polymer. Therefore the important factor is to get more metallic path for heat flux. More precisely this metallic path is the interface at the filler and polymer. Now, if instead of using more bulk to create the extra metallic path for the heat flux, if we transform the same amount of filler by thinning it into extra surfcaes our job is done. Our job is to create extra surfaces. This can be done either by (1) filling more metal in the composite or (2) by transforming the same amoount of the filler into more surface. Definitely the second option serves our purpose.
When we maximize the contact area, i.e., the interface between the polymer and metallic filler, we allow more path and hence extra routes for passage of heat flux through the metal and thus the net heat flux through the whole composite increases. Thus if we transform the same amount of bulk of the filler into more surface, we can generate extra surfaces for the passage of heat flux leading to the enhancement of the effective thermal conductivity. This procedure definitely relies on the minimal usage of filler.

The surface area maximization can be done by making the replica of the filler geometries. In this process the bulk of the filler can be utilized to maximize the surface.  By doing this the new replica of filler are more thinner but they still carry the heat-flux. Thus increasing the number of fillers or the replica by utilizing the bulk we enable more routes. 
In our numerical examples we take simple geometries to show these features. Therefore these simple geometries enable to maximize the surface area just by minimizing the bulk. 
The different geometries which can maximize the surface area and hence the interface between the filler and the base polymer can be be listed by their $S/V$ ratio. These geometries correspond to the shapes whose surface to volume ratio is relatively more. 

\begin{figure}[H]
\begin{center}
\includegraphics*[width=.7\textwidth]{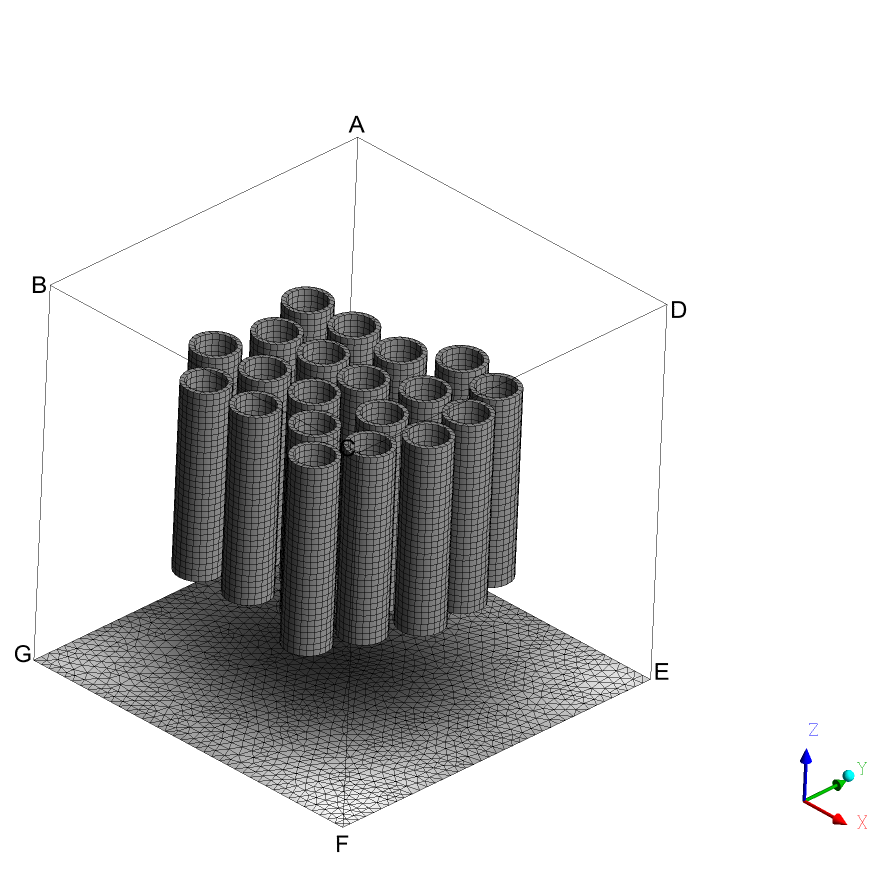}
\end{center}
\caption{ The mesh structure of the composite polymer having varying \textit{hollow cylinderical fillers}. In this case we have $20$ such vertical hollow cylinders. Note that the volume percent of filler content is fixed to a constant value. Thus as we increase the number of these hollow cylinderical fillers the thickness is reduced and contact area between the filler and polymer interface is increased. Note that the the minimum thickness that can be used in these fillers is still macroscopic and has the bulk metallic properties. 
showimg only bottom surface and filler mesh in the cell with $4\%$ volume fraction of $0.011427$ cm thickness with $0.5$cm filler length.}
\label{f37}
\end{figure}

\begin{figure}[H]
\begin{center}
\includegraphics*[width=0.7\textwidth]{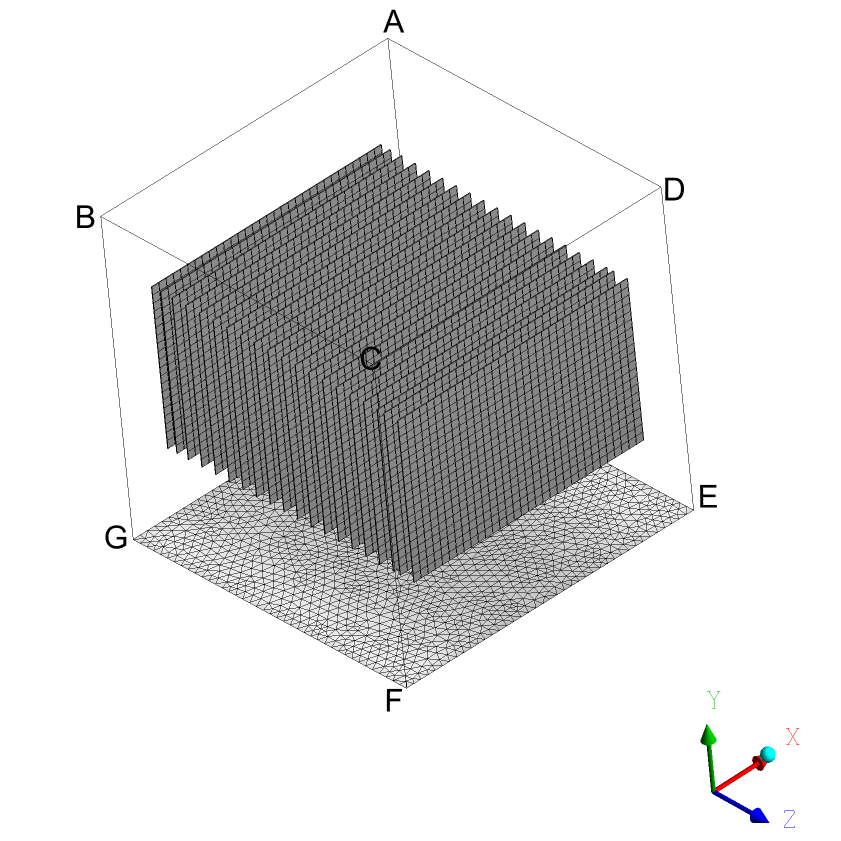}
\end{center}
\caption{ The mesh structure of the composite polymer with fillers in the shape of \textit{parallel sheets kept vertically} and varying in number. In this representative figure we have shown 30 sheets fillers. Again the volume percent of the filler content is fixed to a constant value. By increasing the number of sheets or their dimensions only reduces the thickness of the fillers, but the surface area is maximized in this process. The sheets like fillers are simple to fabricate and easy to implement between the polymer matrix. We also note that if we put the sheets horizontally the symmetry is changed and the it needs much heat flux to pass homogenously to the bottom. 
showimg only bottom surface and filler mesh in the cell with 2\% volume fraction of 0.001667cm thickness with 0.5cm filler length and width 0.8cm.}
\label{f38}
\end{figure}

\begin{figure}[H]
\begin{center}
\begin{tabular}{cc}
\includegraphics*[width=0.67\textwidth]{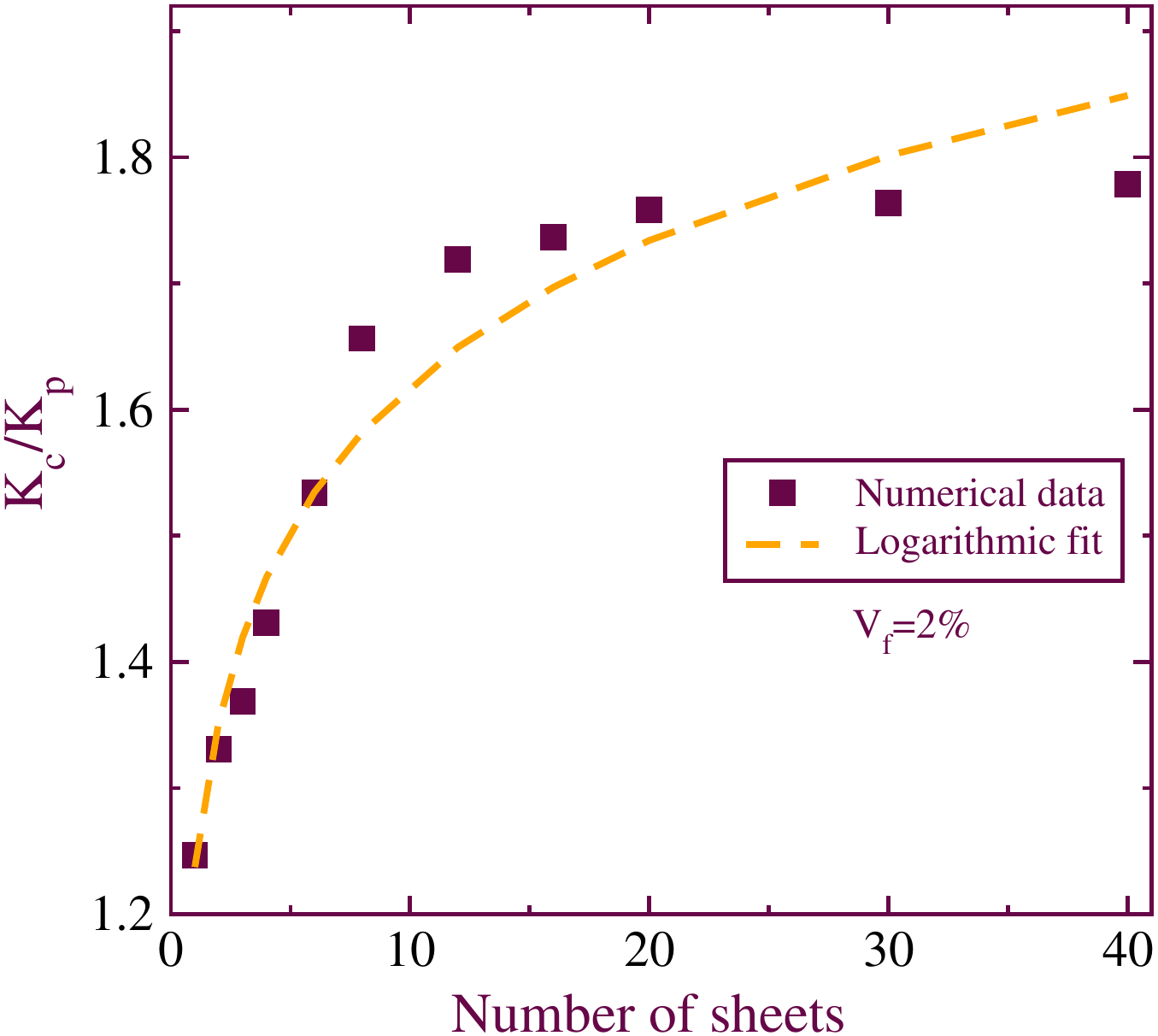} & \ 
 \end{tabular}
\end{center}
\caption {Effect on relative thermal conductivity in the composite with 2\%  multi sheets(30) of size 0.8x0.5(cm xcm). }
\label{f39}
\end{figure}

\begin{figure}[H]
\begin{center}
\begin{tabular}{cc}
\includegraphics*[width=0.67\textwidth]{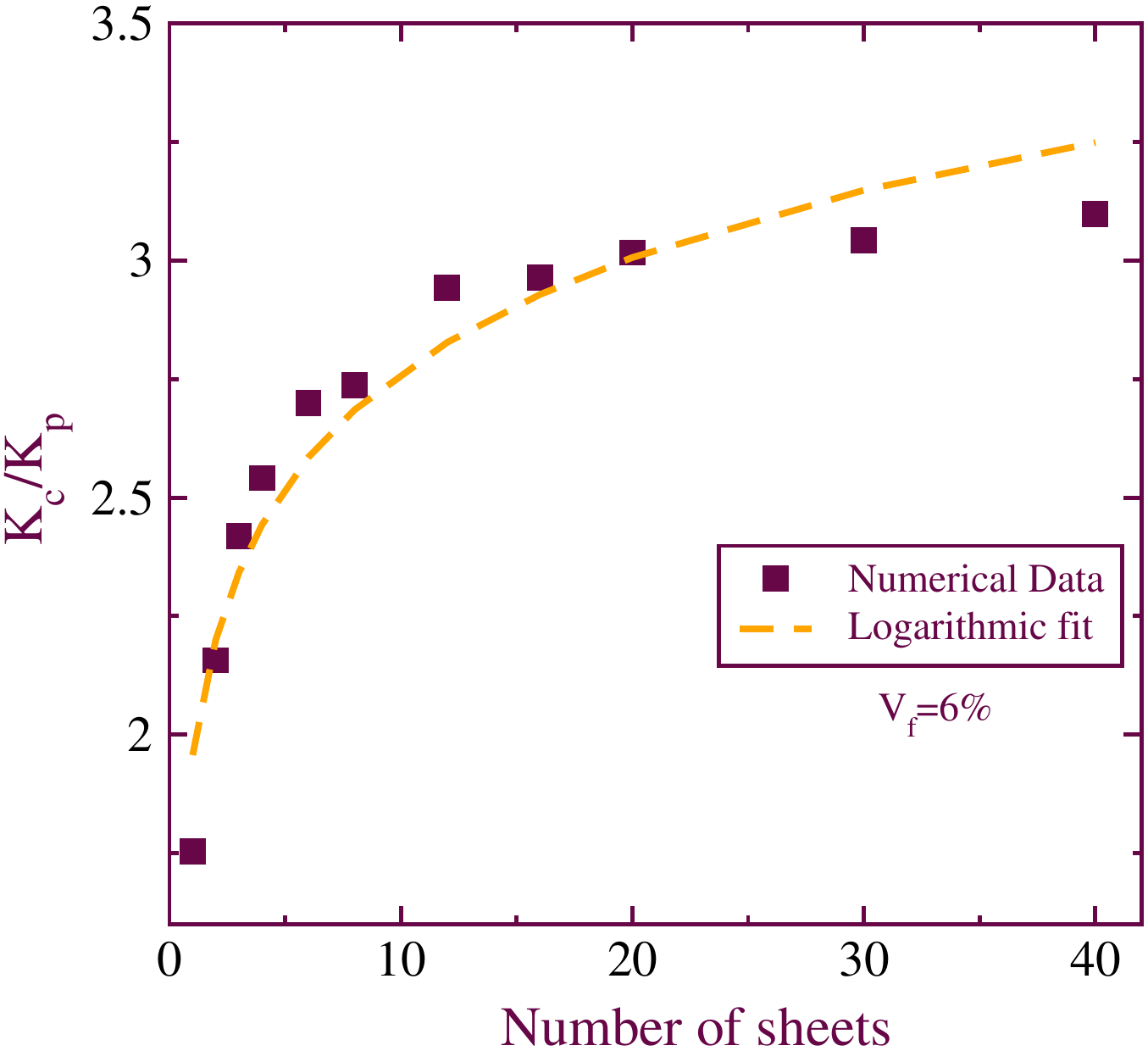} & \ 
\end{tabular}
\end{center}
\caption {Effect on relative thermal conductivity in the composite with 6\%  multi sheets(40) of size 0.5x0.8(cm x cm).}
\label{f40}
\end{figure}

\begin{figure}[H]
\centering
   \includegraphics[width=0.75\textwidth]{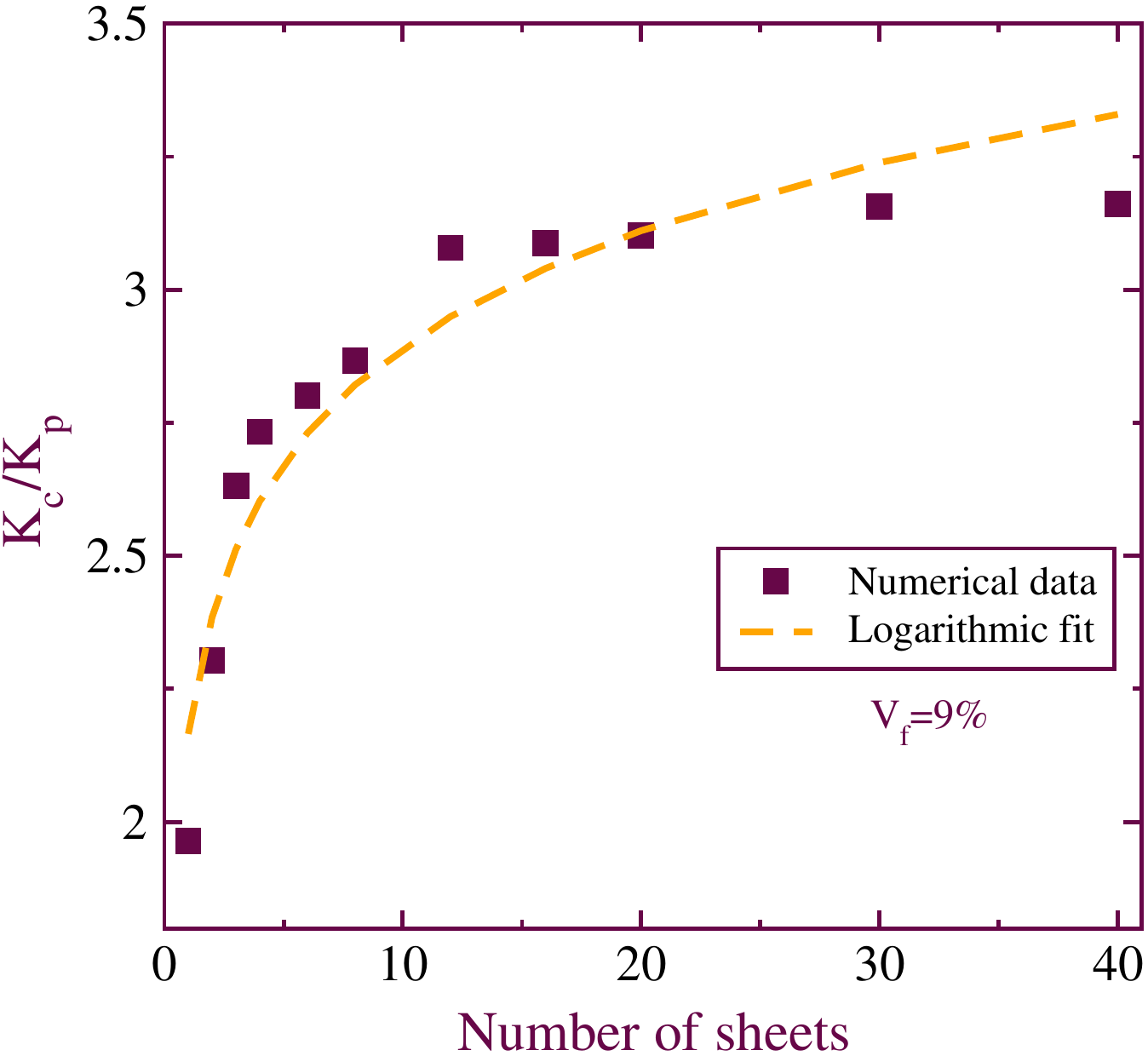}
   \caption{Effect on relative thermal conductivity in the composite with 9\% parralel sheets(40) of size 0.5x0.8(cm x cm).}
   \label{f41} 
   \includegraphics[width=0.9\textwidth]{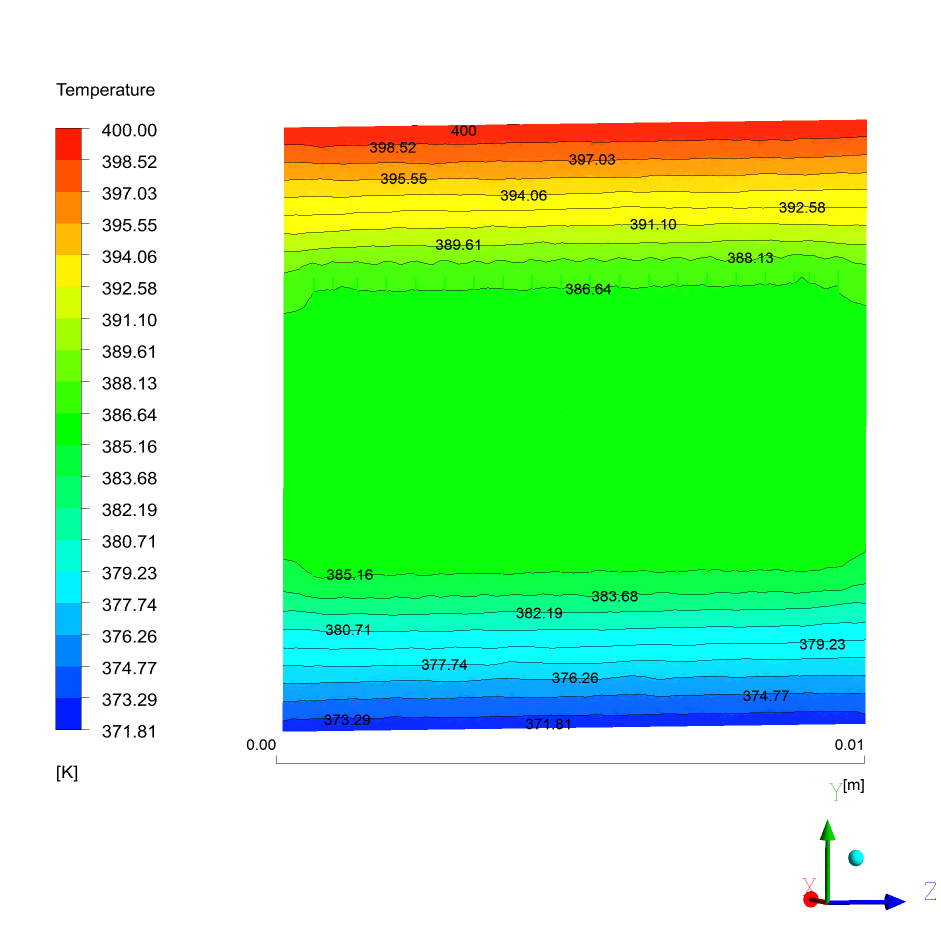}
   \caption{The two dimensional contour of temperature(k) of the cross-section z-y plane ($x = 0.5$ $l_{x}$) for the composite polymer representing the heat flux from the top to the bottom surface. The horizontal lines show the variations of temperature across the surface at different vertical length. The filler content of 2\%  multi sheets(30) of size 0.8x0.5(cm x cm) and thicknees 0.00125cm. }
\label{f42}
\end{figure}

Now we discuss the two important cases to enhance the thermal conductivity by maximizing the interface (i.e., by maximizing the contact surface area in system of composite polymers). 
Note that we have chosen the simple geometries such as planar sheets and cylindercal surfaces to show the representative behaviour of thermal conductivity due to surface maximization. 
For simplicity we consider the following two common geometries (a) multiple planar sheets as shown in fig.~\ref{f37}, and (b) multiple hollow cylinders as shown in fig.~\ref{f38}. In the first example we consider the composite polymers with planar sheets (figure~\ref{f37}) as the metallic fillers. The dimensions (i.e., length and the width) of the each filler is same and fixed. We consider the three distinct cases with volume percent of filler $2\%,~6\%$, and $9\%$ respectively in the composite. In each case the volume percent is always kept constant, i.e., if the number of planar sheets increases the the thickness decreases in same proportion such that the total volume percent does not changes. 
Figures~(\ref{f39}-\ref{f41}) show the corresponding effective thermal conductivity $K_{c}/K_{p}$ for the respective cases. In each of these cases the number of sheets vary upto $40$. These figures clearly indicate that effective thermal conductivity increases with the increasing number of sheets. 
It is also clear that as the volume percent increases the size and hence the surface area of the sheets increases maximizing the thermal conductivity. We also note that we can not put more than a certain number of sheets in the base polymer. We can see that \textit{initially the increase in thermal conductivity is linear, and after then it starts saturating and the growth in the thermal conductivity goes like logarithmic.}

\begin{figure}[H]
\centering
   \includegraphics[width=0.75\textwidth]{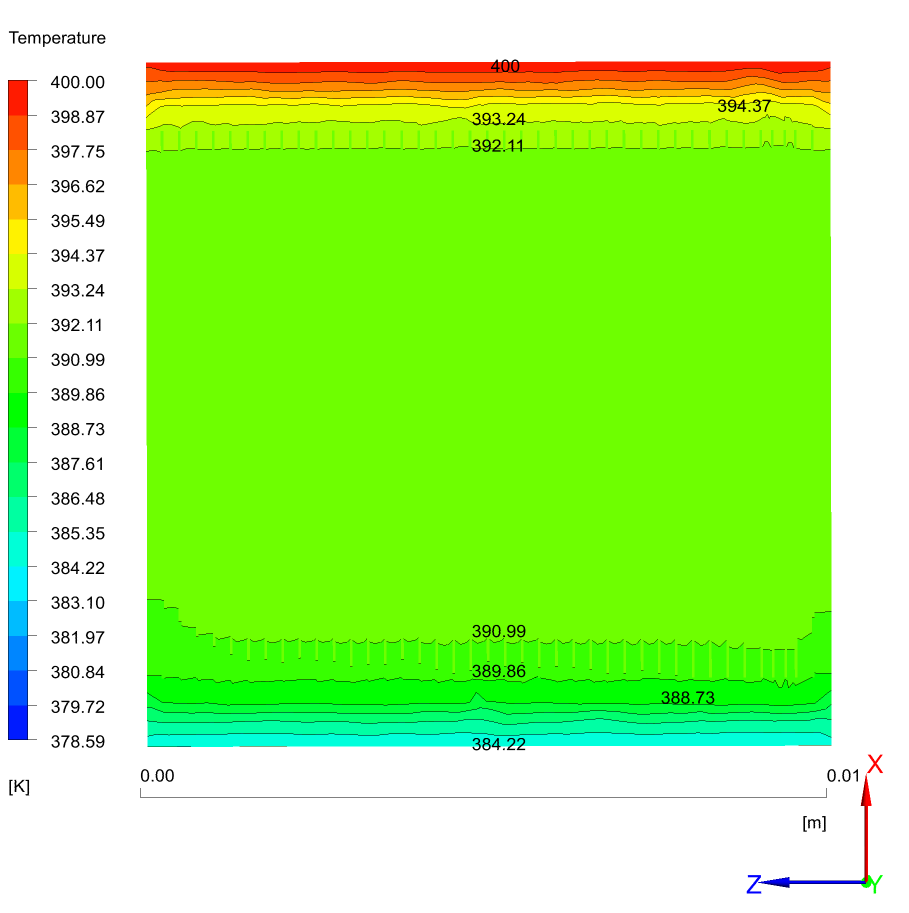}
   \caption{The two dimensional contour of temperature(k) of the cross-section z-x plane ($y = 0.5$ $l_{y}$) for the composite polymer representing the heat flux from the top to the bottom surface. The horizontal lines show the variations of temperature across the surface at different vertical length. The filler content of 6\% multi-sheets(40) of size 0.5x0.8(cm x cm) and thicknees 0.00375cm.}
   \label{f43} 
   \includegraphics[width=0.75\textwidth]{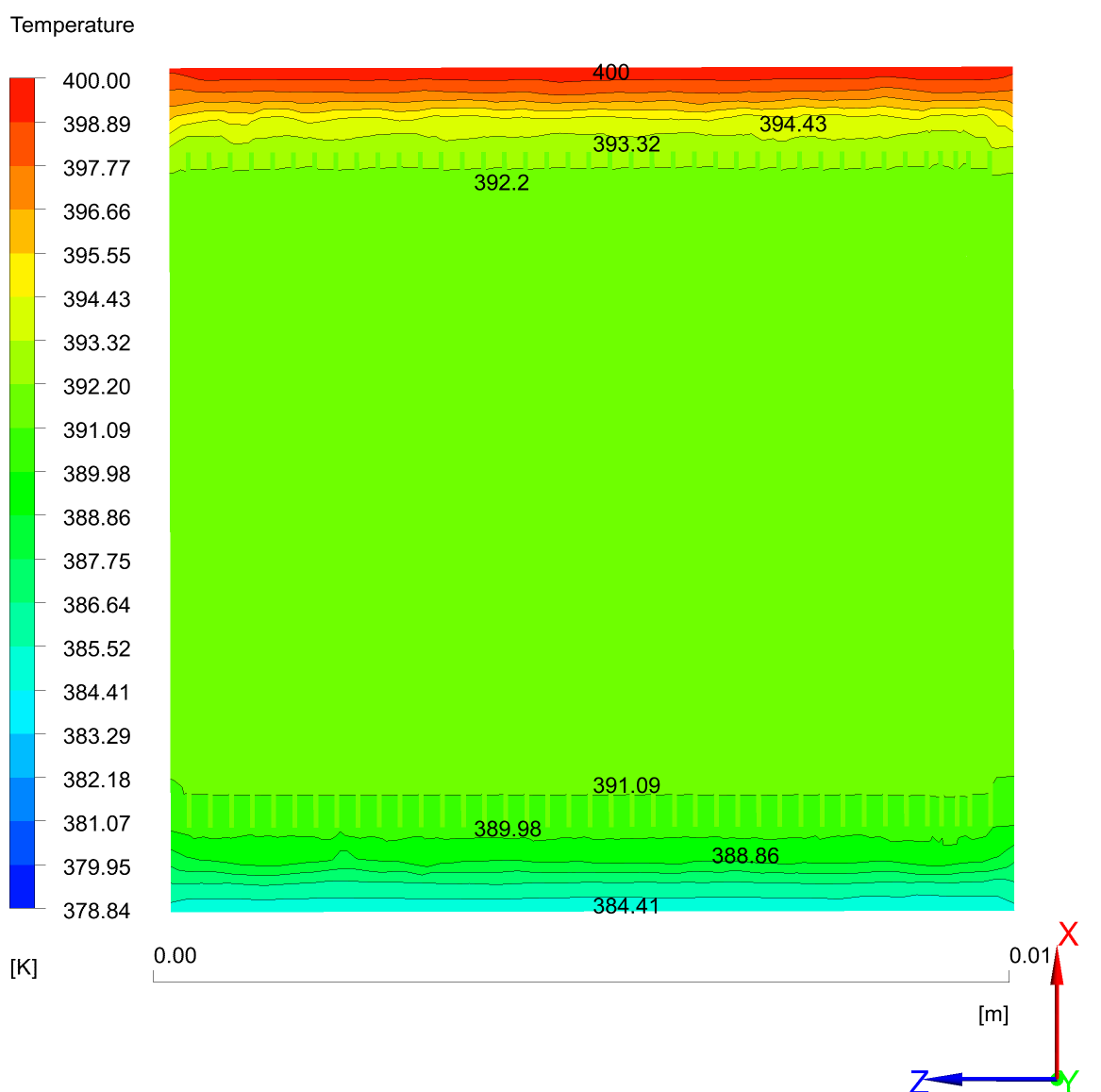}
   \caption{The two dimensional contour of temperature(k) of the cross-section z-x plane ($y = 0.5$ $l_{y}$) for the composite polymer representing the heat flux from the top to the bottom surface. The horizontal lines show the variations of temperature across the surface at different vertical length. The filler content of 9\% parralel sheets(40) of size 0.5x0.8(cm x cm) and thickness 0.005625cm.}
 \label{f44}
\end{figure}

In figs.~(\ref{f42}-\ref{f44}) we show the two dimensional contour plots of temperature gradients for the corresponding three distinct cases. All these contour diagrams correspond to the $40$ planar sheets in all the three cases. For the lesser number of sheet fillers the behaviour is similar but the temperature gradient is low. The horizontal lines show the variations of temperature across the surface at different vertical lengths. These contour lines clearly mark the intensity and rate of heat-flux for different cases mentioned above. We observe that as the number of sheets increases the thermal conductivity increases in all the three cases. This also suggests that for a given volume percent of the filler, by maximizing the interface and contact surface area we can maximize the effective thermal conductivity. 

\begin{figure}[H]
\begin{center}
\begin{tabular}{cc}
\includegraphics*[width=0.65\textwidth]{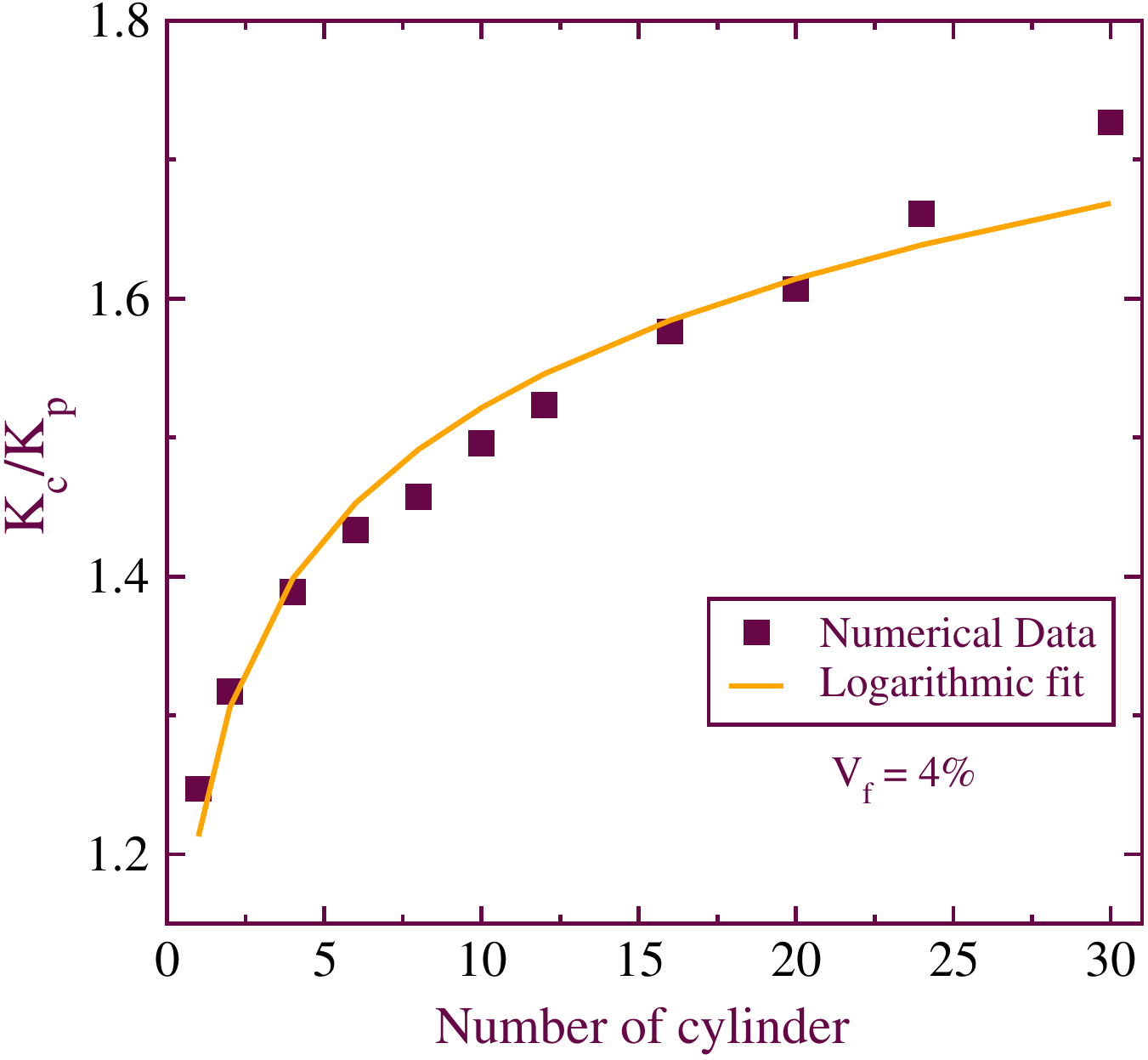} & \ 
 \end{tabular}
\end{center}
\caption {Effect on relative thermal conductivity by decreasing the thickness of multi hollow cylinder(20) and increasing the number of parallel hollow cylinder with constant volume fraction of filler 4\% and filler length 0.5cm in the cell.}
\label{f45}
\end{figure}

\begin{figure}[H]
\centering
 \includegraphics[width=0.7\textwidth]{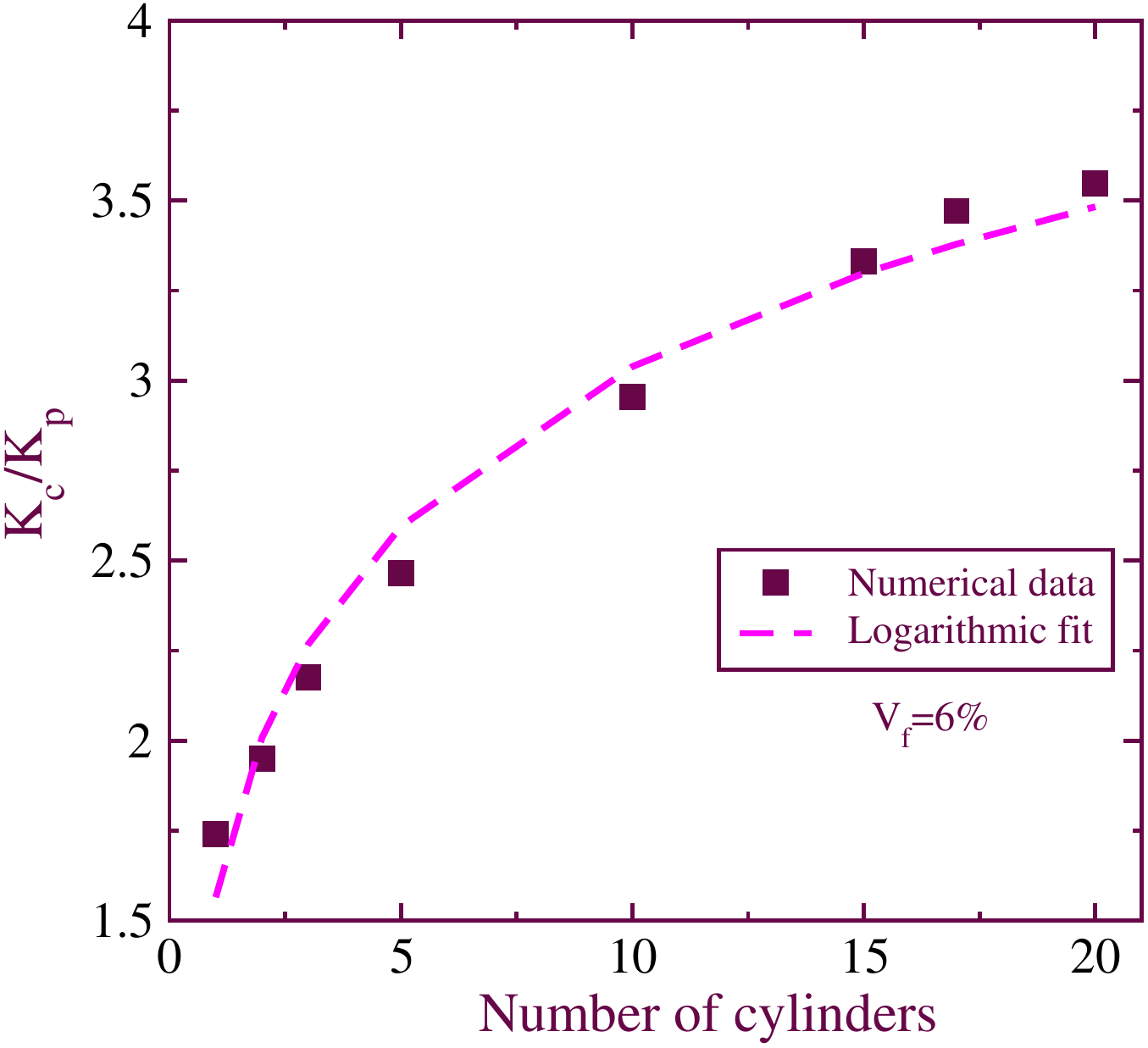}
 \caption{Effect on relative thermal conductivity by decreasing the thickness of multi hollow cylinder(20) and increasing the number of parallel hollow cylinder with constant volume fraction of filler 6\% and filler length 0.8cm in the cell.}
   \label{f46} 
   \includegraphics[width=0.7\textwidth]{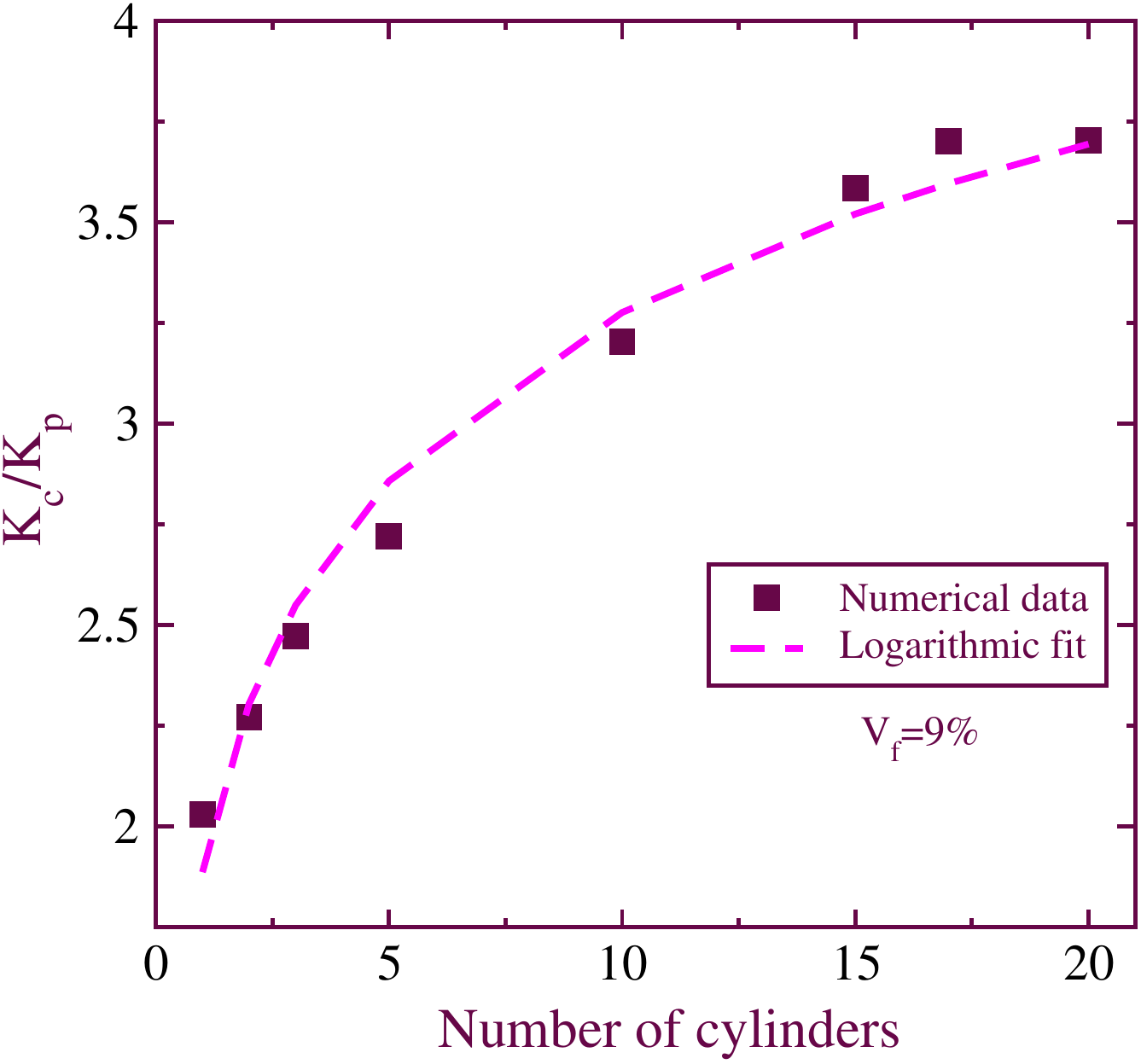}
   \caption{Effect on relative thermal conductivity by decreasing the thickness of multi hollow cylinder(20) and increasing the number of parallel hollow cylinder with constant volume fraction of filler 9\% and filler length 0.8cm in the cell.}
 \label{f47}
\end{figure}

In figs.~(\ref{f45}-\ref{f47}) we show the effective thermal conductivity for the cylinderical fillers. In this case also we have considered the three distinct cases with $4\%,~6\%$, and $9\%$ volume percent of the filler material in the base polymer to compare the effective thermal conductivity. In each of the three cases we have considered maximum upto $20$ identical cylinders.
These figures clearly show that as the number of cylinders increases the effective thermal conductivity also increases. In each case the filler volume fraction is kept constant. With increasing the replica or the number of cylinders only the thickness of the cylinders decreases. We also observe that initial increase in the thermal conductivity is linear and then becomes slow finally behaving as logarthimic growth.  This behaviour is similar to the case of paralle planae sheet fillers. 

\begin{figure}[H]
\centering
   \includegraphics[width=0.78\textwidth]{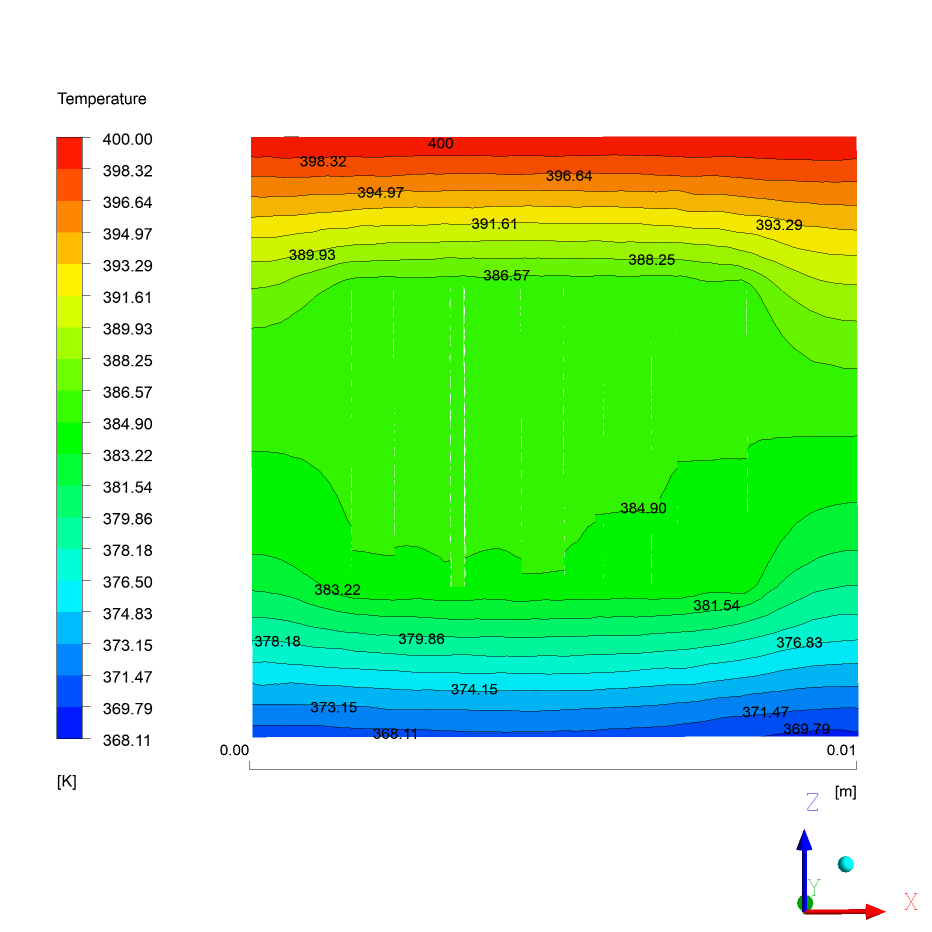}
   \caption{ The two dimensional contour of temperature(k) of the cross-section x-z plane ($y = 0.5$ $l_{y}$) for the composite polymer representing the heat flux from the top to the bottom surface. The horizontal lines show the variations of temperature across the surface at different vertical length. The filler content of $4\%$ hollow multi cylinder(20) of size length 0.5cm and thickness 0.11427cm.}
   \label{f48} 
   \includegraphics[width=0.72\textwidth]{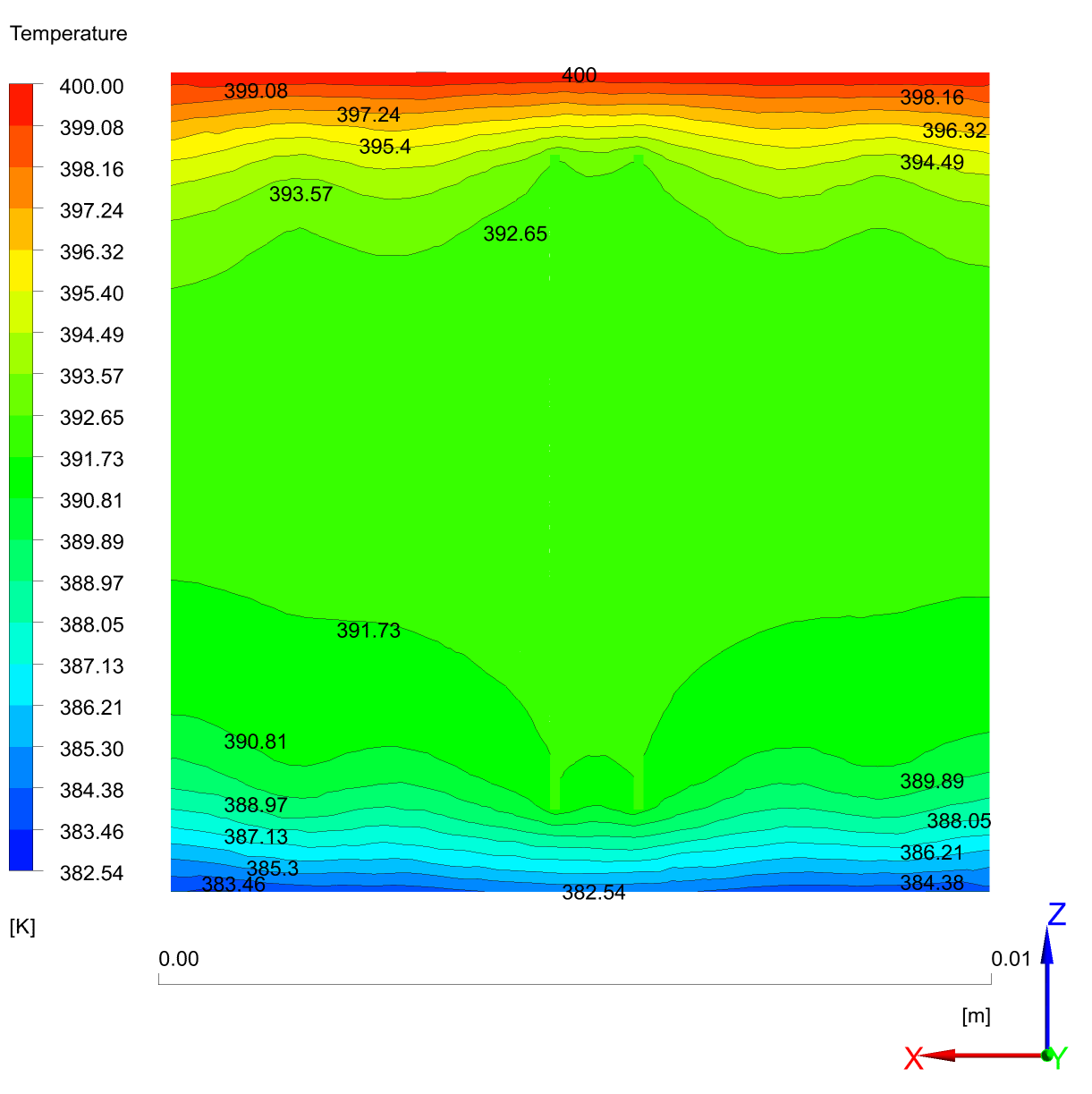}
   \caption {The two dimensional contour of temperature(k) of the cross-section x-z plane ($y = 0.5$ $l_{y}$) for the composite polymer representing the heat flux from the top to the bottom surface. The horizontal lines show the variations of temperature across the surface at different vertical length. The filler content of 6\% hollow multi cylinder(20) of size length 0.8cm and thickness 0.06078cm.}
 \label{f49}
\end{figure}

\begin{figure}[H]
\begin{center}
\includegraphics*[width=.7\textwidth]{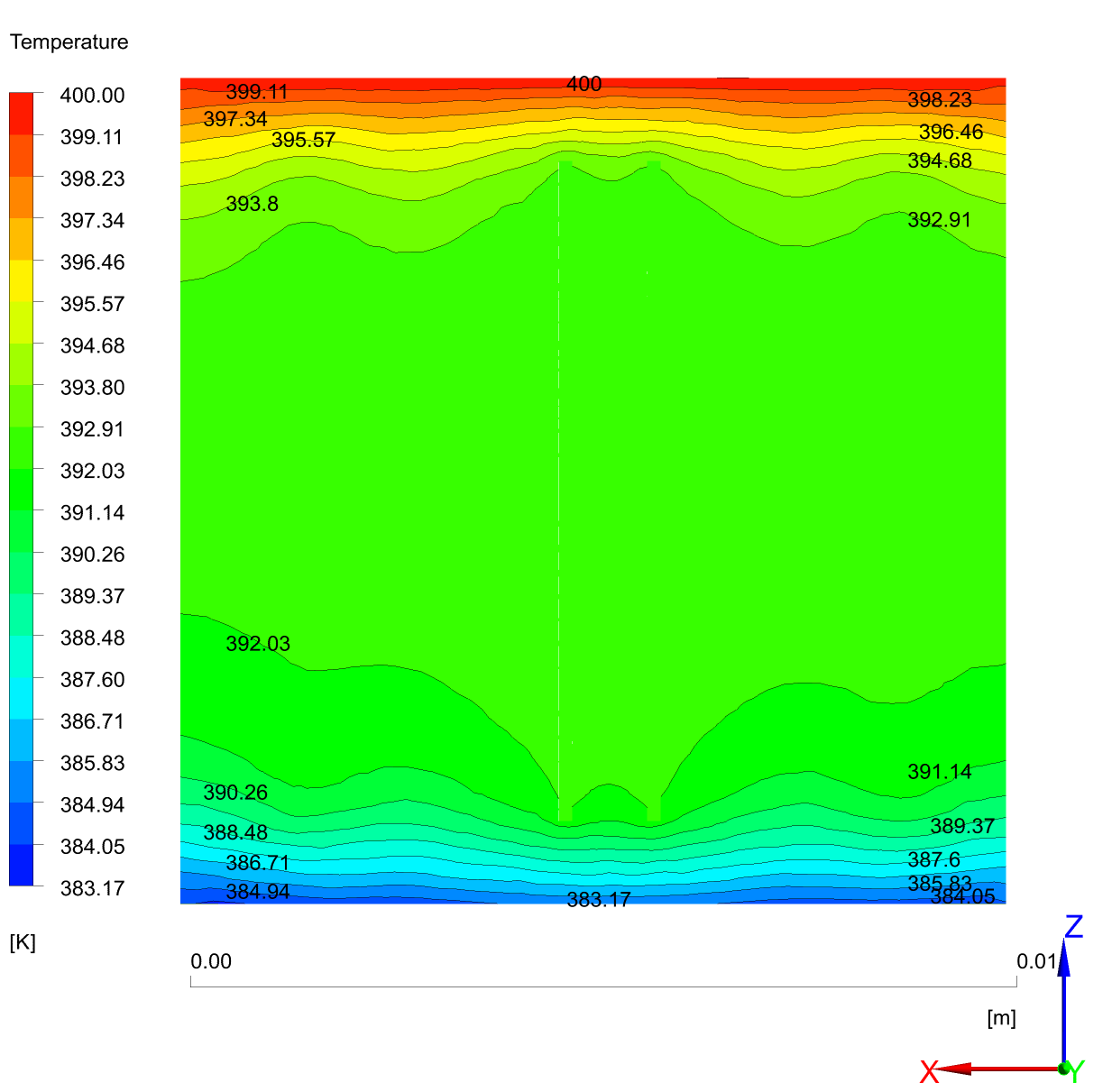}
\end{center}
\caption{  The two dimensional contour of temperature(k) of the cross-section x-z plane ($y = 0.5$ $l_{y}$) for the composite polymer representing the heat flux from the top to the bottom surface. The horizontal lines show the variations of temperature across the surface at different vertical length. The filler content of 9\% hollow multi cylinder(20) of size length 0.8cm and thickness 0.065509cm.}
\label{f50}
\end{figure}

In figs.~(\ref{f48}-\ref{f50}) we show the two dimensional contour plots of temperature gradients for the corresponding three distinct cases for the cylinderical fillers. In all these contour diagrams we have the case for the $20$ identical cylinderical fillers in all the three distinct cases. The horizontal lines show the variations of temperature across the interface of the filler-polymer interface contact surface at different vertical lengths. These contour lines clearly mark the intensity and rate of heat-flux for different cases mentioned above. We observe that as the number of cylindres increases the thermal conductivity increases in all the three cases. This also suggests that for a given volume percent of the filler, by maximizing the interface and contact surface area we can maximize the effective thermal conductivity. The behaviour of effective thermal conductivity for both the fillers, i.e., the parallel sheets and the identical cylinders is similar.

Tables \ref{table:t1}-\ref{table:t3} and tables \ref{table:t7}-\ref{table:t8} describe the dimensions of both the parallel sheets and cylinderical fillers. The correponding thermal conductivity is also shown in each of the cases. In both the situation thermal conductivity increases with increasing surface of the metallic filler although the volume percent is constant. 
In table \ref{table:t4} we also show that if the volume percent of the filler is increased the thermal conductivity does not increase and it becomes constant. This because the contact surface area does not increases anymore even though the bulk is increasing. This celarly establishes the fact that it is just not the bulk which increase the effective thermal conductivity. Our results clearly show that even for lesser volume percent of the filler we can maximize the conductivity which would have been achieved if larger bulk or the volume fraction. Our results are technologically important in fabrication and design of the composite polymers and less cost prone. 
It is clear that both the surface area as well as the bulk of the filler content determine the magnitude of the effective conductivity. The temperature gradient increases with both the surface area as well as the bulk of filler. The exact ratio between these is precisely not defined but as clear from the respective thermal conductivity figures we can make an estimate.  
Our study provides guiding principles for thoughtful consideration for designing process.
It will be desired to get the exact analytical expressions describing the connection between the surface area and the voulme of the filler in the composite polymer.

\newpage
\section{Summary}
In this paper we have outlined the detailed mechanism to maximize the effective thermal conductivity in polymer-composites in which the metallic fillers are used as catalyst for thermal conduction. 
It has been shown that one of the possible route to increase the thermal conductivity without increasing the volume percentage of the filler is to maximize the interfacial area between polymer and filler. Interface represents the surface area in contact between filler and polymer in the composite system. Note that the polymer composite is treated as two phase binary system where both the phases, i.e., the polymer and the metallic filler retain their individual chemical property separated by the interface.  
This study provides a new perspective on the behaviour of effective thermal conductivity based on the interplay between surface-area $(S)$ and the bulk $(V)$ of the filler. We have given the quantitative description of thermal conductivity based on surface-area to volume (i.e., $S/V$) ratio in the composite polymers.

The conventional techniques that have been adopted to get the higher thermal conductivities need a larger amount (e.g.,$>~10\%$ ) of the filler material, but our approach facilitates profound route where the effective thermal conductivity can be increased more than double by using as low as of $6\%$ to $9 \%$ volume of the filler material.
This is remarkable in the sense that we have to utilize very low quantity of the metallic filler which does not alter the inherentence properties of the base polymer. 
Our study shows the following very important facts. (a) The best shape or the best geometries of the fillers which can enhance the thermal conductivity are those which have the high surface-area $S$ to the volume $V$ ratio, i.e., the high specific surface area $S/V$. We have used fillers of different orientations, shape and sizes to confirm the validity of our techniques. 
We show that as the surface-area increases the thermal conductivity can be enhanced more than $100\%$.
Note that composites with the larger amount of filler material does indeed show higher thermal conductivity but then the behaviour of the composite system is more like metallic filler. This clearly violets our basic assumption of the fabrication of composite polymers by using the minimum amount of the filler material. In this perspective our study provides an important procedure to enhance the thermal conductivity by using very minimal usage of the filler.

It is proposed that the ratio of the surface-area to volume of the filler is an important parameter and deciding factor for the enhancement in the thermal conductivity. Although the exact relation betwen these two factors is a subject of further investigation. It can also be proposed that the effective conductivity is a function of length-scale $l$ of the filler which is very clear, since the ratio $S/V$ carries the dimension of the length. It can be observed from our results and discussion that fillers with non-spherical shapes are better catalyst of heat conduction compared to the spherical ones.
(b) Once we choose a specific filler, if we maximize the surface area of the filler (which is done by transforming the bulk into more surface), the thermal conductivity increases.

The current study also shed light on the molecular structure of the polymer for thermal conduction. The heat conduction in solids takes place through the phonons. Due to the molecular structure of polymer chains, phonons can not propagate to a large distatnce and hence the thermal conduction is very poor.
So the basic question is how the surface or interface maximization of the filler enhances the polymer conductivity?. This question can be understood in the perspective of the polymer structure itself. The polymer chain and thier molecular arrangement have important role in this respect. If the surface area of the metallic filler is more then the contact between the polymer molecules and the metal is more leading to large thermal excitations in the polymer molecules. When many polymer molecules are in thermal contact, they resonate and correlate locally to propagate heat upto larger distances. Molecular vibration of the chain is the key factor for the their correlation and hence deciding the enhancement. 

We also observe that the growth in effective thermal conductivity with increase in surface area shows the logaritmic behaviour. It is seen that initially the conductivity increases linearly with an increase in the interface but becomes slow subsequently and then follows the logarithmic behaviour. This behaviour should be accounted again due to the variations in the surface area and exact behaviour can be brought out by the experiments.
We have considered the different geometries for the validity of our claim. The simplest cases which have dealt are identical hollow cylinders, and thin identical planar sheets.
The current procedure can be developed more efficiently by adopting such processes which maximize the interface between the polymer and filler. One such process is mixing of nano fibers or nanosheets of the filler material in the base polymer. Another way that could be used is wrapping the base polymer by filler material in very fine tuned manner.  
We mention explicitly that the minimum thickness of the filler in the composite polymer should be such that it should not lose its macroscopic behaviour. 

\begin{table}[H]
\centering
\caption{Parallel sheets with 2\% filler by volume in composite polymer}
\begin{tabular}{ |p{0.8cm}|p{1.3cm}|p{2.1cm}|p{1cm}|p{1.3cm}|p{1.3cm} | }
 \hline
 \multicolumn{6}{|c|}{} \\ 
  \hline
 No. of Sheets & Thickness (cm) &Temprature(K) &Heat flux & Thermal conductivity (\(K_{c}\))  &Relative thermal conductivity (\(\frac{K_{c}}{K_{p}}\)) \\
 \hline
 1  &0.05 &364.39488	&0.12874	&0.36159	&1.24688 \\
\hline
 2  &0.025 &365.87238	&0.13169	 &0.38589	&1.33066 \\
 \hline
 3  &0.0166667	&366.54367	&0.13278	&0.39690	&1.36863 \\
\hline
 4  &0.0125 &367.48558	&0.13493	&0.41498	&1.43099 \\
 \hline
 6  &0.008333	&368.78974	&0.13885	&0.44490	&1.53416 \\
\hline
 8  &0.00625	&370.60429	&0.14120	&0.48034	&1.65637 \\
\hline
 12 &0.004166	&371.36618	&0.14273	&0.49848	&1.71892 \\
 \hline
 16 &0.003125	&371.57693	&0.14315	&0.50366	&1.73676  \\
 \hline
 20 &0.0025	    &371.82384	&0.14364	&0.50982	&1.75800 \\
 \hline
 30 &0.00125	&371.89195	&0.14378	&0.51154	&1.76394 \\
\hline
 40 &0.001666	&372.05995	&0.14412	&0.51582	&1.77869 \\
 \hline
\end{tabular}
\label{table:t1}
\end{table} 
\begin{table}[H]
\centering
\caption{Parallel sheet with 6\% filler by volume in composite polymer}
\begin{tabular}{ |p{0.8cm}|p{1.25cm}|p{2.1cm}|p{1cm}|p{1.3cm}|p{1.3cm} | }
 \hline
 \multicolumn{6}{|c|}{ } \\
  \hline
 No. of Sheets& Thickness (cm) &Temprature(K) &Heat flux & Thermal conductivity (\(K_{c}\))  &Relative thermal conductivity (\(\frac{K_{c}}{K_{p}}\)) \\
 \hline
 1  &0.15  &372.36 &0.14058 &0.50861 &1.75383 \\
 \hline
 2  &0.075 &376.24 &0.14858 &0.62534 &2.15633 \\
 \hline
 3  &0.05  &378.027 &0.15414 &0.70150 &2.41896 \\
 \hline
 4  &0.0375 &378.35 &0.15954 &0.73691 &2.54105 \\
\hline
 6  &0.025 &379.81 &0.15808 &0.78296 &2.69987 \\
 \hline
 8  &0.01875 &380.06 &0.15825 &0.79363 &2.73666 \\
\hline
 12 &0.0125 &381.18 &0.16062 &0.85345 &2.94294 \\
 \hline
 16 &0.009375 &381.29 &0.1608 &0.85943 &2.96356 \\
 \hline
 20 &0.0075 &381.55 &0.16142 &0.87491 &3.01691 \\
 \hline
 30 &0.005 &381.51 &0.16313 &0.88226 &3.04228 \\
 \hline
 40 &0.00375 &381.78 &0.16367 &0.89830 &3.09758 \\
 \hline
\end{tabular}
\label{table:t2}
\end{table} 
\begin{table}[H]
\centering
\caption{Parallel sheet with 9\% filler by volume in composite polymer}
\begin{tabular}{ |p{0.8cm}|p{1.25cm}|p{2.1cm}|p{1cm}|p{1.3cm}|p{1.3cm} | }
 \hline
 \multicolumn{6}{|c|}{} \\
  \hline
 No. of Sheets &Thickness (cm) &Temprature(K) &Heat flux & Thermal conductivity (\(K_{c}\))  &Relative thermal conductivity (\(\frac{K_{c}}{K_{p}}\)) \\
 \hline
 1   &0.225 &374.04 &0.1479 &0.56972 &1.96456\\
 \hline
 2   &0.1125 &377.72 &0.14883 &0.66800 &2.30344\\
 \hline
 3  &0.075 &379.24 &0.15842 &0.76310 &2.63139\\
 \hline
 4  &0.05625 &379.85 &0.15971 &0.79261 &2.73312\\
\hline
 6  &0.0375 &380.58 &0.15779 &0.81251 &2.80177 \\
 \hline
 8   &0.028125 &380.6 &0.1613 &0.83144 &2.86705 \\
\hline
 12 &0.01875 &381.69 &0.16348 &0.89285 &3.07878 \\
 \hline
 16 &0.014062 &381.70 &0.16349 &0.89288 &3.08781 \\
 \hline
 20 &0.01125 &381.8 &0.16371 &0.89951 &3.10174 \\
 \hline
 30  &0.0075 &382.06 &0.16424 &0.91550 &3.15688 \\
 \hline
 40  &0.005625 &382.08 &0.16427 &0.91669 &3.16098 \\
 \hline
\end{tabular}
\label{table:t3}
\end{table}
  \begin{table}[H]
\centering
\caption{Parallel sheets of different volume fraction in composite polymer}
\begin{tabular}{ |p{0.8cm}|p{1.25cm}|p{2.1cm}|p{1cm}|p{1.3cm}|p{1.3cm} | }
 \hline
 \multicolumn{6}{|c|}{} \\
  \hline
 No. of Sheets &Thickness (cm) &Temprature(K) &Heat flux & Thermal conductivity (\(K_{c}\))  &Relative thermal conductivity (\(\frac{K_{c}}{K_{p}}\)) \\
 \hline
20 (9\%) &0.01125   &381.8   &0.16371 &0.89951    &3.10174 \\
\hline
20 (15\%) &0.0.01875 	&382.33 	&0.16462 	&0.93164 	&3.21254 \\
\hline
20 (20\%) &0.025 	&382.74 	&0.16543 	&0.95846 	&3.30503 \\
\hline
20 (25\%) &0.03125 &384.65 	&0.16925 	&1.10261 	&3.80209 \\
 \hline
\end{tabular}
\label{table:t4} 
\end{table}   
\begin{table}[H]
\centering
\caption{Parallel hollow cylinder with 4\% filler by volume in composite polymer}
\begin{tabular}{ |p{0.8cm}|p{1.3cm}|p{2.cm}|p{1cm}|p{1.3cm}|p{1.3cm} | }
 \hline
 \multicolumn{6}{|c|}{} \\ 
  \hline
 No. of Cylinder &Thickness (cm) &Tempratur(K) &Heat flux & Thermal conductivity (\(K_{c}\)) &Relative thermal conductivity
 (\(\frac{K_{c}}{K_{p}}\)) \\
 \hline
 1  &0.111723       &364.39816  &0.12875	&0.36166	&1.24712  \\
 \hline
 2  &0.0732         &365.63782	&0.13125	&0.38197	&1.31716  \\
 \hline
 4  &0.0416	        &366.82758	&0.13361	&0.40278	&1.38892  \\
 \hline 
 6	&0.03212        &367.51928	&0.13500	&0.41563    &1.43322  \\
 \hline
 8  &0.02538	    &367.69355	&0.13654	&0.42264	&1.45740  \\
 \hline
 10 &0.02103	    &368.29438	&0.13755	&0.43384	&1.49603  \\
 \hline
 12	&0.01797	    &368.84399	&0.13764	&0.44178	&1.52339  \\
 \hline
 16	&0.01396	    &369.86072	&0.13778	&0.45715	&1.57638  \\
 \hline
 20 &0.11427        &369.98228	&0.13988	&0.46601	&1.60694  \\
 \hline
 24 &0.009675       &370.55131	&0.14186	&0.48172	&1.66113  \\
 \hline
 30 &0.007869       &371.46276	&0.14290	&0.50077	&1.72680  \\
 \hline
\end{tabular}
\label{table:t5}
\end{table}
\begin{table}[H]
\centering
\caption{Parallel hollow cylinder with 6\% filler by volume in composite polymer}
\begin{tabular}{ |p{0.8cm}|p{1.25cm}|p{2.1cm}|p{1cm}|p{1.3cm}|p{1.3cm} | }
 \hline
 \multicolumn{6}{|c|}{} \\
  \hline
 No. of Cylinder &Thickness (cm) &Temprature(K) &Heat flux & Thermal conductivity (\(K_{c}\)) &Relative thermal conductivity
 (\(\frac{K_{c}}{K_{p}}\)) \\
 \hline
 1   &0.162436   &371.68   &0.14293   &0.50470   &1.74033    \\
 \hline
 2   &0.120178   &373.91   &0.14747   &0.56524   &1.94909    \\
 
 3   &0.102283   &375.98   &0.15154   &0.63089   &2.17549    \\

 5   &0.085306   &378.18   &0.15605   &0.71517   &2.46610     \\

 10  &0.069918   &381.11   &0.16191   &0.85712   &2.95559      \\

 15  &0.063972   &382.86   &0.16556   &0.96593   &3.33079     \\
 
 17  &0.06249   &383.43   &0.16681   &1.00670   &3.47138    \\

 20  &0.06078   &383.73   &0.1674    &1.02889   &3.54789      \\
 \hline
\end{tabular}
\label{table:t6}
\end{table}
\begin{table}[ht]
\centering
\caption{Parallel hollow cylinder with 9\% filler by volume in composite polymer}
\begin{tabular}{ |p{0.8cm}|p{1.25cm}|p{1.6cm}|p{1cm}|p{1.3cm}|p{1.3cm} | }
 \hline
 \multicolumn{6}{|c|}{} \\
  \hline
 No. of Cylinder &Thickness (cm) &Temprature (K) &Heat flux & Thermal conductivity (\(K_{c}\)) &Relative thermal conductivity (\(\frac{K_{c}}{K_{p}}\)) \\
 \hline
 1    &0.195775  &374.7  &0.14894  &0.58870  &2.02999  \\
 \hline
 2    &0.142878  &376.72 &0.15332  &0.65859  &2.27100  \\
 \hline
 3    &0.120178  &378.22 &0.15617  &0.71703  &2.47253   \\
\hline
 5    &0.098314  &379.81 &0.15934  &0.78920  &2.72139   \\
\hline
 10   &0.077992  &382.31 &0.1643   &0.92877  &3.20267    \\
\hline
 15   &0.069918  &383.88 &0.16759  &1.03964  &3.58497    \\
 \hline
 17   &0.067879  &384.29 &0.16846  &1.07249  &3.70042    \\
\hline
 20   &0.065509  &384.3 &0.16858 &1.07376    &3.70261     \\
 \hline
\end{tabular}
\label{table:t7}
\end{table}
\begin{table}[H]
\begin{center}
\caption{The different correlation models used often in the literature for the calculation of effective thermal conductivity.}
\begin{tabular}{ | l | c | c | p{0.5cm} |}
\hline
Author & Correlations & Application Conditions & Ref. \\ \hline
Maxwell & \(\frac{K_{c}} {{K_{p}}} = \bigg[\frac{{{K_{f}}}+2{{K_{p}}}+2{{V_{f}}}({{K_{f}}}-{{K_{p}}})}{{{K_{f}}}+2{{K_{p}}}-2{{V_{f}}}({{K_{f}}}-{{K_{p}}})}\bigg]\) & particle filler & 47 \\
& & low content \(({V_{f}}<10\%)\)& \\ \hline
Russel & \(\frac{K_{c}} {{K_{p}}} = \bigg[\frac{{{V_{f} ^{2/3}+\frac{K_{p}} {{K_{f}}}(1-{V_{f} ^{2/3}})}}}{{V_{f} ^{2/3}-{{V_{f}}}+\frac{K_{p}} {{K_{f}}}(1-{V_{f} ^{2/3}})}}\bigg]\) & particle filler & 48  \\ 
& & low content \(({V_{f}}<10\%)\)& \\ \hline
Baschirow-& \(\frac{K_{c}} {{K_{p}}}=1-\frac{a^{2}\pi}{4}+\frac{a\pi p}{2}\bigg[1-\frac{p}{a}\ln\bigg(1+\frac{a}{p} \bigg)\bigg]\) & particle filler  & 49\\ 
Selenew & \( p=\frac{{{K_{f}}}}{{{K_{p}}}-{{K_{f}}}}\), \( a=\bigg(\frac{6{{V_{f}}}}{\pi}\bigg)^{1/3} \)& low content\(({V_{f}}<10\%)\)&\\ \hline
Nielsen & \(\frac{{K_{c}}}{{K_{p}}}=\bigg[\frac{1+AB{V_{f}}}{1-B{V_{f}}\Psi}\bigg]\) & Particle  filler, short fiber filler, & 56-57 \\
& \(B=\bigg[\frac{\frac{{K_{f}}}{{K_{p}}}-1}{\frac{{K_{f}}}{{K_{p}}}+A}\Psi\bigg]\)= \(1+\bigg(\frac{1-V_{r}}{V_{r}^{2}}\bigg)\)\(V_{f}\), \(A={K_{E}}-1\)& \({K_{E}}\) difficult to determine & \\ \hline 
Bruggeman & \(1 ={{V_{f}}} + \frac{{{K_{p}}}-{{K_{c}}}}{{{K_{f}}}-{{K_{p}}}} \bigg(\frac{{{K_{p}}}}{{{K_{c}}}}\bigg)^{1/3}\) & particle filler & 50 \\
& &low content\(({V_{f}}<10\%)\)&\\ \hline
Halpin-Tsai & \(\frac{{K_{c}}}{{K_{p}}}= \bigg[\frac{1+\xi{V_{f}}\eta}{1-\eta{V_{f}}}\bigg] \)&short fiberfiller, &\\
& \(\eta=\frac{\frac{{K_{f}}}{{K_{p}}}-1}{\frac{{K_{f}}}{{K_{p}}}}+\xi\) ,\(\xi=\sqrt{3}\lg\frac{a}{b}\) & direction transverse to fiber & 51  \\ \hline   
Springer- & \(\frac{{K_{c}}}{{K_{p}}}= \bigg[1-2 \sqrt{\frac{V}{\pi}}+ \frac{1}{B}\bigg(\pi-\frac{4}{\sqrt{1}-(\frac{B^{2}V}{\pi})}\bigg)A\bigg]\) & short fiber filler, & 52\\
Tsai &\(\frac{B}{2}= \bigg(\frac{{K_{p}}}{{K_{c}}}-1 \bigg)\) & direction perpendicular to fiber & \\  \hline
Agari.Y   & \(\lg {K_{c}}={V_{f}}{C_{2}}\lg {K_{f}}+(1-{V_{f}})\lg ({C_{1}}{K_{p}})\) & Particle filler, & 53-55 \\ 
& & C difficult to determine & \\ \hline
\end{tabular}
\label{table:t8}
\end{center}
\end{table}
\newpage





\end{document}